\pdfoutput=1
\documentclass[pdftex,useAMS,usenatbib,usegraphicx,dvipdfmx]{pasj01}
\usepackage{lscape}
\usepackage{color}
\usepackage{comment}
\usepackage{natbib, aas_macros}
\newcommand{\simgt}{\lower.5ex\hbox{$\; \buildrel > \over \sim \;$}}
\newcommand{\simlt}{\lower.5ex\hbox{$\; \buildrel < \over \sim \;$}}

\def\h70kpc{\mathrel{h_{70}^{-1}{\rm kpc}}}

\def\Msol{\mathrel{M_\odot}}
\def\hMsol{\mathrel{h^{-1}M_\odot}}
\def\h70Msol{\mathrel{h_{70}^{-1}M_\odot}}

\usepackage{version}	
\begin{document} 
\Received{}
\Accepted{}


\title{Halo Concentration, Galaxy Red Fraction, and Gas
Properties of Optically-defined Merging Clusters
\thanks{Based on data collected at Subaru Telescope, which is operated
by the National Astronomical Observatory of Japan.}}

\author{Nobuhiro \textsc{Okabe}\altaffilmark{1,2,3}}
\altaffiltext{1}{Department of Physical Science, Hiroshima University,
1-3-1 Kagamiyama, Higashi-Hiroshima, Hiroshima 739-8526, Japan}
\altaffiltext{2}{Hiroshima Astrophysical Science Center, Hiroshima University, 1-3-1 Kagamiyama, Higashi-Hiroshima, Hiroshima 739-8526, Japan}
\altaffiltext{3}{Core Research for Energetic Universe, Hiroshima University, 1-3-1, Kagamiyama, Higashi-Hiroshima, Hiroshima 739-8526, Japan}

\email{okabe@hiroshima-u.ac.jp}

\author{Masamune \textsc{Oguri}\altaffilmark{4,5,6}}
\altaffiltext{4}{Research Center for the Early Universe, University of Tokyo, Tokyo 113-0033, Japan}
\altaffiltext{5}{Department of Physics, University of Tokyo, Tokyo 113-0033, Japan}
\altaffiltext{6}{Kavli Institute for the Physics and Mathematics of the Universe (Kavli IPMU, WPI), University
of Tokyo, Chiba 277-8582, Japan}

\author{Hiroki \textsc{Akamatsu}\altaffilmark{7}}
\altaffiltext{7}{SRON Netherlands Institute for Space Research, Sorbonnelaan 2, 3584 CA Utrecht, The Netherlands}

\author{Akinari \textsc{Hamabata}\altaffilmark{5}}

\author{Atsushi J. \textsc{Nishizawa}\altaffilmark{8}}
\altaffiltext{8}{Institute for Advanced Research, Nagoya University Furocho, Chikusa-ku, Nagoya, 464-8602
Japan}

\author{Elinor \textsc{Medezinski}\altaffilmark{9}}
\altaffiltext{9}{Department of Astrophysical Sciences, Princeton University, Princeton, NJ 08544, USA}

\author{Yusei \textsc{Koyama}\altaffilmark{10,11}}
\altaffiltext{10}{Subaru Telescope, National Astronomical Observatory of Japan, National Institutes of Natural Sciences, 650 North A'ohoku Place, Hilo, HI 96720, USA}
\altaffiltext{11}{Graduate University for Advanced Studies (SOKENDAI), Osawa 2-21-1, Mitaka, Tokyo
181-8588, Japan}

\author{Masao \textsc{Hayashi}\altaffilmark{12}}
\altaffiltext{12}{National Astronomical Observatory of Japan, Osawa 2-21-1, Mitaka, Tokyo 181-8588, Japan}

\author{Taizo \textsc{Okabe}\altaffilmark{5}}

\author{Shutaro \textsc{Ueda}\altaffilmark{13}}
\altaffiltext{13}{Institute of Astronomy and Astrophysics, Academia
Sinica, P.O. Box 23-141, Taipei 10617, Taiwan}

\author{Ikuyuki \textsc{Mitsuishi}\altaffilmark{14}}
\altaffiltext{14}{Department of Physics, Nagoya University, Aichi 464-8602, Japan}

\author{Naomi \textsc{Ota}\altaffilmark{15}}
\altaffiltext{15}{Department of Physics, Nara Women's University, Kitauoyanishi-machi, Nara, Nara 630-8506, Japan}




\maketitle

\begin{abstract}
 We present multi-wavelength studies of optically-defined merging
 clusters, based on the Hyper Suprime-Cam Subaru Strategic Program.
 Luminous red galaxies, tracing cluster mass distributions,
 enable to identify cluster subhalos at various merging stages,
 and thus make a homogeneous sample of cluster mergers, which is
 unbiased with respect to the merger boost of the intracluster medium (ICM).
  We define, using a peak-finding method, merging clusters with
 multiple-peaks and single clusters with single-peaks from the CAMIRA cluster catalog.  
 Stacked weak-lensing analysis indicates that our sample of the
 merging clusters is categorized into major mergers.
  The average halo concentration for the merging clusters is $\sim70\%$ smaller than
 that of the single-peak clusters, which agrees well with predictions of numerical simulations.
The spatial distribution of subhalos is less centrally concentrated than the mass distribution of the
 main halo. 
 The fractions of red galaxies in the merging clusters are not higher than
 those of the single-peak clusters. 
 We find a signature of the merger boost of the ICM from stacked {\it Planck} Sunyaev-Zeldovich
 effect and {\it ROSAT} X-ray luminosity, but not in optical richness. 
 The stacked X-ray surface brightness distribution, aligned with the
 main-subhalo pairs of low redshift and massive clusters, shows that the
 central gas core is elongated along the merger axis and overall gas
 distribution is misaligned by $\sim60$ deg.
The homogeneous, unbiased sample of cluster mergers
 and multi-wavelength follow-up studies provide a unique opportunity to make a complete picture of merger physics over the whole process.
 \end{abstract}

\section{Introduction}

Based on hierarchical structure formation scenario of the Cold
Dark Matter (CDM) paradigm, small structures form first, which grow into more
massive objects through matter accretion along their
surrounding filamentary structure. Galaxy clusters sometimes merge with
less massive clusters, groups, and galaxies.
Minor mergers with a mass ratio $<0.1$ are ubiquitous,
whereas major mergers with similar-mass are an intermittent event. 
Even so, various gas phenomena are induced by major mergers, specifically
merger shocks and turbulence, which gives a significant
impact on both the thermal history and cluster evolution
\citep[e.g.][]{Takizawa00,Ricker01,Sarazin02,ZuHone11}. 
The merging process can be broadly divided into three phases;
pre-merger (or early phase), on-going, and post-merger.
The early phase of cluster mergers is the phase before the two cluster cores
start to interact with each other. No significant effect on the
intracluster medium (ICM) is appeared. 
On-going merger is the phase when the two cores are passing through.
The close interaction triggers merger shocks, resulting in heating the
temperature of the ICM and increasing the X-ray luminosity of clusters.  
The increase of the temperature and X-ray luminosity last on about half
of the sound crossing time, less than a few Gyrs \citep[e.g.][]{Ricker01}. 
The feature is well-known as ``{\it merger boost}'', giving rise to a
significant effect on cluster cosmology \citep[e.g.][]{Krause12,Kay12,Yu15}.
At the post merger phase after cluster passages, gas temperature and X-ray
luminosity decrease because of adiabatic expansion.
In some cases, they become below the initial quantities and regarded as
X-ray underluminous clusters.

During cluster mergers, the physical state of merging clusters is instantly
and dramatically deviated from virialization. 
Furthermore, the merger-induced shocks and turbulence generate relativistic particles, and diffuse radio emission,
so-called radio relics and halos, are observed in various merging clusters
\citep[e.g.][]{Govoni04,vanWeeren10,Feretti12},
although the physics of particle acceleration has not yet been fully
understood \citep[e.g.][]{Fujita03,Brunetti08,Fujita15,vanWeeren17,Akamatsu17}.
It is therefore of critical importance to understand the microphysical
processes underlying cluster mergers and these impacts on cosmological applications.

Our understanding of gas physics in merging clusters, by X-ray
and radio observations, is limited in the context of the whole process of cluster mergers.
This is because most of observing targets were chosen from X-ray luminous clusters. 
Since X-ray luminosity is non-linearly increased at on-going phase, 
we now face to a serious issue of a selection bias toward on-going phase.
Indeed, most of well-known merging clusters are at the on-going phase,
and X-ray observations of early-/late-phase merging clusters are very limited
\citep[e.g.][]{Belsole04,Owers13,Akamatsu16}.
Hence, a next study of merging clusters is demanded to add new insights
into the pool of knowledge on cluster mergers over the whole process
including the pre- and post-merger phase.
For this purpose, it is of vital importance to define a homogeneous merging cluster
sample with a selection function independent of the merger boost of the ICM.

The total mass distribution of galaxy clusters is also affected by cluster mergers. 
For instance. the concentration parameter, which represents the degree of the concentration
of the interior mass density, becomes lower due to the presence of
massive subhalos at some distance \citep[e.g.][]{Neto07,Duffy08,
Bhattacharya13,Child18}.
Recent weak gravitational lensing studies
\citep[e.g.][]{Oguri12,Okabe13,Umetsu14,Merten15,Okabe16b,Cibirka17,Miyazaki18}
revealed that the average concentration parameter as a function of
cluster masses is in
an excellent agreement with predictions of numerical simulations 
\citep[e.g.][]{Bhattacharya13,Meneghetti14,Ludlow14,Diemer15,Child18,Ragagnin18}.
The relation between mass and concentration holds for X-ray selected and
shear-selected clusters.
However, it has not yet been explored by weak-lensing studies
how much the concentration parameter is changed by cluster mergers,
because the number of merging clusters was insufficient for
detailed studies. Therefore, we need a large sample of merging clusters for weak-lensing analysis.

Several theoretical work discussed how the star formation activity is enhanced
by cluster mergers
\citep[e.g.][]{Dressler83,Lavery88,Evrard91,Fujita99b,Bekki99,Kapferer06,Kronberger08}. 
One of the key processes is that a tenuous, hot gas in infalling galaxies is
compressed by the ram-pressure or shock fronts of the ICM, and triggers
a one-time star-burst. Subsequently, continuous gas stripping in
galaxies quenches the star formation during cluster collisions.
Since the duration of the star-burst is order of Gyr or less, 
it is difficult to witness the overall, rapid evolution of star
formation and quenching activities. 
Recent spectroscopic observations of on-going mergers at $z\simlt0.4$
\citep[e.g.][]{Stroe15b,Stroe17,Deshev17} have shown both high and low
star formation rates. 
A large sample of clusters with various merging stages provides a
straightforward means of overcoming the relatively large intrinsic scatter of
star formation rates.

In short, a construction of a large sample of merging clusters
comprising various merging stages provides us with a unique opportunity
for follow-up multi-wavelength observations,
which will open-up a statistical approach to conduct a more
comprehensive study of cluster mergers from the early to late phase.
The advantages of the approach are to    
\begin{itemize}
 \item fill in the parameter spaces of cluster mergers, such as merging
       stages and mass ratios, 
\item  characterize the impact of merging phenomena on the cluster
       evolution via a phase-separated ``anatomy'',
\item  control the systematics of outliers in scaling relations and
       construct better mass proxy calibration for cluster cosmology study,       
\item  take a snapshot of various stages of
       transient phenomena such as shock heating, adiabatic
       expansion, star formation, quenching process, and particle acceleration.
\end{itemize}
This kind catalog would be timely in incoming new era of cluster studies based on big data.

The galaxy distribution contains a unique and ideal information to construct a
homogeneous, unbiased sample of cluster mergers, thanks to the following
five reasons.
First, the number of luminous red galaxies is almost
conserved during cluster mergers, in contrast to the ICM merger boost.
Second, spectroscopic observations of all member galaxies in thousands
clusters are not realistic, and photometric information is more efficient
to search cluster subhalos.
Third, the lifetime of galaxy substructure is much larger than that of
the collisional gas \citep[e.g.][]{Tormen04}. Thus, galaxy subhalos survive even after the core passage
and can be easily identified at any merging phases.
Forth, the angular resolution of weak-lensing mass reconstruction is too
poor to search cluster substructures in normal and small clusters at
wide redshift ranges at $0.1<z<1.1$ \citep[e.g.][]{Okabe14a}. Furthermore, it suffers from misidentification of background and/or
foreground objects accidentally superposed with the targeting clusters.
Fifth, as reported in previous studies \citep[e.g.][]{Okabe08}, the galaxy
distribution of luminous red galaxies is very similar to the dark matter
distribution of targeting clusters regardless of dynamical states.
Thus the galaxy distribution provides a powerful means of identifying cluster mergers.
There are, however, two disadvantages of this approach.
First, from the photometric information alone it is hard to distinguish between internal
subhalos and surrounding halos because of the degeneracy of peculiar
velocity and physical separation along the line-of-sight.
Second, the galaxy distribution cannot identify dynamical states of
cluster mergers, that is, the state of being either before or after the
core passage. Therefore, multi-wavelength follow-up studies,
especially for individual clusters, are
essential for understanding physical properties of defined merging subsamples.
The morphology of the ICM  which is a compressible fluid
provides us with helpful information to distinguish between pre- and
post- mergers.

In this paper, we present multi-wavelength studies for optically-defined merging
clusters, based on the Hyper Suprime-Cam Subaru Strategic Program
\citep[HSC-SSP;][]{HSC1stDR,HSC1styrOverview,Miyazaki18HSC,Komiyama18HSC,Kawanomoto18HSC,Furusawa18HSC,Bosch18HSC,Haung18HSC,Coupon18HSC}.
The HSC-Survey is an ongoing wide-field imaging survey
using the HSC \citep{Miyazaki15,HSCcam} which is a new prime
focus camera on the 8.2m-aperture Subaru Telescope. 
The HSC-SSP survey is composed of three layers of different depths (Wide, Deep and UltraDeep). 
The Wide layer is designed to obtain five-band ($grizy$) imaging over $1400$~deg$^2$.
The HSC-SSP survey has both excellent imaging quality ($\sim$$0.''7$
seeing in $i$-band) and deep observations ($r\simlt26$~AB~mag).
Recently \cite{Oguri18} constructed a CAMIRA cluster catalog from HSC-SSP S16A
dataset covering $\sim 240$~deg$^2$ using the CAMIRA algorithm \citep{Oguri14b} which is a
red-sequence cluster finder based on the stellar population synthesis model fitting.
The catalog contains $\sim 1900 $ clusters at $0.1<z<1.1$ with richness larger than $N_{\rm cor}=15$.
Photometric redshifts of the clusters are shown to be accurate at
$\sigma_z/(1+z)\sim 0.01$ for the whole redshift range,
which are estimated by comparing the photometric redshifts to
spectroscopic redshifts of brightest cluster galaxies (BCGs).

We construct a catalog of merging cluster candidates from the CAMIRA cluster catalogue
\citep{Oguri18}. We carry out multi-wavelength studies for the catalog,
compiling multi-band dataset from the HSC-SSP photometry
\citep{HSC1stDR,HSCPhotoz17} and weak-lensing \citep{HSCWL1styr}, Sloan
Digital Sky Survey (SDSS) spectroscopic data \citep{Abolfathi18}, 
 {\it ROSAT} X-ray \citep{Truemper82}, {\it Planck} Sunyaev-Zeldovich effect \citep[SZE;][]{Planck16ymap}, the NRAO VLA Sky Survey
\citep[NVSS;][]{Condon98NVSS}, and the TIFR GMRT
Sky Survey \citep[TGSS;][]{Intema17TGSS}.
As a first demonstration of the power of optically-defined subsamples, we in
this paper focus on stacking analyses of multi-wavelength dataset.
We describe our optical-selected merging clusters and multi-wavelength
analyses in Sec. \ref{sec:data}.
The main results and discussion are presented in Sec. \ref{sec:result}.
We summarize our results in Sec. \ref{sec:sum}.
Throughout the paper we assume a flat $\Lambda$CDM cosmology with $H_0=70~{\rm 
km~s^{-1}Mpc^{-1}}$, $\Omega_{m,0}=0.28$ and $\Omega_\Lambda=0.72$.

\section{Data Analysis} \label{sec:data}

\subsection{Definition of subsamples}\label{subsec:sample}

We identify, using a peak-finding method, merging cluster candidates from the CAMIRA cluster catalogue
\citep{Oguri18} constructed from the HSC-SSP Survey \citep{HSC1stDR,HSC1styrOverview,HSCcam}.
We use the HSC-SSP S16A data \citep{HSC1stDR,HSCPhotoz17,HSCWL1styr}.

We select red-sequence galaxies in color-magnitude plane following \cite{Nishizawa18}.
The band combinations are $g-r$, $r-i$, and $i-y$
in the redshift ranges of $0.1 < z < 0.4$, $0.4 < z < 0.7$,
and $0.7 < z< 1.1$, in order to cover the $4000{\rm \AA}$ break at these
redshifts, respectively.
The median and width of colors of red-sequence galaxies are the same as \cite{Nishizawa18}.
We use red-sequence galaxies of which apparent $z$-band magnitudes are brighter
than the observer-frame magnitude with the constant $z$-band absolute
magnitude $M_z=-18$ ABmag. We also apply the apparent magnitude cut $m_z<25.5$.
We here adopt K-correction in conversion between
apparent and absolute magnitudes taking into account passive evolution \citep{Nishizawa18}.
We then construct Gaussian smoothed maps (FWHM$=200h_{70}^{-1}$ kpc) of
number densities of red-sequence galaxies within each box size of
$1.5r_{\rm 200m}$ 
centered on the CAMIRA positions. In order to avoid photometric outliers
in the survey catalog, we do not use luminosity densities which can be significantly affected by these outliers. 
Here, $r_{\rm 200m}$ denotes the radius within which the mean density is $200$ times the mean mass density of
the Universe at the cluster redshift. To estimate $r_{\rm 200m}$, We assume a scaling relation
between the richness, $N_{\rm cor}$, and the mass,
$M_{\rm 200m}=10^{14}h^{-1}(N_{\rm cor}/15)M_\odot$, because
\citet{Oguri18} have found that richness $N_{\rm cor}=15$ roughly
corresponds to $M_{\rm 200m}=10^{14}h^{-1}M_\odot$ by a comparison
between the observed cluster abundance and the expected halo of \citet{Tinker08}.
The $r_{\rm 200m}$ spans from $\sim 1h_{70}^{-1}$ Mpc to
$\sim 2h_{70}^{-1}$ Mpc. Therefore, with the assumed FWHM of the
Gaussian smoothing we can resolve galaxy
concentrations with one-tenth or one-twentieth of cluster sizes.

We define the threshold of peak-identifications in the smoothed number
density map based on solely by observational
data, because both the number of red galaxies and angular sizes of smoothing
scale depend on cluster redshifts.
It also has the advantage of avoiding the assumption of baryonic physics
input in numerical simulations.
We use an average of stacked galaxy map at each
cluster redshift as the threshold of peak heights.
The average peak heights appeared in the stacked maps
at a slice of $\Delta z=0.05$ smoothly change as a function of
redshifts, and correspond to 5 galaxies at $z\simlt0.8$. 
We interpolate the average peak heights as a function of cluster redshifts.

Our peak-finding algorithm is composed of three steps. We first identify
peaks above the redshift-dependent threshold. Second, we subtract
a component of extended galaxy distributions around the highest peak
from the sub-peaks, in order to mis-select sub-peaks
which have peak-heights contaminated by the smoothed distribution of the main halo.
We here assume that the extended galaxy distribution is the same as
the average distribution appeared in the stacked galaxy maps.
Third, if there is the third highest peak, we also subtract contamination of higher peaks.
The peak findings are limited to within $r=1.2r_{\rm 200m}$ from the CAMIRA center.
With this procedure we identified 1561 single-peak clusters, 175 two-peaks clusters and 14 three-peaks clusters.
The total number of the clusters is slightly smaller than the total number of
CAMIRA clusters, because imaging data for our subhalo search are missing in some clusters.
Since we cannot resolve internal structures within the FWHM, there may
be closed-pair clusters whose separations are smaller
than the FWHM in the single-peak cluster catalog, although they are not the
majority considering its low chance probability.
Furthermore, since all real clusters have internal structures, 
the single-peak clusters potentially contain less massive subhalos
whose peak heights are under the detection limit. 
We therefore refer to them as ``{\it single}'' clusters rather than
``relaxed'' clusters. 
We also refer to two- and three- peaks clusters as ``{\it merging}''
clusters for simplicity.
This method has a problem of the projection effect that the
distributions of the main and sub clusters along the line-of-sight is
highly degenerate. 
Therefore, the galaxy distribution alone cannot distinguish
whether the main clusters and the subhalos are really interacting or not.
The purity of the subhalos is discussed in Secs \ref{subsec:Sigma_n} and \ref{subsec:v-s}.

\begin{figure*}
\begin{center}
\includegraphics[width=\hsize]{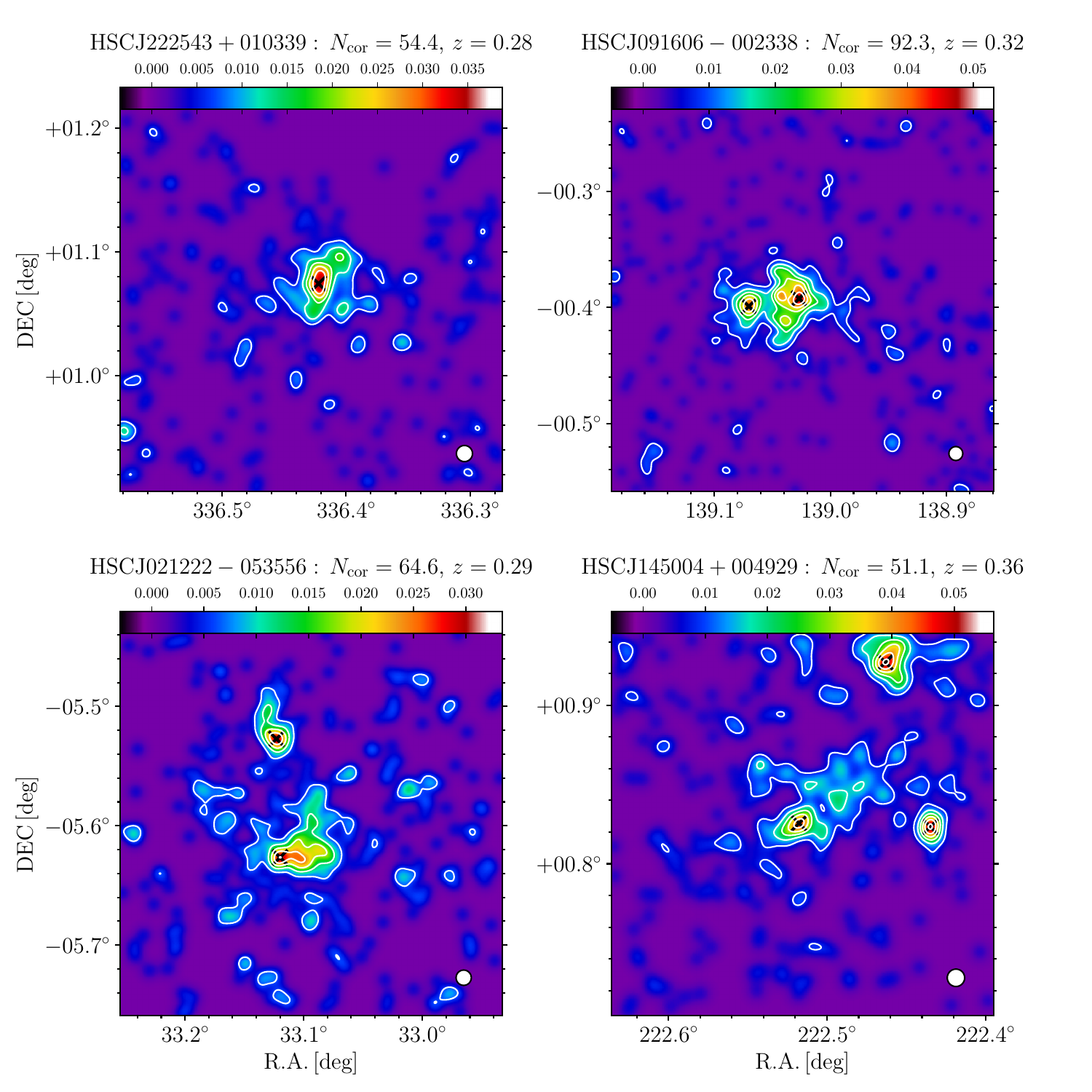}
\end{center}
 \caption{Examples of galaxy maps centered the CAMIRA positions. 
Smoothing scale, FWHM=$200h_{70}^{-1}$ kpc, are shown as white
 circles in the lower-right corners.
 Overlaid are contours corresponding to the number of galaxies,
 starting from 2 galaxies per pixel and stepping by a galaxy per pixel
 at each redshift. Black crosses denote galaxy peak positions.
 The color bar denotes the number of galaxies per a pixel. 
{\it Top-left};
 the single cluster, HSCJ222543+010339. {\it Top-right}; the merging
 cluster, HSCJ091606-002338, as an example of peak-separations of
 $0.2h_{70}^{-1}~{\rm kpc}<d_{\rm sh}<0.6h_{70}^{-1}~{\rm kpc}$. Here,
 $d_{\rm sh}$ is the projected distance of galaxy subhalo from the highest peak.
 {\it
 Bottom-left}; the merging cluster, HSCJ021222-053556, as an example of peak-separations of
 $0.6h_{70}^{-1}~{\rm kpc}<d_{\rm sh}<1.2h_{70}^{-1}~{\rm kpc}$. {\it
 Bottom-right}; the merging-cluster, HSCJ145004+004929, as an example of peak-separations of
 $1.2h_{70}^{-1}~{\rm kpc}<d_{\rm sh}<2.4h_{70}^{-1}~{\rm kpc}$.}
 \label{fig:galmap}
\end{figure*}

Given the subsamples, we carry out galaxy, weak-lensing, SZE, X-ray,
and radio analyses to statistically understand their physical properties.
Throughout the paper, all the observables will be
discussed using the overdensity radius $r_{200}\sim0.7r_{\rm 200m}$,
where $r_{200}$ is the radius at which the mean enclosed density is
$200$ times the critical mass density of the universe at a cluster redshift.
Fig. \ref{fig:galmap} shows typical examples of galaxy maps of single
and merging clusters at $z\sim0.3$.

\subsection{Weak-lensing Analysis} \label{sec:WL}

We measure average weak-lensing masses of the merging and single clusters
using a method of Point Spread Function (PSF) correction
known as re-Gaussianization \citep{Hirata03}, which is implemented in 
the HSC pipeline \citep[see details in][]{HSCWL1styr}. 
For the weak-lensing analysis, we use only galaxies satisfying with
full-color and full-depth condition from the HSC galaxy catalog 
to achieve both precise shape measurement and photometric redshift
estimation. Therefore, the weak-lensing analysis is carried out
subsets of the full sample of CAMIRA clusters. 
We select background galaxies behind each cluster in the four-bands magnitude ($griz$) plane, 
following \cite{Medezinski18}.

In order to measure an average mass distribution over a subsample, we
 compute stacked lensing signals. For this purpose, 
we measure the reduced tangential shear $\langle \Delta \Sigma_{+}\rangle$
computed by azimuthally averaging the measured tangential ellipticity,
$e_+$, for a pair of $i$-th galaxy and $j$-th cluster \citep{Mandelbaum06,Johnston07}, 
\begin{eqnarray}
\langle \Delta \Sigma_{+}\rangle (r_k) = 
\frac{\sum_{ij} e_{+,i} w_{i,j} \langle \Sigma_{{\rm cr}}(z_{l,j}, z_{s,i})^{-1}\rangle^{-1}}{2 R (1+K) \sum_{ij} w_{i,j}}, \label{eq:g+}
\end{eqnarray}
where tangential ellipticity is
\begin{eqnarray}
e_{+}&=&-(e_{1}\cos2\varphi+e_{2}\sin2\varphi).
\label{eq:gt}
\end{eqnarray}
The radius positions, $r_k$, are computed by the weighted harmonic mean
\citep{Okabe16b} of radial distances from peak positions of the number density maps. 
The inverse of the mean critical surface mass density for individual pairs between $i$-th galaxy and $j$-th cluster is computed by the probability function $P(z)$ 
from the machine learning method \citep[MLZ;][]{MLZ14} calibrated
with spectroscopic data \citep{HSCPhotoz17},
\begin{eqnarray}
 \langle \Sigma_{{\rm cr}}(z_{l,j},z_{s,i})^{-1}\rangle =
  \frac{\int^\infty_{z_{l,j}}\Sigma_{{\rm cr}}^{-1}(z_{l,j},z_{s,i})P(z_{s,i})dz_{s,i}}{\int^\infty_{0}P(z_{s,i})dz_{s,i}}.
\end{eqnarray}
Here, $z_{l,j}$ and $z_{s,i}$ are the cluster and source redshift, respectively.
The critical surface mass density is $\Sigma_{{\rm cr}}=
c^2D_{s}/4\pi G D_{l} D_{ls}$, where $D_s$ and $D_{ls}$ are the
angular diameter distances from the observer to the sources and from the
lens to the sources, respectively.  
The weighting function is given by
\begin{eqnarray}
w_{i,j}=\frac{1}{e_{\rm rms,i}^2+\sigma_{e,i}^2}\langle \Sigma_{{\rm cr}}(z_{l,j},z_{s,i})^{-1}\rangle^2\label{eq:weight}.
\end{eqnarray}
where $e_{\rm rms}$ and $\sigma_e$ are the root mean square of
intrinsic ellipticity and the measurement error per component
($\alpha=1$ or $2$), respectively.
The shear responsivity, $R$, and the calibration factor, $K$, are
computed by $R=1-\sum_{ij} w_{i,j} e_{{\rm rms},i}^2/\sum_{ij} w_{i,j}$ and
$K=\sum_{ij} m_i w_{i,j}/\sum_{ij} w_{i,j}$, respectively. Here $m_i$ is
the multiplicative shear calibration factor for individual objects \citep{Mandelbaum17,HSCWL1styr,Miyatake18}.
We also conservatively subtract an additional, negligible offset term for
calibration factor \citep{Miyaoka18}.

In order to investigate the dynamical dependence of the average halo concentration, 
we use an NFW profile \citep{NFW96} for model fitting.
The NFW profile is expressed in the form:
\begin{equation}
\rho_{\rm NFW}(r)=\frac{\rho_s}{(r/r_s)(1+r/r_s)^2},
\label{eq:rho_nfw}
\end{equation}
where $\rho_s$ is the central density parameter and $r_s$ is the scale
radius.
We measure weak-lensing masses $M_{200}$ rather than
$M_{200\rm m}$.
The spherical mass and the halo concentration are defined by
$M_{200}=4\pi 200 \rho_{\rm cr} r_{200}^3/3$ and
$c_{200}=r_{200}/r_s$, respectively.

The reduced shear model, $f_{\rm model}$, is expressed in terms of the differential surface mass density
$\Delta \tilde{\Sigma}$ and the local surface mass density $\Sigma$ as follows,
\begin{eqnarray}
f_{\rm model}=\Delta \tilde{\Sigma}(1+\mathcal{L}_z \Sigma),
\end{eqnarray}
where $\mathcal{L}_z=\sum_{ij} \langle \Sigma_{{\rm cr},ij}^{-1}\rangle w_{i,j}/\sum_{ij} w_{i,j}$.
To keep the linearity for the ensemble average, 
we use $(1+\mathcal{L}_z \Sigma)$ as the non-linear correction term instead of $(1-\mathcal{L}_z \Sigma)^{-1}$.
Given the mass model, the log-likelihood is described by
\begin{eqnarray}
-2\ln {\mathcal L}&=&\ln(\det(C_{nm})) +  \label{eq:likelihood} \\
 &&\sum_{n,m}(\Delta \Sigma_{+,n} - f_{{\rm model}}(r_n))C_{nm}^{-1} (\Delta
 \Sigma_{+,m} - f_{{\rm model}}(r_m)), \nonumber
\end{eqnarray}
where the covariance matrix, $C$, is composed of the uncorrelated large-scale structure (LSS) $C_{\rm LSS}$, along the line-of-sight \citep{Schneider98} and the shape noise $C_g$.
In model fitting, we adopt the Markov chain Monte Carlo (MCMC) method
with flat priors of $\log 10^{-5}<\log
\left(M_{200}/10^{14}\h70Msol\right)< \log 50$ and $\log 10^{-5}<\log\left(c_{200}\right)< \log 30$.
Although all quantities are the average values over the subsamples, we express them without
the bracket, $\langle\rangle$, in this paper.

We also implement the lensing signal from the merging subhalos in the
modeling in order to estimate their average masses, $M_{200}^{\rm (sh)}$.
We fix the positions of the subhalos and compute the lensing signal from the
cluster center using an off-centering effect \citep{Yang06}. 
In computations of the lensing model from the subhalos,
there are two important considerations. 
First, all the optically-identified clusters do not always fulfill 
the full-color and full-depth condition for weak-lensing analysis in
the HSC-SSP survey. Second,
the number of background galaxies and its lensing
efficiency are different for different clusters. In order to consider the
lensing contributions in the stacked lensing signal, 
we introduce a weighting function of each subhalos associated with the
$j$-th cluster, specified by
\begin{eqnarray}
 W_j=\frac{\sum_i w_{i,j} }{\sum_{ij} w_{i,j}}, \label{eq:W_cl}
\end{eqnarray}
where $w_{i,j}$ is defined by equation (\ref{eq:weight}).
We mention that the weighting function is also considered in
estimation of other observables (eqs. \ref{eq:y} and \ref{eq:N}; Secs
\ref{subsec:Psz}, \ref{subsec:Lx} and \ref{subsec:Ncor}) for a
comparison of weak-lensing masses.
Given radial distances of the subhalo, $r_{{\rm sh},j}$, in the $j$-th
cluster, the surface mass density for the subhalos is computed by
\begin{eqnarray}
 \Sigma_{\rm NFW}^{\rm (sh)}(r)&=& \frac{1}{2\pi \sum_j W_j} \nonumber \\
 &&\sum_j \int^{2\pi}_0 d\theta  W_j \Sigma(\sqrt{r^2+r_{{\rm sh},j}^2-2rr_{{\rm sh,j}}\cos \theta}). \label{eq:offsetmodel}
\end{eqnarray}
By integrating the above equation from $0$ to $r$,
we calculate the averaged mass density, $\bar{\Sigma}_{\rm NFW}^{\rm
(sh)}(<r)$, and subsequently obtain the reduced shear $f_{\rm NFW}^{\rm (sh)}$.
It is very difficult to constrain the concentration parameter for the
merging subhalos because of a degeneracy with the lensing signal from
the main cluster.
We therefore fix $c_{200}$ for the subhalos based on the best-fit concentration parameter
for the single clusters and adopt a single parameter $M_{200}^{\rm (sh)}$ for the subhalos.
The model for the total mass components can be expressed by 
\begin{eqnarray}
f_{\rm model} = f_{\rm NFW}+f_{\rm NFW}^{\rm (sh)}. \label{eq:fWLmodel}
\end{eqnarray}
We carry out weak-lensing analysis in physical unit rather than comoving
unit in order to avoid a $(1+z)$ stretch of subhalo distributions found in
physical unit and contamination of unknown structures in lensing profiles
as much as possible. Since all other observables that are compared with lensing masses
correctly take account of the lensing weight (eq. \ref{eq:W_cl}), the
unit in weak-lensing measurements does not matter in this paper.


\subsection{Red fraction} \label{subsec:fred_ana}

We estimate the fraction of red cluster member galaxies to all cluster
member galaxies (the red fraction) of each cluster using MLZ photometric
redshifts of galaxies \citep{HSCPhotoz17}.
We define member galaxies within 1Mpc of each cluster center
so that 95\% confidence interval of the photometric redshifts falls in
the cluster photometric redshift obtained by the CAMIRA catalog.
In our analysis, we use only member galaxies
whose $z$-band magnitudes are brighter than that of
red sequence galaxies with stellar mass $M_*=10^{10.45}M_\odot$,
computed by the CAMIRA algorithm. This magnitude cut
enables us to consistently determine the red
fraction over all redshift range $0.1<z<1.1$.
We divide member galaxies into red and blue
cluster members based on the HSC photometry using the criteria shown
in Fig. 2 of \citet{Oguri18}. In order to subtract the foreground
and background contamination, for each cluster we estimate the average
number densities of red and blue galaxies at the cluster redshift from
the numbers of red and blue galaxies in the annulus of
1--3$h_{70}^{-1}$ Mpc around the cluster center, and use them to subtract
the foreground and background contamination for estimating the red fraction
(i.e., the number ratio of red to total member galaxies after the
subtraction) of each cluster.
We stress that the peak-finding method and its galaxy selection in our definition of the merging
clusters are different from the estimation of the red fractions,
and thus reasonably assume that there is no correlation between the red
fractions and the subsample definitions. 
The result is shown in Sec. \ref{subsec:fred_result}.

\subsection{Stacked {\it Planck} SZE analysis} \label{subsec:Psz}

We use the NILC \citep[Needlet Internal Linear
Combination;][]{Remazeilles13} and MILCA \citep[Modified Internal Linear
Combination Algorithm;][]{Hurier13} 
all-sky Compton parameter maps ($y$-map, hereafter) released by \citet{Planck16ymap}. 
This $y$-map was constructed from the multiple Planck frequency channel
maps, by characterizing in terms of noise properties and residual
foreground contamination.
We apply 60\% Galactic and point-source masks provided by the Planck
Collaboration to reduce contamination from galactic emission and point sources.
Both the $y$-map and mask are provided in the HEALPix format \citep{HEALPix05}. 

We compute a noise-weighted $y$-map \citep{Planck11SZE} around the highest peaks in physical scale unit.
In the stacking procedure, we additionally use a weight function of
lensing contributions (eq \ref{eq:W_cl})
in order to make a fair comparison between SZE observables and weak-lensing masses, $M_{200}$.
This is because the full-color and full-depth condition is not always satisfied with all
the subsamples of the CAMIRA clusters \citep[see][for details]{Oguri18} and the lensing contributions of
individual clusters in the average mass measurements are different (Sec \ref{sec:WL}).
Given the $y$-map ($y$) and the standard deviation map ($\sigma_y$),
we project each $y$-map of the subsample onto a two dimensional rectangular
grid using a nearest neighbour interpolation, 
\begin{eqnarray}
 \langle y \rangle_i=\frac{\sum_{j} y_{i,j} \sigma_y^{-2} W_j}{\sum_{j}
  \sigma_y^{-2}W_j}, \label{eq:y}
\end{eqnarray}
where the subscript, $j$, denotes the cluster.
Since a $y$-map resolution is poor with the circular Gaussian beam of 
FWHM$=10$ arcmin, $Y_{\rm SZ}$ are measured within a cylindrical circle
$2r_{200}$ centered at peak positions weighted with the stacked $\langle y\rangle$.
The measurement errors of $Y_{\rm SZ}$ include the statistical noise,
bootstrap sampling errors and aperture sizes caused by weak-lensing
measurement uncertainty. We represent the result and discussion in
the Sec \ref{subsec:Ysz}.

\subsection{Stacked RASS analysis} \label{subsec:Lx}

We stack the X-ray emission from the position-sensitive proportional
counter (PSPC) of ROSAT all-sky survey (RASS) in
the 0.1−2.4 keV band \citep{Truemper82}, in a similar manner to the
stacked {\it Planck} SZ images (Sec. \ref{subsec:Psz}).
We exclude regions of point sources, from the second ROSAT all-sky survey source
catalogue \citep[the 2RXS catalogue;][]{Bolle16}, which are not
statically associated with galaxy density peaks. 
The excluded radius is four times the $1\sigma$ size of the PSF. 
A point spread function is empirically estimated by stacking the RASS images of 
2RXS point sources located within one degree centering main peaks of
individual clusters. 
We stack count images as well as the corresponding RASS exposure map
\citep{Anderson15} and then compute count rate images.
Since our targets are located at low Galactic column density region, the
galactic absorption at each cluster field is not significantly changed. 
As a weighing function, we use $D_{\rm L}^2(z_{l})/D_{\rm L}^2(z_{\rm ref})$ and
the lensing contribution $W$ (eq. \ref{eq:W_cl}) in order to standardize the flux to its
expectation at the average redshifts of weak-lensing samples, for the
purpose of the $L_X$ and $M_{200}$ scaling relation study
\citep{Miyazaki18}. 
Here, $D_L$ is the luminosity distance to the clusters and $z_{\rm ref}$
is the reference redshift.
To avoid a positional fluctuation of the background component caused by
unknown sources, we estimate the background component by fitting the radial profiles. 
We measure the aperture photometry for the stacked images within
$r_{200}$ and subtract the background components. Assuming the
metalicilty $Z=0.2$ and $k_BT=1.9$ keV \citep{Vikhlinin09a}, 
we covert the estimated count rate to the X-ray luminosity, $L_X$.
The measurement uncertainty is composed of the statistical noise assuming Poisson errors,
bootstrap sampling errors, and aperture sizes caused by weak-lensing mass uncertainty.
We represent the result and discussion in the Sec
\ref{subsec:Ysz}, including the {\it Planck} SZE and mass scaling relation.


\section{Result and Discussion} \label{sec:result}

\subsection{Halo concentration and mass ratio}\label{subsec:C200}

\begin{figure*}
\begin{center}
\includegraphics[width=0.48\hsize]{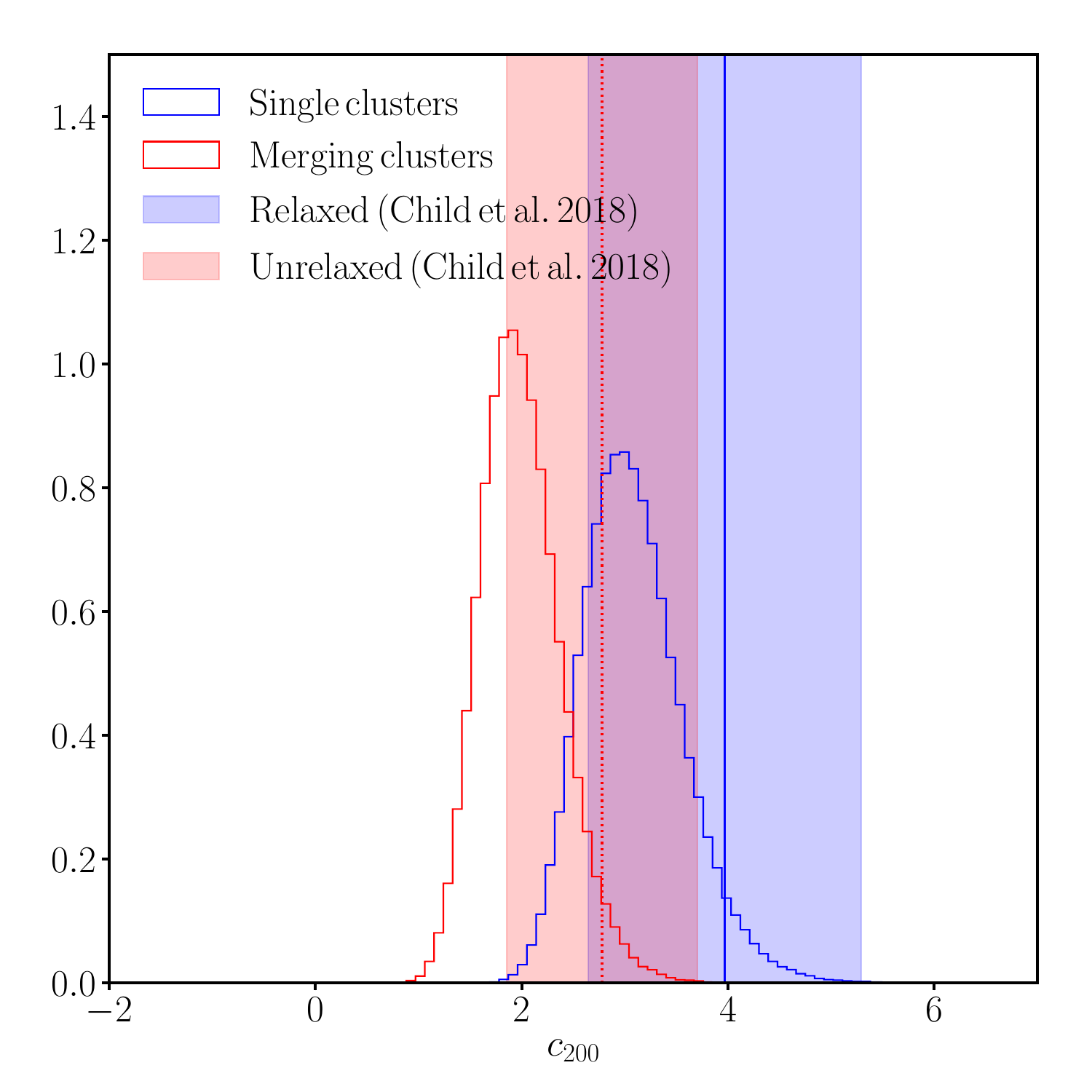}
\includegraphics[width=0.48\hsize]{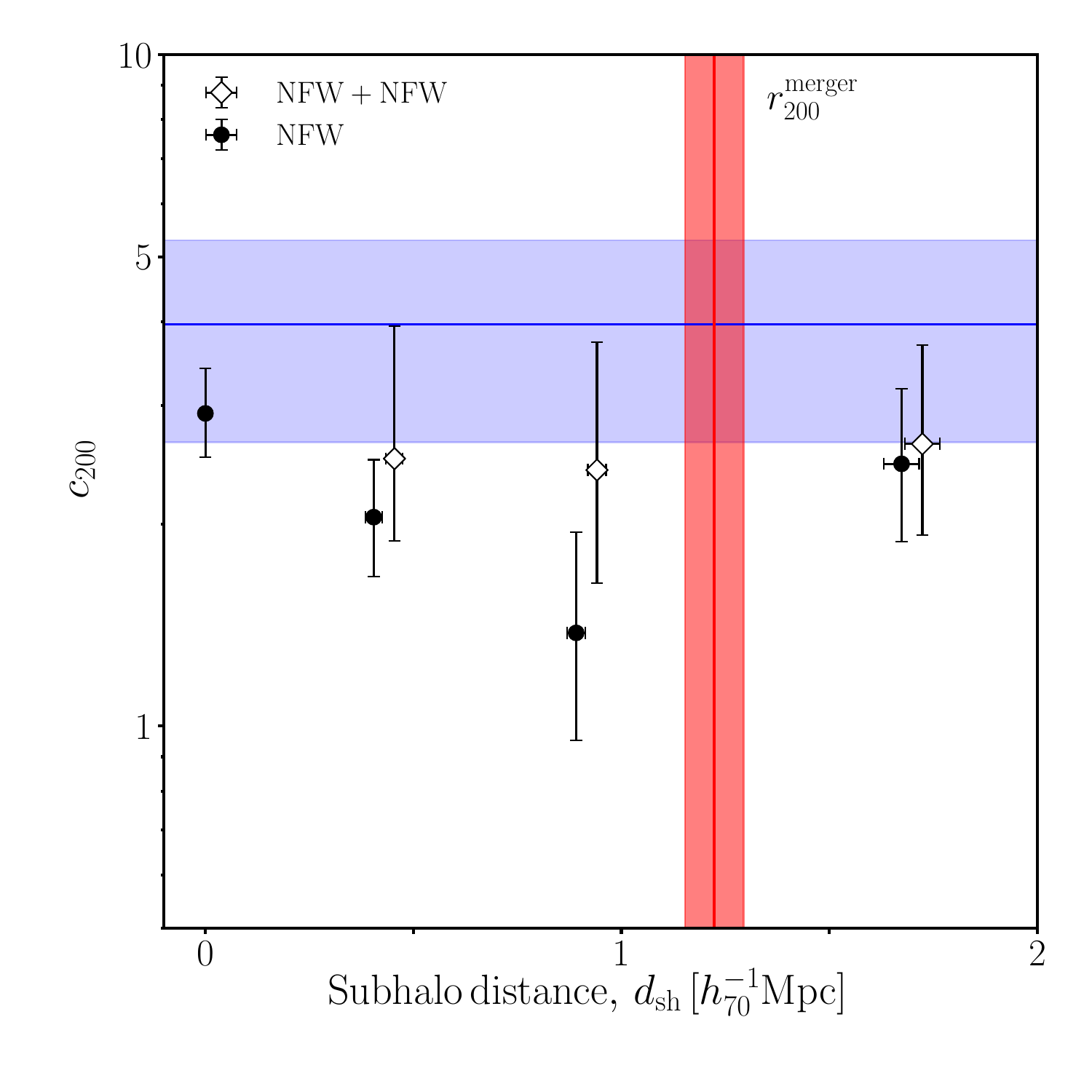}
\end{center}
 \caption{{\it Left}: Posterior distributions of halo concentrations for single
 (blue) and merging (red) clusters. The solid blue line and
 filled blue area are a prediction and $1\sigma$ scatter of relaxed clusters at $z=0.38$
 \citep{Child18}, respectively. The dotted red line denotes a prediction
 of a unrelaxed sample of clusters of \cite{Child18}, $c_{200}^{\rm
 unrelaxed}/c_{200}^{\rm relaxed}=0.69$, expected from their result at $z=0$. 
The filled red area is $1\sigma$ scatter of unrelaxed clusters.
 {\it Right}: The halo concentration, $c_{200}$, as a function of projected
 distances of subhalos, $d_{\rm sh}$.
 The vertical red line and region are the best-fit $r_{200}$ of
 the total mass of all the merging clusters and its $1\sigma$
 uncertainty, respectively.
 The horizontal blue region is
 the same as the left panel. Black circles denote the best-fit NFW model. 
 When the subhalos are located inside $r_{200}$ on the sky, the concentration
 parameter decreases as the subhalo distance increases. In
 contrast, the concentration of the merging clusters with subhalos of
 $d_{\rm sh}>r_{200}$ is comparable to that of the single clusters. 
 White diamonds are the concentration parameter of the main halo
 modeled with the two NFW components for the main and sub halos.
 The concentration parameters are recovered when the
 subhalo component is considered.}
\label{fig:Chist}
\end{figure*}

We study a difference of halo concentrations between the single and merging
clusters based on weak-lensing analysis (Sec \ref{sec:WL}).
Left panel of Fig. \ref{fig:Chist} shows posterior distributions of the two
subsamples by weak-lensing analysis.
Assuming a single NFW component for both the single and merging
clusters, the best-fit concentration parameters for the single and merging clusters are
$c_{200}^{\rm single}=2.92_{-0.41}^{+0.48}$ and $c_{200}^{\rm merger}=1.98_{-0.35}^{+0.41}$,
respectively. The average redshift of the two samples are similar
($z^{\rm single}=0.38$ and $z^{\rm merger}=0.37$).
The best-fit mass for the single clusters, $M_{200}^{\rm
single}=1.48_{-0.25}^{+0.29}\times10^{14}\h70Msol$, is about half of the
merging clusters, $M_{200}^{\rm merger}=2.98_{-0.48}^{+0.55}\times10^{14}\h70Msol$. 
Although the concentration parameter depends weakly on the halo mass
\citep[e.g.][]{Bullock01, Duffy08, Ludlow12,Bhattacharya13,Meneghetti14,
Ludlow14,Diemer15,Child18}, the difference of the two masses is negligible
in the following discussion of the halo concentration.

A recent numerical simulation \citep{Child18} defined relaxed clusters
and unrelaxed clusters using a distance between the halo center and the
center of mass of all particles.
In the left panel of Fig. \ref{fig:Chist}, 
the solid blue line and filled blue area denote the average concentration for the
simulated, relaxed clusters and its $1\sigma$ scatter computed at $M_{200}^{\rm single}$ and $z^{\rm
single}$, respectively. 
Our best-fit concentration for the single clusters is
somewhat lower than a prediction of the numerical simulation, though
they are consistent within intrinsic scatter. 
\cite{Child18} have also shown that the concentration of the
unrelaxed clusters is $69$ per cent lower than that of the relaxed
clusters at $z=0$ (Figure 3 in their paper).
The dotted red line and filled blue area are the average concentration
for the unrelaxed clusters and its $1\sigma$ scatter, assuming $c_{200}^{\rm
 unrelaxed}/c_{200}^{\rm relaxed}=0.69$ at $z=0$.
Our result gives $c^{\rm merger}_{200}/c^{\rm
single}_{200}=0.68\pm0.17$, in agreement with \cite{Child18}.
\cite{Neto07} show that the concentration of the unrelaxed clusters is
$\sim66$ per cent of the relaxed clusters at our best-fit mass range.
Similarly, other numerical simulations \citep[e.g.][]{Duffy08,Bhattacharya13} have pointed out that
the halo concentration of the full sample of simulated
clusters is lower than that of the relaxed clusters.

We study whether the halo concentration depends on the peak separation or
not. Since the overdensity radius of the merging clusters is
$r_{200}=1.22\pm0.07h_{70}^{-1}$ Mpc, we divide the merging cluster
sample into three subsamples
with the projected distances of subhalos, $d_{\rm sh}$, in $[0.2,0.6], [0.6,1.2]$, and
$[1.2-2.4]h_{70}^{-1}$ Mpc. The number of the subsamples and the actual
number of the WL clusters are summarized in
Table \ref{tab:subsample}. Five clusters out of the 189 merging
clusters fall from the radial ranges. We adopt
the closest distance for the three-peaks clusters in the subsample selection.
The right panel of Fig. \ref{fig:Chist} shows the concentration parameter as a function of
projected distances of subhalos, $d_{\rm sh}$. When the projected
distance of subhalos are less than $r_{200}$, the concentration
parameters are underestimated by $\sim50-70$ per cent compared to the
single clusters. Furthermore, the best-fit concentration parameter decreases as the subhalo
distance increases. When the subhalos are located outside $r_{200}$, the
concentration parameter is comparable to that of the single clusters.
This trend can be naturally explained by the position of the subhalos,
because merging clusters, which have subhalos at larger radii, tend to
possess a less centrally concentrated profile.
On the other hand, mass components outside clusters do not affect the
internal structure of the targeting clusters.


\begin{table}
\caption{Subsamples: $^a$ The signal-to-noise ratio of weak-lensing
 signal computed by $(S/N)_{\rm WL}=\left(\Delta \Sigma_{+,n}
 C_{nm}^{-1} \Delta \Sigma_{+,m}\right)^{1/2}$ where $C$ is the error
 covariance matrix $C_g+C_s+C_{\rm LSS}$.} \label{tab:subsample}
\begin{center}
 \begin{tabular}{cccc}
\hline
merging clusters \\
$d_{\rm sh}$ [$h_{70}^{-1}$Mpc] & $(S/N)_{\rm WL}$$^a$ 
  & \small{Number of WL/total clusters} \\
\hline
  $0.2-0.6$ & 13.0 & 46/53   \\
  $0.6-1.2$ & 10.9 & 61/68   \\
  $1.2-2.4$ &  8.8 & 57/63   \\
\hline
  single clusters \\
$N_{\rm cor}$ & $(S/N)_{\rm WL}$$^a$ & \small{Number of WL/total clusters} \\
 \hline
 $15-18$  &   11.3 & 464/552  \\
 $18-21$  &   11.6  & 272/316\\
  $21-25$  &   12.6 & 245/283 \\
  $25-30$  &  12.9 & 164/191  \\
  $30-40$  &  12.9 & 137/156  \\
  $40-100$ & 16.7 & 53/63   \\
 \hline
 \end{tabular}
\end{center}
\end{table}

\begin{figure*}
\begin{center}
\includegraphics[width=\hsize]{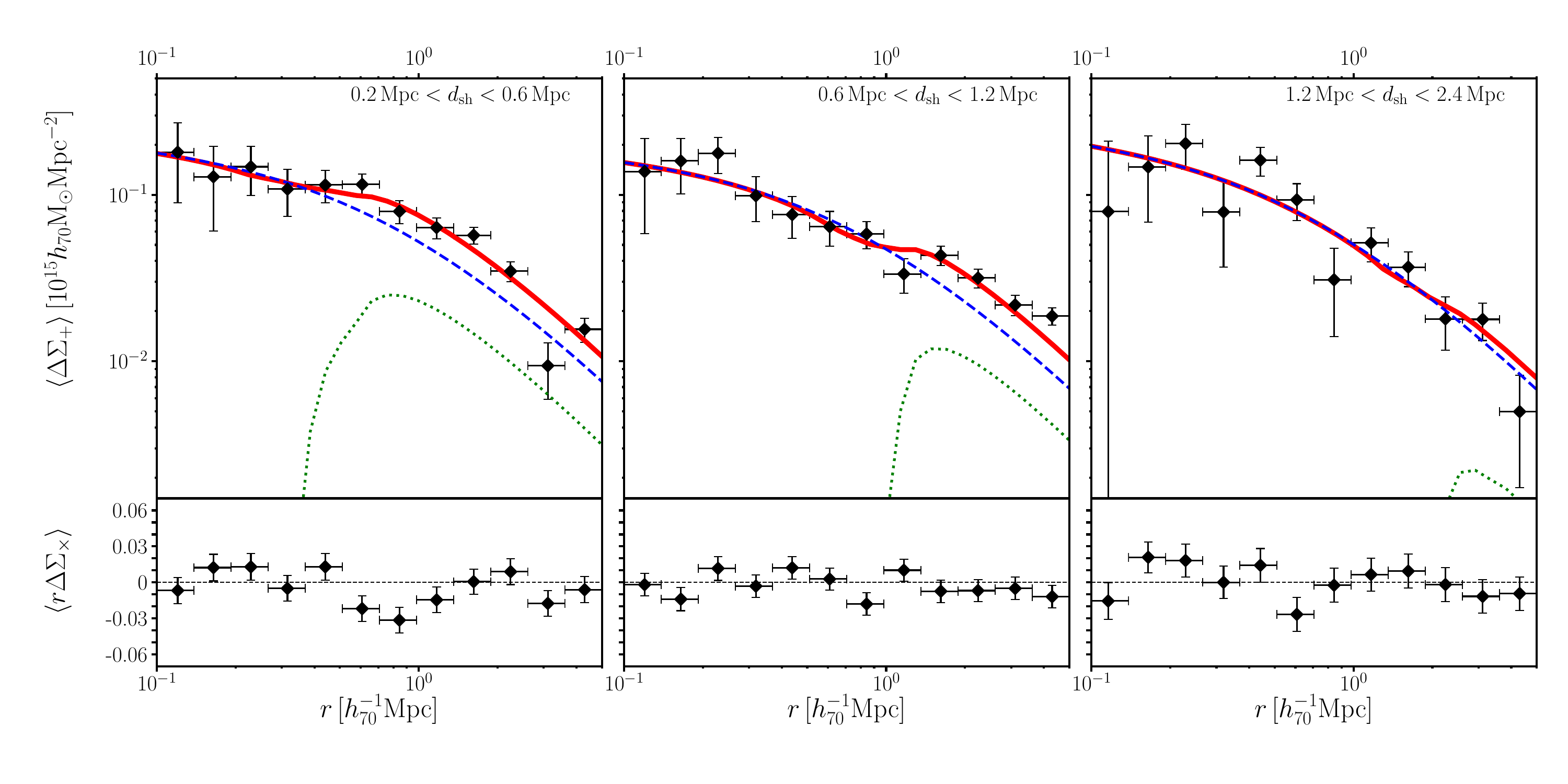}
\end{center}
\caption{{\it Top}: The stacked tangential shear profile for the
 three subsamples of the merging clusters. The subsamples are defined by the subhalo
 distances of $0.2h_{70}^{-1}\,{\rm Mpc}<d_{\rm sh}<0.6h_{70}^{-1}\,{\rm
 Mpc}$, $0.6h_{70}^{-1}\,{\rm Mpc}<d_{\rm sh}<1.2h_{70}^{-1}\,{\rm Mpc}$,
 and $1.2h_{70}^{-1}\,{\rm Mpc}<d_{\rm sh}<2.4h_{70}^{-1}\,{\rm Mpc}$,
 respectively. The errors are composed of the diagonal component of $(C_g+C_s+C_{{\rm LSS}})^{1/2}$.
 Solid red, dashed blue, and dotted green lines are
    the total mass model, the NFW model for the main halo and the
 subhalo mass, respectively. {\it Bottom} : The product of the $45$
 degree rotated component, $\Delta \Sigma_\times$, as a function of the cluster-centric radius, $r$.}
\label{fig:stackedWL}
\end{figure*}

Next we model the two NFW components as the main and subhalos for the
merging clusters (eqs \ref{eq:offsetmodel} and
\ref{eq:fWLmodel}).
Since the concentration parameter of the subhalos is not well
determined, we assume $c_{200}=3$ from the best-fit of the single
clusters.
When we change $c_{200}^{\rm (sh)}$ by $\pm1$, the subhalo masses are changed by
only by $\ltsim10\%$.
Fig. \ref{fig:stackedWL} shows stacked lensing profiles of
the there subsamples of the merging clusters. The total mass component
is shown by red thick lines. The main and subhalo components are shown
by dashed blue and dotted green lines, respectively. Clear bumps in the
stacked lensing profiles are found associated with the positions of the
subhalos. It indicates that the mass is indeed associated with the
galaxy subhalos detected by the peaks in the galaxy maps (Fig. \ref{fig:galmap}).
The bump is caused by the off-centering effect \citep{Yang06}. The
concentration parameters for the main halo components are shown by white
diamonds in the right panel of Fig. \ref{fig:Chist}. In contrast to one NFW component
analysis, the estimated concentration parameter is almost constant
against the subhalo distances. The values are in good agreement with
that of the single clusters. Thus, once the subhalo component is
considered in the modeling, the halo concentration is recovered.

We also estimate the mass ratios for the three subsamples; 
$M_{\rm sh}/M_{\rm main}=0.37_{-0.26}^{+1.02}$, $0.47_{-0.28}^{+0.91}$
and $0.14_{-0.11}^{+0.27}$ from small to large separations,
respectively.
The measurement uncertainty of the mass ratio takes into account the
covariance error matrix between two mass components.
Since the mass estimations of the main and subhalos are highly
correlated, the errors of the mass ratio are relatively large.
Similarly, we repeat the two-halo modeling for the all merging
clusters at $0.2\,h_{70}^{-1}{\rm Mpc}<d_{\rm sh}<2.4\,h_{70}^{-1}{\rm Mpc}$ and obtain $M_{\rm
sh}/M_{\rm main}=0.31^{+0.68}_{-0.20}$. The average value of the mass
ratios for the three subsamples is $0.33$, where the lensing weight
(eq. \ref{eq:W_cl}) is considered. Since we fix the positions of
the subhalos and the halo concentration, the subhalo mass is sensitive to
the lensing amplitude and thereby the mass ratios with and without
subdivisions are consistent with each other. 
The mass ratio is smaller than the ratio, $0.75\pm0.18$, of peak heights
appeared in the galaxy maps.
Considering measurement uncertainty, 
the majority of our sample of the merging clusters is major mergers with the
mass ratio at order of $\gtsim 0.1$.

We note that the projection effect of surrounding halos is not large at
$r\simlt r_{200}$ as we will discuss in Secs \ref{subsec:Sigma_n} and
\ref{subsec:v-s}. Thus, the weak-lensing mass and halo
concentration for the merging clusters at $d_{\rm sh}<r_{200}$ can be
interpreted as results affected by the internal subhalos.

We ignored mis-centering effect \citep[e.g.][]{Yang06,Johnston07,Okabe13}
on the halo concentration in the above discussion.
The density peaks do not necessarily correspond to the center of halo
mass. If there were offsets between density peaks and mass centers,
concentration parameters would be underestimated by
mis-centering effect. However, since the concentration
parameters of the main halos of the merging clusters (two NFW modeling)
agree with that of the single clusters (the right panel of Fig. \ref{fig:Chist}),
both the merging and single clusters would have similar distribution of
offset distances if they exist.
Although the mis-centering effect might affect the absolute value of
concentration parameters, it does not significantly affect the
comparison of how much halo concentrations are relatively affected by the subhalos. 
Modeling with the mis-centering effect would be important to precisely
measure the mass-concentration relation, which is left for future works.

\subsection{Projected number density} \label{subsec:Sigma_n}

\begin{figure*}
\begin{center}
 \includegraphics[width=0.48\hsize]{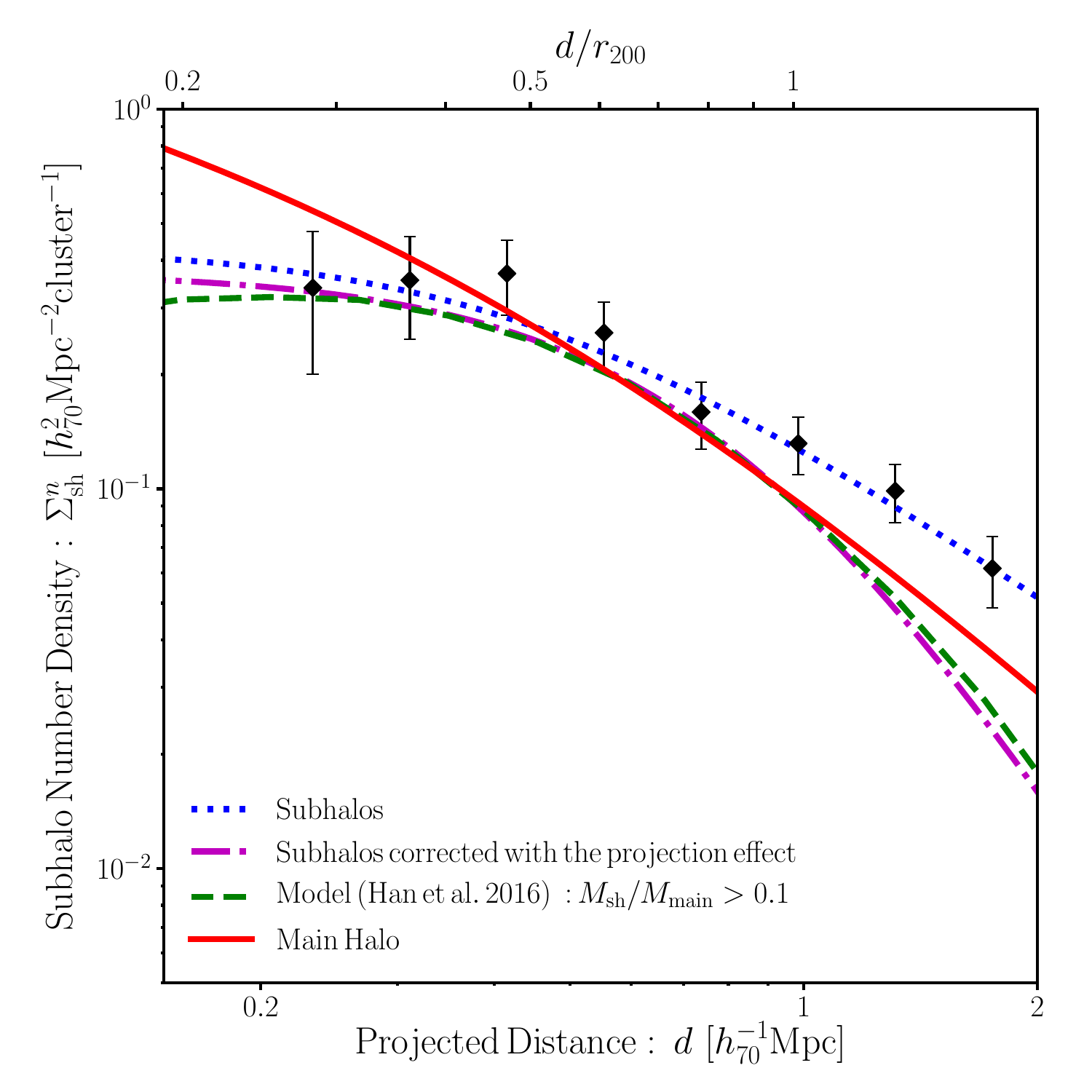}
 \includegraphics[width=0.47\hsize]{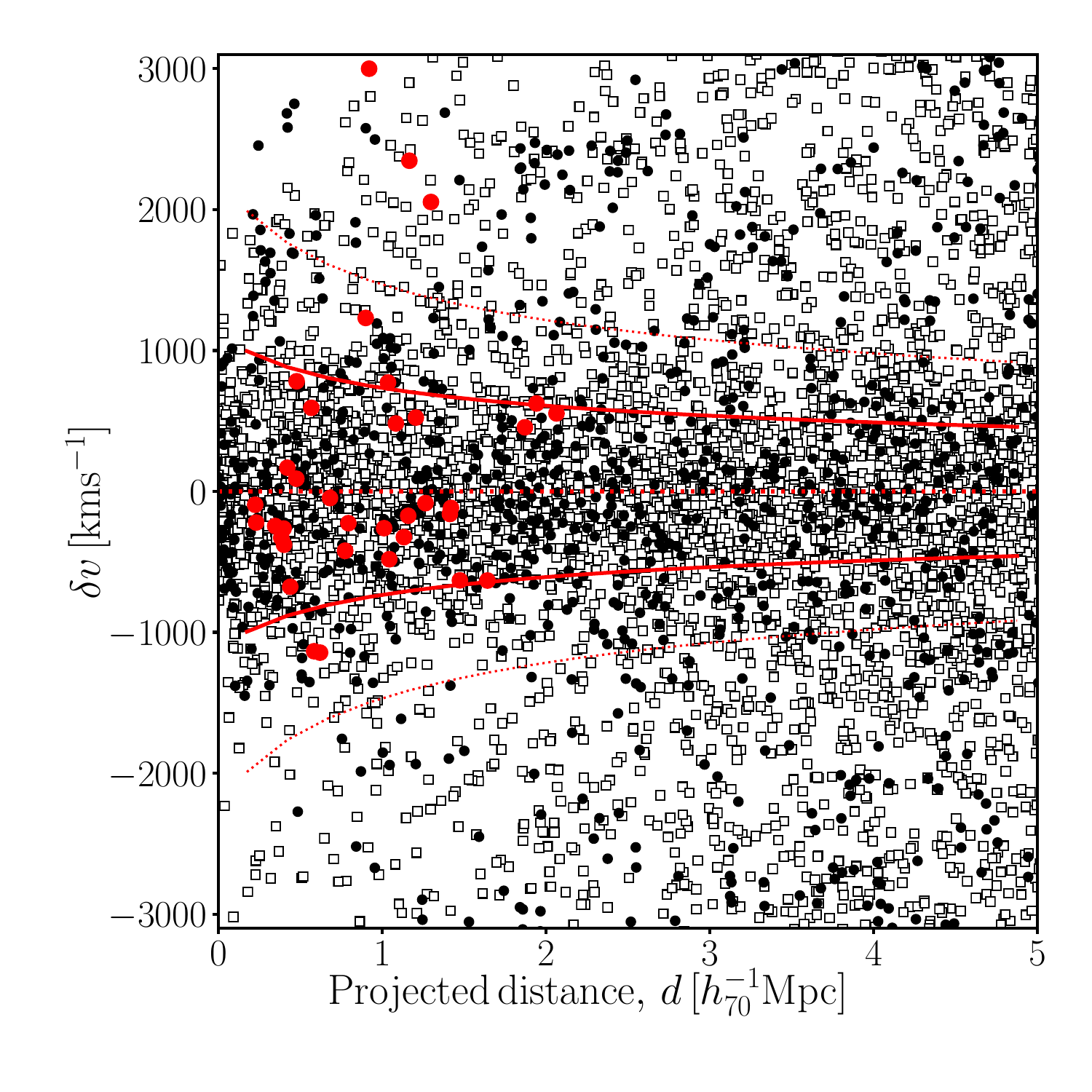}
\end{center}
 \caption{{\it Left:}The radial profile of the averaged projected number
 density of the sub halos per
 a cluster. The dotted blue and solid red lines show the best-fit
 $\beta$ profile of the subhalo radial profile and the NFW profile for
 the main cluster normalized by $\int_0^{r_{200}}dr 2\pi r \Sigma_{\rm
 NFW}=1$, respectively.
 The dashed-dotted magenta line is the best-fit profile for the
 subhalos, corrected with the projection effect.
 The dashed green line is the normalized project distribution for
 subhalos with $M_{\rm sh}/M_{\rm main}>0.1$ \citep{Han16}. {\it Right:}
 The stacked velocity-space diagram from the SDSS. White squares and black circles  are
 spectroscopic galaxies around the single and merging clusters, respectively.
 Large red circles denote subhalos with spectroscopic counterparts. The thick, red solid lines are the caustic
 amplitude estimated by the weak-lensing mass for the main halo of the
 merging clusters. The thin, dotted red line denotes twice the caustic amplitude.
  }
\label{fig:N_sub}
\end{figure*}

The projected average number density of cluster subhalos as a function of
projected distances from the main peaks is shown in the left panel of 
Fig. \ref{fig:N_sub}.
As expected, the optically-identified, massive subhalos
are widely distributed from cluster centers to outskirts.
The errors for the projected number density are assumed to be Poisson noise. 
The projected number density is well described by a $\beta$ model,
formed by $\Sigma_{\rm sh}^n\propto (1+(r/r_c)^2)^{0.5-\beta}$.
The corresponding three-dimensional number density is written as $n\propto
(1+(r/r_c)^2)^{-\beta}$. 
The best-fit parameters (the blue curve) are $r_c=0.45_{-0.22}^{+0.52}h_{70}^{-1}$ Mpc
and $\beta=1.20_{-0.16}^{+0.45}$, giving an outer slope of $n\propto
r^{-2.4}$.

We note that our selection of the subhalos may suffer
from the projection effect.
Hence, the projection of galaxy concentrations along the
line-of-sight appears in the radial profile.
As a simple test, we introduce an additional constant parameter, $\Sigma_{\rm
proj}^{(n)}$ as the projection-effect component.
The best-fit parameters are $r_c=0.85_{-0.44}^{+0.62}h_{70}^{-1}$ Mpc,
$\beta=2.18_{-0.76}^{+2.56}$ and $\Sigma_{\rm
proj}^{n}=0.045_{-0.027}^{+0.023}h_{70}^2\,{\rm Mpc}^{-2}\,{\rm
cluster}^{-1}$. The best-fit radial distribution of subhalos excluding
the projection effect is shown in the dashed-dotted magenta line.
The project effect is small at $r \ltsim r_{200}$, whereas it accounts for
about a half of detected subhalos at $r\gtsim r_{200}$.

The projected mass density profile of the main halo of the merging
clusters is shown by red line.
We use the best-fit NFW model for the full sample of the merging clusters
modeled with two-halo components. The profile is normalized by
$\int^{r_{200}}_0 dr 2 \pi r \Sigma_{\rm NFW}=1$.
Here, $r_{200}$ for the main cluster is derived from the two-halo
component fit described in Secs. \ref{sec:WL} and \ref{subsec:C200}.
The radial profile of the
subhalos is significantly shallower than the total matter distribution,
as predicted by theoretical works
\citep[e.g.][]{Gao04,Diemand04,Nagai05,Taylor05b,Ludlow09,Han16}.
The radial distribution of subhalos within their
parent halos is less centrally concentrated than that of dark matter.
Subhalos captured by more massive halos are subject to
dynamical friction, losing their angular momentum and subsequently falling inward the center.
Simultaneously, their masses are reduced by the tidal force which increases 
with an increasing radius from the cluster center.
\cite{Diemand04} and \cite{Gao04} find that the radial distribution of
subhalos is largely independent of their self-bound mass, which is
caused by an efficient mixing of the above physical process of the evolution of subhalos
within the potential of the main halo. 
\cite{Han16} propose an analytic model which simultaneously predicts both the mass function and spatial
distribution of subhalos, considering statistical description of the
amount of mass tidally stripped from individual subhalos.
They find that the model radial profile agrees with
results of numerical simulations.
We compute a surface number density of subhalos using
SUBGEN \citep{Han16} with their final mass higher than 10 per cent of the
main halo with $M_{200}=10^{14}\hMsol$ and $c_{200}=3$. 
The model profile normalized by the total number of subhalos is shown by
green dashed line. The observed profile at $r<r_{200}$ is in an
agreement with the prediction, but is somewhat shallower outside
$r_{200}$. On the other hand, the dashed-dotted magenta profile, for
which the projection effect is corrected,
is in a good agreement with the theoretical prediction over a wide radial range.

Numerical simulations \citep[e.g.][]{Neto07,Ludlow12} defined
relaxed and unrelaxed clusters using the following three criteria; 
subhalo mass fraction, center of mass displacement, and ratio of kinetic
energy to gravitational energy.
They found that the fraction
of unrelaxed halos are roughly $\sim30$ percent and increases as the
halos masses increase. The fraction of unrelaxed clusters also depends
on resolutions of numerical simulations. Our sample of the merging
clusters account for $\sim10$ percent of the total sample.
Since the two definitions are different, a fair comparison is difficult
but our fraction is lower than expected by numerical simulations.
We investigate mass dependence of the merger fraction and find $\sim50$
percent for $M_{200}>5\times10^{14}h_{70}^{-1}M_\odot$,
$\sim25$ percent for $M_{200}\sim3\times10^{14}h_{70}^{-1}M_\odot$, and
$\sim10$ percent for $M_{200}\sim10^{14}h_{70}^{-1}M_\odot$. This trend
would be explained by the peak-finding technique.
Less massive clusters have smaller number of galaxies, and thus
sparse distributions of a small number of galaxies give rise to the possibility that 
peak-heights of some subhalos are below the threshold.
Therefore, some subhalos of the order of $\sim10^{13}h_{70}^{-1}M_\odot$
in halos of $\sim10^{14}h_{70}^{-1}M_\odot$ halos might be not identified by the current technique.
Using a larger number of clusters in the future HSC-SSP dataset,
we will develop more sophisticated detection schemes and discuss the
purity of the merger sample especially for less massive clusters.

\subsection{Velocity-space diagram} \label{subsec:v-s}

We compute a velocity-space diagram of the subhalos stacked over the
merging clusters (right panel of Fig. \ref{fig:N_sub}).
The spectroscopic data is retrieved from the SDSS DR14
\citep{Abolfathi18}.
We compute the average line-of-sight velocity around the each galaxy
peak within $200h_{70}^{-1}$ kpc and regard it as the velocity of the main and sub
halos. The deviations of the line-of-sight
velocities of the subhalos from the main halo are shown by large red circles.
As a reference, spectroscopic galaxies around the single and merging
clusters are overlaid. Many galaxies are confined
within the characteristic, ``trumpet'', shape expected in the spherical
infall model \citep[e.g.][]{Kaiser87,Diaferio97,Mamon13}. The caustic amplitude computed by the
weak-lensing mass of the main halo for the merging clusters is shown in
the solid red line.
Most subhalos (30 subhalos) are located within the caustics surfaces, indicating that they
are indeed physically interacted with the main halo.
On the other hand, a few subhalos are located outside the caustic
surface, and thus contaminating components caused by the projection
effect. The fraction of the contamination is consistent with that
estimated from the subhalo radial profile (the left panel of
Fig. \ref{fig:N_sub}).

\subsection{Red fraction}\label{subsec:fred_result}

We statistically investigate, using the optically-selected cluster sample,
whether the red fraction is changed by cluster mergers or not.
As mentioned in Sec \ref{subsec:fred_ana},
we use the massive galaxies, with $M_*>10^{10.45}M_\odot$, which are
centrally concentrated inside the clusters.
We subtract the numbers of red and blue galaxies in the annulus
($1-3h_{70}^{-1}{\rm Mpc}$) to compute the red fraction (Sec. \ref{subsec:fred_ana}).
The inner radius is comparable to
$r_{200}^{\rm single}= 0.95\pm0.06h_{70}^{-1}{\rm Mpc}$ for the single
clusters and $r_{200}^{\rm merger}=1.22\pm0.07h_{70}^{-1}$ Mpc for the
merging clusters ( $r_{200}^{\rm merger,main}=
1.07\pm0.10h_{70}^{-1}{\rm Mpc}$ for the main halo), determined by the weak-lensing
analysis (Sec. \ref{subsec:C200}). We do not find significant structures
of massive galaxies outside $r_{200}$ of the clusters except for merging clusters with large peak separations.
We thus exclude the merging clusters from the three subsamples with peak
separation of $d_{\rm sh}>1.2h_{70}^{-1}\,{\rm
Mpc}$. (Sec. \ref{subsec:C200}) in order to avoid over-subtraction in
the following results.

\begin{figure*}
 \begin{center}
 \includegraphics[width=0.48\hsize]{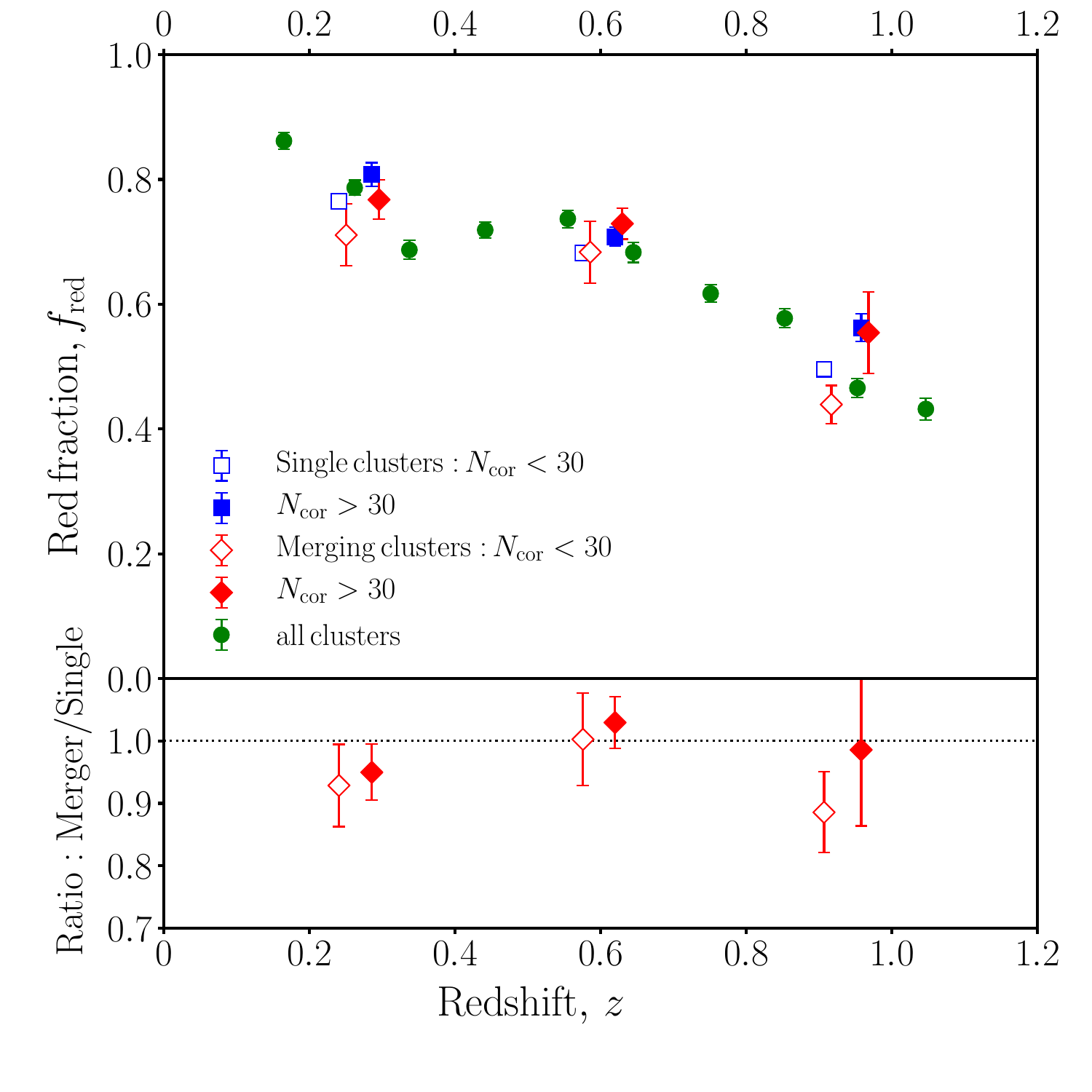}
 \includegraphics[width=0.48\hsize]{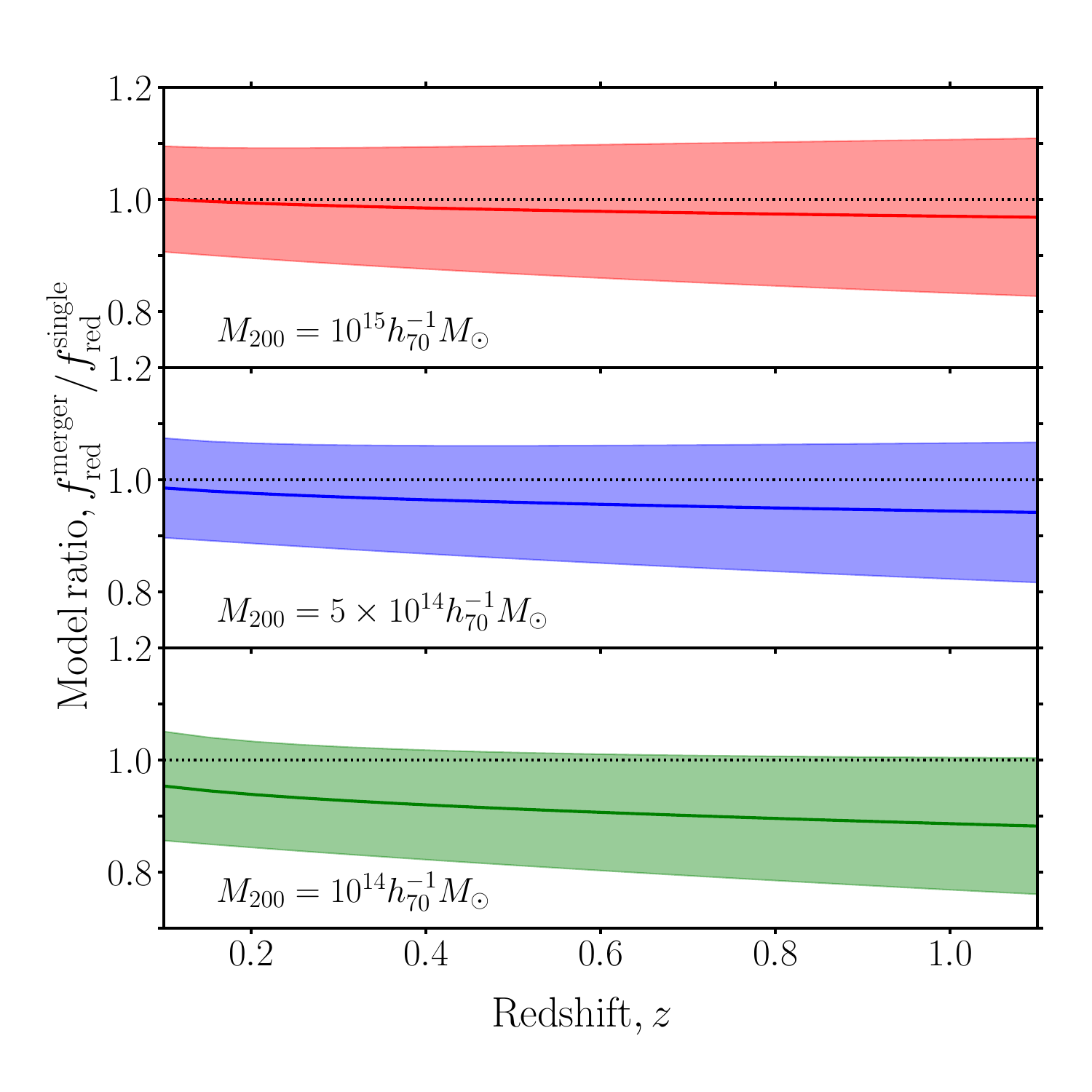}
 \end{center}
\caption{{\it Left:} The red fraction as a function of the cluster redshift (upper
 panel). Green circles are the red fractions of all the clusters excluding
 the merging clusters of which peak separations are $d_{\rm sh}>1.2h_{70}^{-1}\,{\rm Mpc}$. 
 Filled blue squares and red diamonds denote the red fractions of the single and merging
 clusters with $N_{\rm cor}>30$, respectively.
 Open blue squares and red diamonds are those with $N_{\rm cor}<30$.
 The lower panel shows ratios between red fractions of the merging and
 single clusters.  {\it Right:} The model-dependent estimations of the
 ratio of red fractions between the merging and single clusters. The top,
 middle, and bottom panels are the red-fraction ratio of clusters with
 masses of $10^{15}$, $5\times10^{14}$ and $10^{14}h_{70}^{-1}\Msol$, respectively. The
 solid line and color area are the best-fit and its $1\sigma$ uncertainty, respectively.
 The black dotted lines denote the case that the ratio is equal to unity.
 }
 \label{fig:fred}
\end{figure*}

The left panel of Fig. \ref{fig:fred} shows the redshift evolution of
the average red fraction which is the ratio of red member galaxies to
all member galaxies selected by their photometric redshifts.
We adopt the biweight estimation in order to down-weight outliers in a
small number of subsample. The uncertainties are the errors of the biweight mean.  
The red fractions for all the clusters gradually decreases as cluster
redshifts increase, as discovered by \citet{Butcher84}.

We next compare, in a model-independent way, red fractions for the merging and single clusters (the
left panel of Fig. \ref{fig:fred}).
Since the red fraction weakly depends on cluster richness and 
the richness population of the merging clusters is different from
that of the single clusters, we define subsamples in a redshift and
richness space.
The red fractions for the merging clusters are comparable to those for
the single clusters, and some cases show lower values.
The deviation is not statistically significant at $\simlt 2\sigma$ level.
Nevertheless, in any subsamples, the red fraction of the merging clusters does not
exceed the value of the single clusters.
It implies that star formation would be triggered by cluster mergers.

As a complementary approach, we fit red fractions of the merging and single clusters as a function of
cluster richness and redshift and intrinsic scatter of red fractions.
A functional form is simply assumed to be $f_{\rm red}=f_{\rm
red,0}+a_zz^{b_z}+a_NN_{\rm cor}^{b_N}$. The powers of redshift and
richness, $b_z$ and $b_N$, are determined by fitting all the sample.
We then obtain the normalizations and slopes of the single and merging
clusters by fixing the powers. The red-fraction ratio between the
merging and single clusters is computed by taking account of the
correlation between the errors on each parameter by calculating the
error covariance matrix. The right panel of Fig. \ref{fig:fred} shows
the red-fraction ratio as a function of cluster redshift, for a fixed 
cluster mass. We here convert from the richness to the cluster mass based
on stacked weak-lensing analysis (Sec. \ref{subsec:Ncor}).
Overall, the red-fraction ratios agree with unity,
although the ratio for the low mass cluster of $M_{200}=10^{14}h_{70}^{-1}\Msol$ is somewhat lower.
The general feature is consistent, within errors, with the
aforementioned model-independent estimations.
The red-fraction ratio typically decreases as the cluster redshift
increases and the mass decreases. If the star formation is triggered by cluster mergers,
its time-scale can be roughly estimated by the cluster size and the collision
velocity and thus is $2r_{200}/v\sim2\,{\rm Gyr}(v/1000\,{\rm kms^{-1}})^{-1}$.
As mentioned above, the observational evidence is not significant. 
Therefore, more precise observational constraints on red fractions and
their properties require larger cluster samples.

Various processes to trigger star formation during cluster mergers are
proposed; the ram-pressure enhancement
\citep[e.g.][]{Dressler83,Evrard91,Fujita99b},
a change of the gravitational tidal field \citep{Bekki99}, and the encounters and
interactions between galaxies \citep{Lavery88}.
The ram-pressure scenario advocates that
the high ram-pressure of the ICM compresses the hot, tenuous inter-stellar
medium (ISM) in gas-rich galaxies and trigger a one-time burst of star
formation, but 
gas stripping weaken the star-burst phenomenon during the cluster collision
\citep[e.g.][]{Fujita99b}.
\citet{Ruggiero17} investigated through numerical simulations how much of an initial gas mass in
Milky-Way like galaxies is turned into stars after a single crossing of
idealized clusters of $10^{14}$ and $10^{15}\Msol$.
They found that the initial gas mass is more efficiently converted to
stars in less massive clusters, because the ram-pressure stripping process is more efficient
in more massive clusters, making the galaxy lose its gas faster, and
consequently prevent the steady star formation over the entire orbit.
The general feature predicted by the numerical simulations
is similar to our results of mass dependence of the red-fraction ratio.
\citet{Dressler83} suggested that the star formation is triggered by gas-rich
galaxies first infall into the ICM, so called the first infall model.
It does not contradict with the trend that the red-fraction ratio
decreases as the cluster redshift increases.
Even if a dusty star formation activity \citep{Koyama10} is triggered by
cluster mergers, it does not change the interpretation that the star
formation in the merging clusters is likely to be more active.

The activities of star formation in on-going merging clusters at
$z\ltsim0.4$ is studied through spectroscopic observations or narrow band
filters. 
For instance, \citet{Stroe15b} studied star-formation rate in two merging clusters with prominent radio relics (CIZA J2242.8+5301
and 1RXS J0603.3+4213 ). They found that the star-formation rate density for CIZA J2242.8+5301 is
on the order of 15 times the peak of the star-formation history, and 
the overall star-formation rate density in 1RXS J0603.3+4213 is consistent
with blank fields at the same redshift. In CIZA J2242.8+5301,
the projected distribution of highly star-forming galaxies is localized
along the merger axis. 
\citet{Stroe17} carried out spectroscopic observations of 19 clusters
and found that the normalization of $H\alpha$ luminosity function
in merging clusters is much higher than that of relaxed clusters.
\citet{Ferrari05} showed a lack of bright post-star-forming objects in A3921.
\citet{Deshev17} found a depletion of star forming galaxies within the
central $\sim1.5$ Mpc region of the merging clusters A520, in contrast
to relaxed clusters. Therefore, the results appear to be not converged
due to the large intrinsic scatter of the abundance of star forming galaxies.
Indeed, our estimation of intrinsic scatter of the red fraction is large, $\sigma_{\rm red}=0.18$.
Since our estimations are based on photometric redshifts,
further systematic follow-up studies of clusters compiling various merging stages,
cluster masses, and redshifts are essential to understand star
formation activities in merging clusters, in order to witness short time
scale variations in star formation and understand their mass and redshift dependence.

\subsection{Mass - richness relation}\label{subsec:Ncor}

We study the correlation between the mass and richness.
Based on the weak-lensing analysis (Sec \ref{subsec:C200}), we define
the three subsamples of the merging clusters (Table \ref{tab:subsample})
using peak separations of $d_{\rm sh}$ in $[0.2,0.6],[0.6,1.2]$, and $[1.2,2.4]$ Mpc. 
As for the single clusters, we adopt the same definition of the subsamples as that in the {\it Planck} SZ
 and mass scaling relation (Sec \ref{subsec:Ysz})
 because of a requirement for the stacking analysis of the {\it Planck} data.

\begin{figure}
\begin{center}
\includegraphics[width=\hsize]{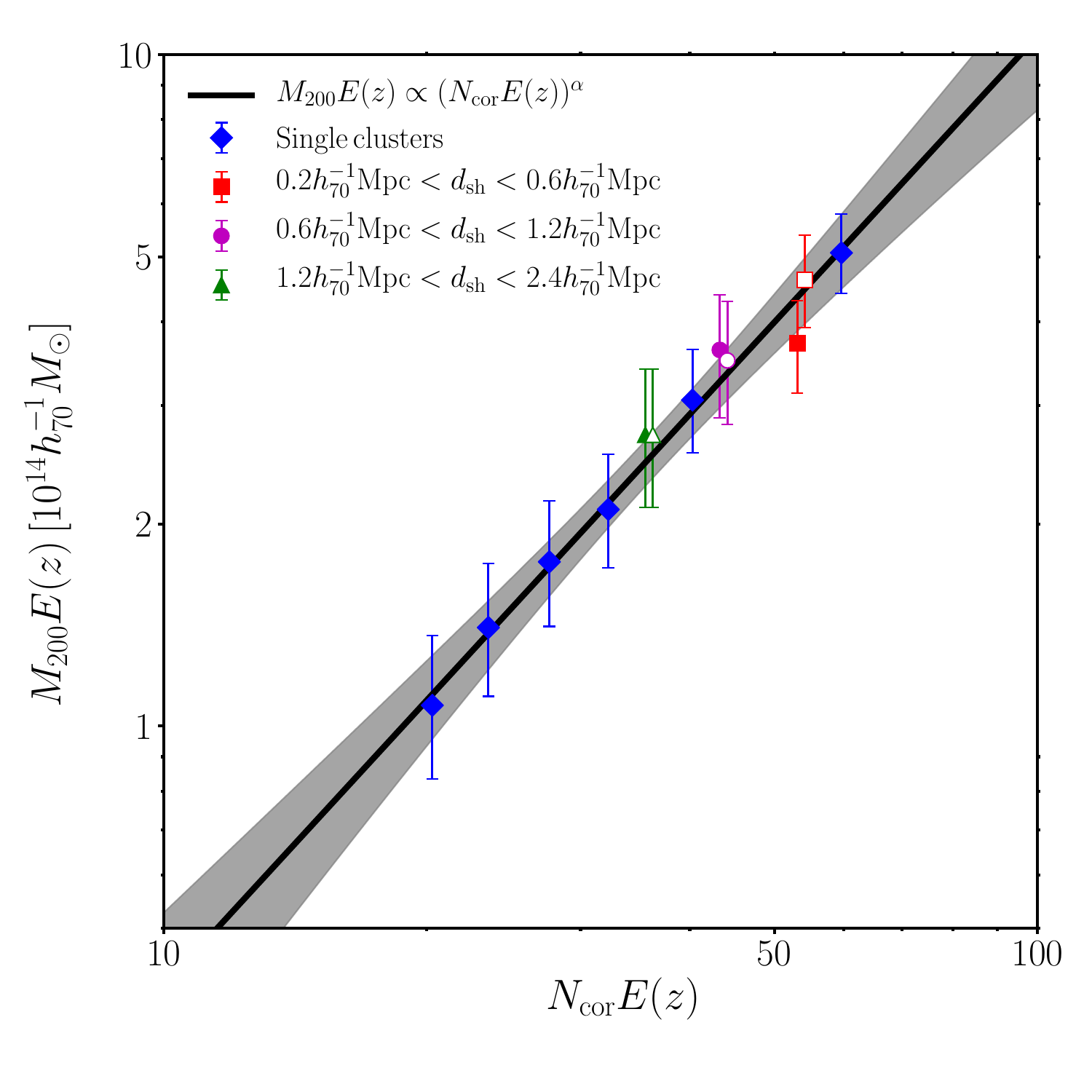}
\end{center}
 \caption{Mass-richness relation. Blue diamonds denote the single clusters. 
 Red square, magenta circle, and green triangle are the merging clusters
 divided by peak separations. As for the merging clusters, 
 the solid and open symbols are the total mass and the single NFW mass, respectively.
 The solid and gray area denote the best-fit
 and its $1\sigma$ error of the scaling relation for the single clusters with a free slope.} \label{fig:N-M}
\end{figure}

The average richness for each subsample is computed with a weight assigned to
the lensing contribution of the $j$-th cluster (eq. \ref{eq:W_cl}), in
the same analogy as Secs \ref{subsec:Psz} and \ref{subsec:Lx},
\begin{eqnarray}
 N_{\rm cor}= \frac{\sum N_{\rm cor,j} W_j}{\sum W_j}. \label{eq:N}
\end{eqnarray}
Fig. \ref{fig:N-M} shows the mass and richness relation. 
We use the NFW masses for the single clusters (Secs \ref{sec:WL} and
 \ref{subsec:C200}). Both the total masses of the two NFW components and
 the single NFW masses for the merging clusters are represented in the
 figure. We find a tight correlation between the mass and richness 
 regardless of the mass definitions.

We assume that the scaling relation is described by the following power-law relation,
\begin{eqnarray}
  M_{200} E(z) = M_{N} \left(\frac{N_{\rm cor} E(z)}{N_0}\right)^\alpha, \label{eq:M-N}
\end{eqnarray}
where $E(z)=(\Omega_{m,0}(1+z)^3+\Omega_\Lambda)^{1/2}$ is the evolution
factor and $N_0=15$.
We perform linear regression using their log-space variables without correcting for selection function. 
The intrinsic scatter of the mass, $\sigma_{\ln M}$, is considered in
the fitting \citep{Okabe10c}. We obtain 
$M_N=0.72_{-0.14}^{+0.18}\times 10^{14}h_{70}^{-1}M_\odot$,
$\alpha=1.41_{-0.22}^{+0.23}$, and $\sigma_{\rm ln M}<0.08$ for the single clusters.
When we include the richness and the total mass of the merging clusters, the best-fit parameters are
$M_N=0.76_{-0.14}^{+0.18}\times 10^{14}h_{70}^{-1}M_\odot$,
$\alpha=1.35_{-0.20}^{+0.20}$, and $\sigma_{\rm ln M}<0.07$, respectively.
We find in both cases a steeper slope than $M_{200}\propto N_{\rm cor}$,
which conflicts with our first assumption to make smoothed maps (Sec
\ref{subsec:sample}). 
However, the discrepancy does not significantly affect the
subhalo search, because our maps sufficiently covered the wide area.
There is no significant deviation of the merging clusters in the
mass and richness relation, as seen in the small upper limit of the
intrinsic scatter. It indicates that the richness is insensitive to the
merger boost, as your expectations.

For future relevant papers, we explicitly note the best-fit
normalizations and slopes for $M_{500}$ and $M_{200\rm m}$ in the form of eq. \ref{eq:M-N};
$M_N=0.46_{-0.07}^{+0.08}\times 10^{14}h_{70}^{-1}M_\odot$ and
$\alpha=1.43_{-0.17}^{+0.17}$ for $M_{500}$ and
$M_N=0.95_{-0.19}^{+0.24}\times10^{14}h_{70}^{-1}M_\odot$ and
$\alpha=1.41_{-0.24}^{+0.24}$ for $M_{200\rm m}$, respectively.

\subsection{{\it Planck} SZ and {\it ROSAT} $L_X$ scaling relations} \label{subsec:Ysz}

In this subsection, we study scaling relations between the weak-lensing
masses and {\it Planck} SZ (Sec \ref{subsec:Psz}) and {\it ROSAT}
X-ray luminosity (Sec \ref{subsec:Lx}) and the merger boost of the ICM.

\begin{figure*}
\begin{center}
\includegraphics[width=0.48\hsize]{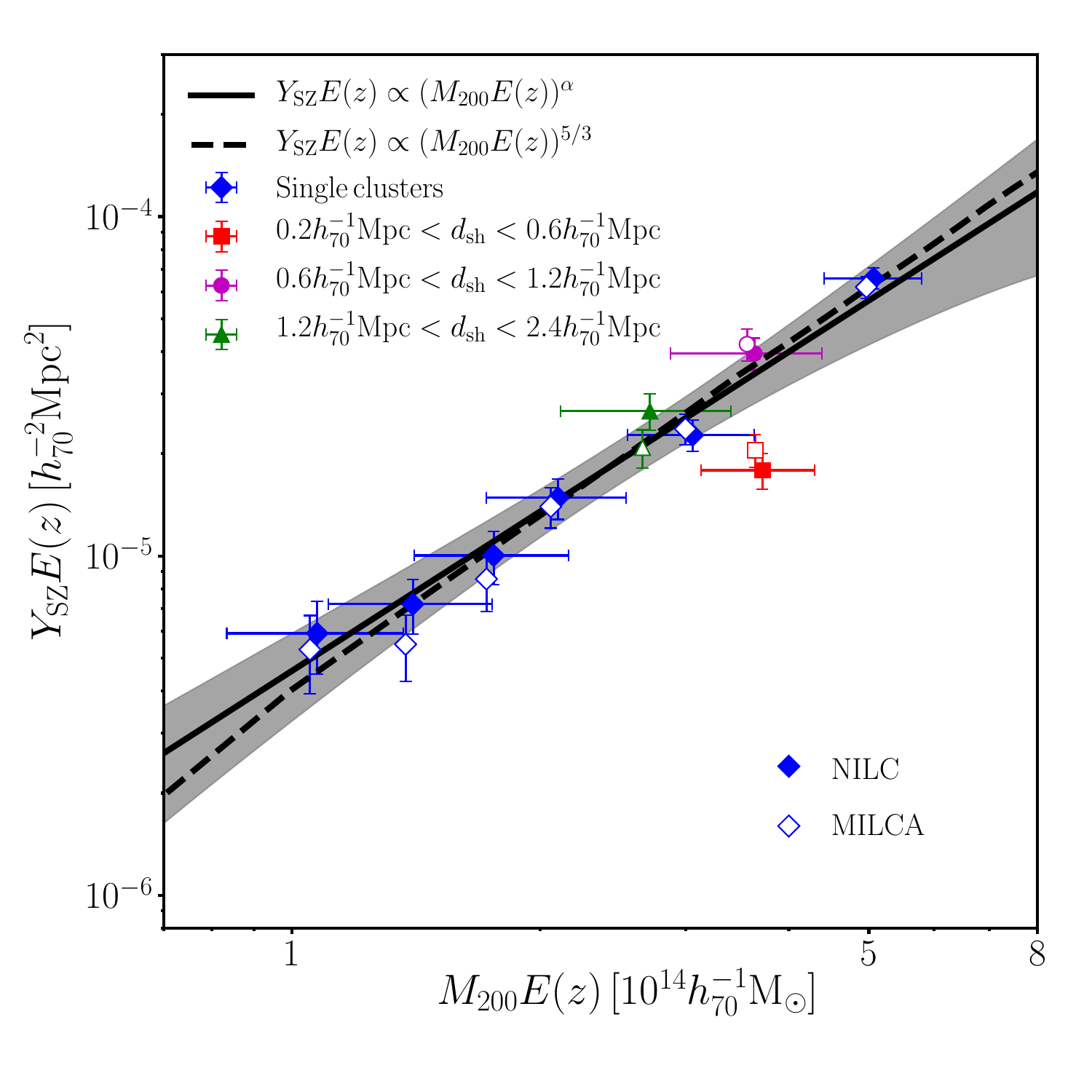}
\includegraphics[width=0.48\hsize]{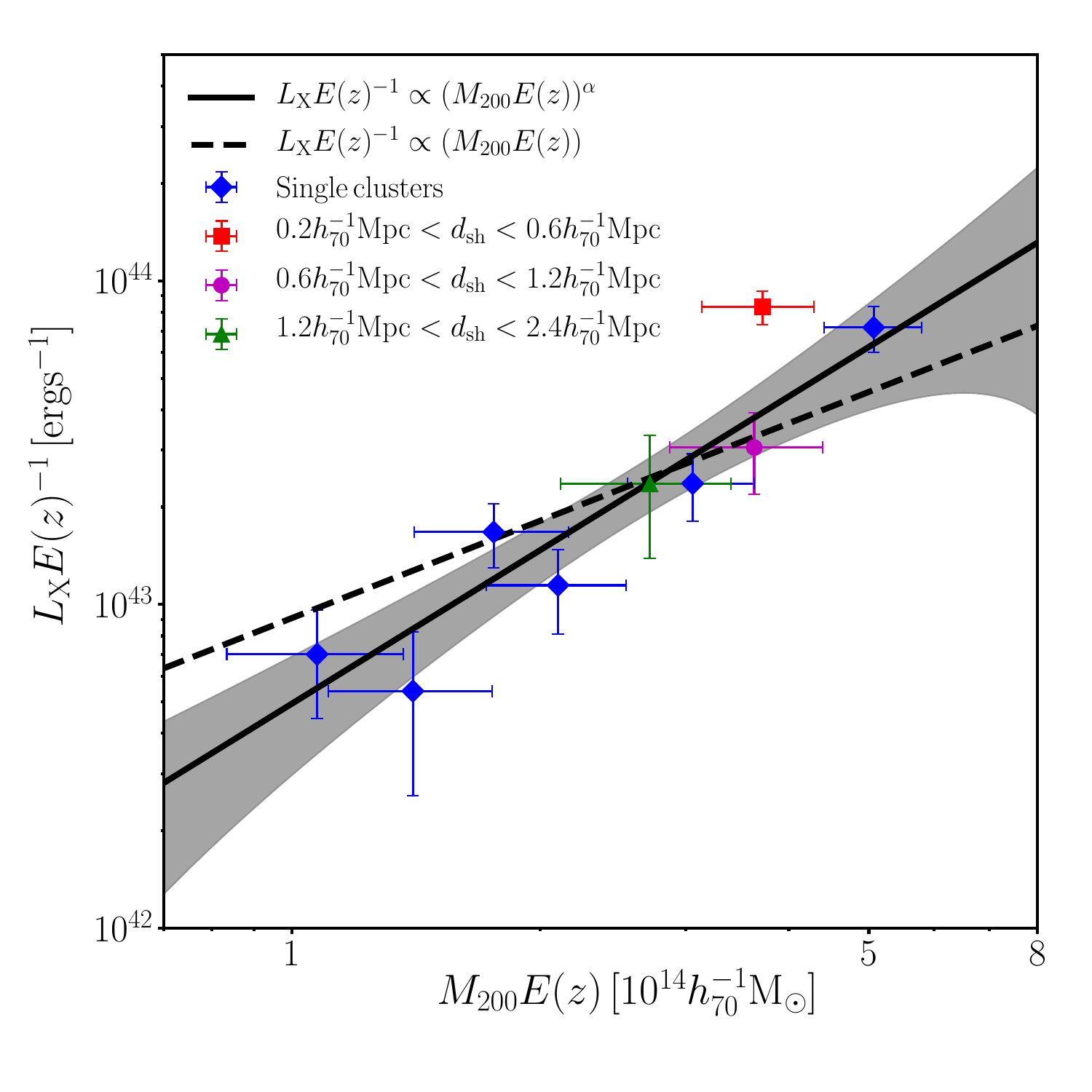}
\end{center}
 \caption{{\it Left}: The {\it Planck} $Y_{\rm SZ}$ and $M_{200}$ scaling
 relation. Blue diamonds denote the single clusters. 
 Red square, magenta circle, and green triangle are the merging clusters
 divided by peak separations. The solid and open symbols are NILC and
 MILCA data, respectively. The solid and gray area denote the best-fit
 and its $1\sigma$ error of the scaling relation for the single clusters with a free slope. The
 dashed line is the best-fit of the scaling relation with a fixed slope.
{\it Right}: Similar to the left panel, but for the {\it ROSAT} $L_{X}$ and $M_{200}$ scaling relation.
}
 \label{fig:Ysz-M}
\end{figure*}

We compute a cylindrically integrated quantity $Y_{\rm SZ}$ using two maps of public $y$ map; MILCA
\citep{Hurier13} and NILC \citep{Remazeilles13}.
We split the merging sample into three subsamples using peak
separations of $d_{\rm sh}$ in $[0.2,0.6],[0.6,1.2]$, and $[1.2,2.4]$ Mpc,
based on the weak-lensing analysis (Sec \ref{subsec:C200}).
As for the single clusters, we use the richness, $N_{\rm cor}$, to
define subsamples so that the signal-to-noise ratio of the tangential
profiles for the subsamples is
as uniform as possible and the number of each subsample is higher than fifty.
The second condition is required to compile the sufficient number of
clusters in stacked $y$ maps.
The definition of the subsample is summarized in Table \ref{tab:subsample}.
Since all the clusters do not always fulfill the full-depth and
 full-color condition for weak-lensing analysis, the stacked $y$
 measurements are computed with a weight assigned to
the lensing contribution. The $Y_{\rm SZ}$ are measured within
$2r_{200}$ determined by weak-lensing mass measurements, taking into
account the large {\it Planck} FWHM. For the merging clusters, we use the
overdensity radii of the total masses of the main and sub clusters.
We obtain consistent results for the MILCA and NILC $Y_{\rm SZ}$
measurements (Fig. \ref{fig:Ysz-M}).

We find that the mass scaling relation with the {\it Planck} SZ
measurements is in excellent agreement with a self-similar solution
(dashed line) 
\begin{eqnarray}
 Y_{\rm SZ} E(z)=Y_0 \left(\frac{M_{200}E(z)}{M_0}\right)^{5/3}.
\end{eqnarray}
We fit the scaling relation for the single clusters with a free slope parameter, $Y_{\rm SZ}
\propto (M_{200}E(z))^{\alpha}$, and obtain
$\alpha=1.56_{-0.23}^{+0.29}$ (NILC) and $\alpha=1.65_{-0.23}^{+0.29}$
(MILCA), consistent with the self-similar slope.
We use the lognormal quantities in the linear regression with the intrinsic
scatter. The normalization is $Y_0=4.58_{-1.23}^{+1.41}
\times10^{-6}\,[{\rm Mpc}^2]$ at $M_0=10^{14}h_{70}^{-1}\Msol$ for NICL and 
$Y_0=3.91_{-1.10}^{+1.29}\times10^{-6}\,[{\rm Mpc}^2]$ for MILCA, respectively.
The intrinsic scatter is constrained with upper limits of $\sigma_{\ln
Y}<0.12$ (NILC) and $\sigma_{\ln Y}<0.16$ (MILCA).

The {\it Planck} SZ measurement for the merging clusters is in overall agreement with the
scaling relation. We repeat the linear regression for the full sample,
and find that the results do not significantly change;
$\alpha=1.51_{-0.23}^{+0.27}$, $\sigma_{\ln Y}<0.24$ (NICL)
and
$\alpha=1.59_{-0.22}^{+0.28}$, $\sigma_{\ln Y}<0.16$ (MICLA). The
deviations from the best-fit scaling relation is discussed in detail later.

\cite{Planck16ymap} have computed the stacked $Y_{\rm SZ}$ around
optically-selected clusters from the SDSS III \citep{Wen12}.
Through the scaling-relation between the richness and mass for a
subsample of galaxy clusters, they obtain the power-law slope,
$\alpha=1.92\pm0.42$ in the $Y_{\rm SZ}-M_{200}$ relation,
although the assumed functional form of the scaling relation is $Y_{\rm
SZ}E(z)^{-2/3} \propto M^\alpha$ is different from ours. 
A fair comparison of the normalization is very difficult because of the
different choices of the measurement radius and the adopted scaling
relation. Furthermore, their mass calibration is not carried out
consistently for the whole sample clusters in contrast to our method.
As the zero-th order comparison, we obtain $Y_0\sim 4
\times10^{-6}\,[{\rm Mpc}^2]$ in our definition from
their values. Therefore, their measurements does not necessarily conflict
with our results.

We next compute the stacked X-ray luminosity from the RASS in the energy band
of $0.1-2.4$ keV. The X-ray luminosity scales as $L_X \propto \rho_{\rm
gas}^2 V \Lambda(T_X)$ by thermal bremsstrahlung emission, where
$\rho_{\rm gas}$ is the gas mass density and $\Lambda$ is the
emissivity. Since the emissivity is almost constant in the soft band,
the scaling relation between the X-ray luminosity and cluster mass have
the form of 
\begin{eqnarray}
 L_X^{\rm RASS} E(z)^{-1}\propto M_{200} E(z),
\end{eqnarray}
under an assumption of constant gas mass fraction.
However, we find that our result prefers a steeper slope (the right panel of
Fig. \ref{fig:Ysz-M}). When we fit the scaling relation for the single
clusters with a free slope,
\begin{eqnarray}
L_X E(z)^{-1} =L_0 \left(\frac{M_{200}E(z)}{M_0}\right)^{\alpha},
\end{eqnarray}
we obtain $\alpha=1.58_{-0.30}^{+0.36}$ and $L_0=4.93_{-1.66}^{+2.01}\times10^{42}\,{\rm [ergs^{-1}]}$. 
We obtain the constraint on the log-normal scatter of $L_X$ as $\ln
\sigma_L<0.21$, which is slightly larger than that of $Y_{\rm SZ}$.
We refit including the merging clusters, and find
$L_0=4.46_{-1.50}^{+1.96}\times10^{42}\,{\rm [ergs^{-1}]}$,
$\alpha=1.79_{-0.30}^{+0.35}$ and $\ln \sigma_L<0.29$.
The slope is slightly steeper and intrinsic scatter is larger, albeit
consistent within $1\sigma$ errors.

\cite{Rykoff08} studied the scaling relation between the X-ray luminosity in
0.1-2.4 keV and the total mass for the maxBCG galaxy clusters
\citep{Koester07} in the redshift range $0.1\le z \le 0.3$. 
They find $\alpha=1.65\pm0.13$ and $L_0=3.4\pm0.4(\rm stat)\pm0.4(\rm
sys)\,10^{42}[{\rm ergs^{-1}}]$, which are in agreement with our results.
\cite{Anderson15} measured the stack X-ray emission around locally brightest
galaxies from the SDSS, using the RASS data. They measure X-ray luminosity
in $0.5-2.0$ keV and study a scaling relation assuming $L_X\propto E(z)^{7/3}
M_{500}^\alpha$, where the mass $M_{500}$ is estimated by their internal technique.
They obtain a steep slope, $\alpha=1.85^{+0.15}_{-0.16}$, which is in good
agreement with our results. We convert the normalization from their
scaling relation to ours, taking into account the difference of the
energy band and $M_{500}=0.64M_{200}$ based on our weak-lensing
measurement. The resulting normalization, $L_0\sim4.4\times10^{42}{\rm
ergs^{-1}}$, is in a good agreement with ours.
\cite{Arnaud10} studied the $L_X$ and mass scaling relation,
at the overdensity $500$,
for a representative sample of 33 X-ray clusters from the REFLEX
catalogue \citep{Bohringer07}. The best-fit slope for X-ray selected
clusters, $\alpha=1.64$, is consistent with that of the optically
selected clusters.
We convert the X-ray luminosity from within $r_{500}$ to within
$r_{200}$ using $L_{500}/L_{200} = 0.96$ \citep{Piffaretti11}. The
resulting normalization is $L_0=1.5\times10^{43},{\rm ergs^{-1}}$, which
is about two times higher than that for the CAMIRA clusters.
Similar results are found in shear-selected clusters.
\citet{Miyazaki18} found that the X-ray luminosity for the
shear-selected clusters is about half of that of X-ray clusters drawn
from the MCXC clusters \citep{Piffaretti11}, using the same selection
criteria on the mass and redshift.

We next study the difference of the scaling relations between
the merging and single cluster samples.
In order to discuss the deviation of the scaling relation due to the merging phenomena,
we consider an ideal case that the two clusters of which mass and SZ or X-ray
observables of two merging clusters perfectly follow scaling relations,
$X=g(M)\equiv A M^{\alpha}$,
where $M=M_{200}E(z)$ and $X=Y_{\rm SZ}E(z)$ or $X=L_X E(z)^{-1}$ in
this study.
If $X$ for the two clusters is conserved during mergers or by the
projection effect, the total $X_{\rm tot}$ measured
within a certain radius can be derived by
\begin{eqnarray}
 X_{\rm tot}=X_{\rm 1}+X_{\rm 2}= A \frac{1+(M_2/M_1)^{\alpha}}{(1+M_2/M_1)^{\alpha}} (M_1+M_2)^{\alpha},
\end{eqnarray}
where the subscripts ($1$ and $2$) label two clusters.
Therefore, after the merger or the projection, $X$ is deviated by a factor of
$f_M=(1+(M_2/M_1)^{\alpha})/(1+M_2/M_1)^{\alpha}$ from the scaling relation.
Since our sample is contaminated by the projection effect,
we adopt a statistical treatment in order to quantify the merger boost on
the SZ and X-ray observables.
Based on the projected number density of the subhalos (Sec. \ref{subsec:v-s}), 
a chance probability of the projection effect for each cluster in our
sample can be calculated by integrating $\Sigma_{\rm proj}^{(n)}$ from
the innermost to outermost radii on the sky ;
$P_{\rm proj}=0.046_{-0.026}^{+0.023}$ for $0.2-0.6h_{70}^{-1}$ Mpc,
$P_{\rm proj}=0.154_{-0.088}^{+0.079}$ for $0.6-1.2h_{70}^{-1}$ Mpc,
and 
$P_{\rm proj}=0.614_{-0.353}^{+0.314}$ for $1.2-2.4h_{70}^{-1}$ Mpc,
respectively. We here introduce the boost factor of $b_X$ of the ICM, caused by
cluster mergers, in the scaling relation. The scaling relation can be statistically expressed by 
\begin{eqnarray}
 X_{\rm tot} = A b_X f_M M_{\rm tot}^{\alpha} (1-P_{\rm proj}) + A
  f_M M_{\rm tot}^{\alpha} P_{\rm proj}. 
\end{eqnarray}
Consequently, the boost factor of cluster mergers is obtained by
\begin{eqnarray}
b_X=\frac{X/g(M)-f_MP_{\rm proj}}{f_M(1-P_{\rm proj})}.
\end{eqnarray}
If $b_X$ is higher than unity, the gas observables are enhanced by
cluster mergers.

\begin{figure}
\begin{center}
 \includegraphics[width=\hsize]{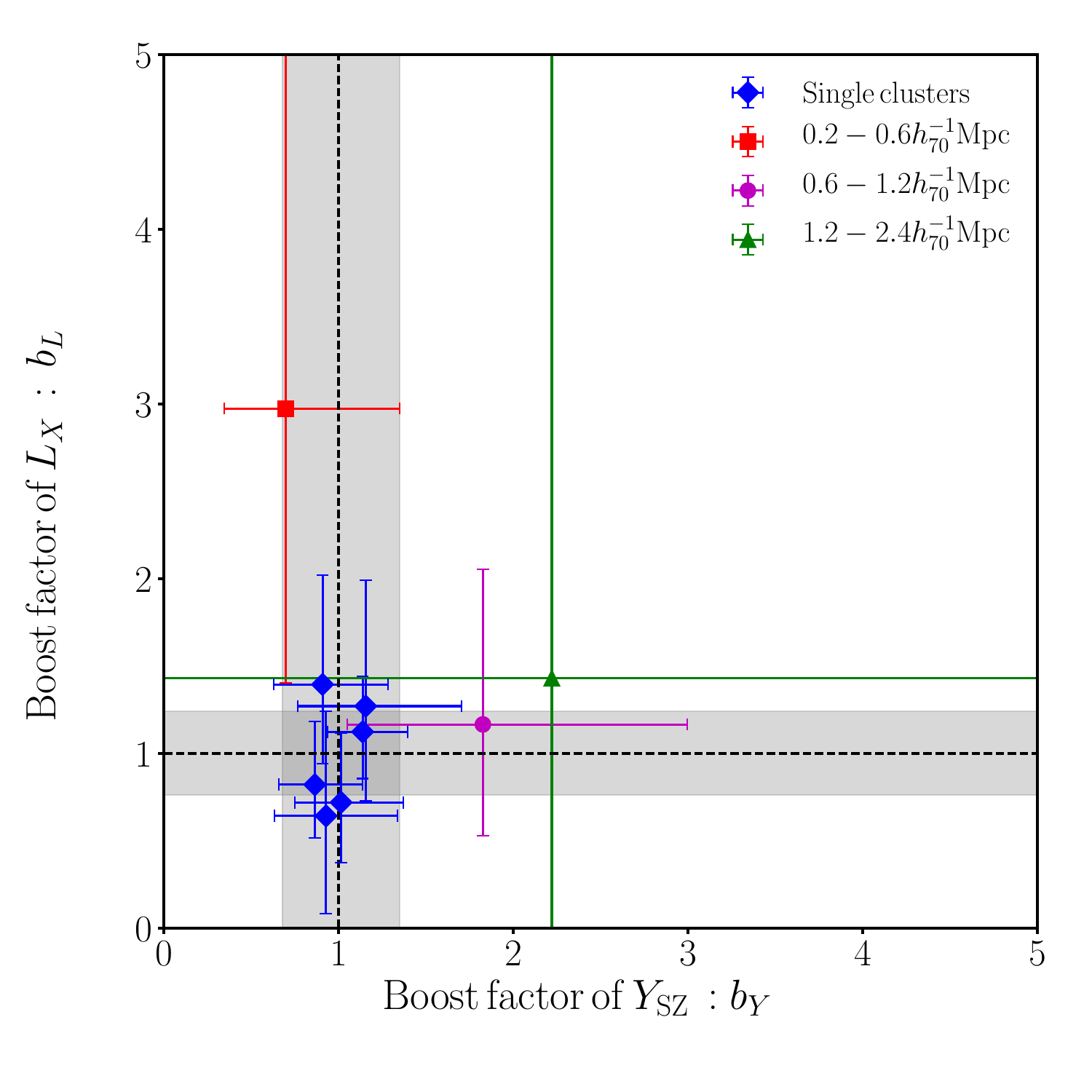}
\end{center}
\caption{Boost factors of $Y_{\rm SZ}$ and $L_X$ caused by cluster
  mergers.
  Point symbols are the same as in Fig. \ref{fig:Ysz-M}.
  The gray areas denote the regions combined with the $1\sigma$
  uncertainty of the normalization and the upper limit of the intrinsic
  scatter. The boost factors of the merging clusters with peak
 separations of $0.2-0.6h_{70}^{-1}$
  Mpc and $0.6-1.2h_{70}^{-1}$ Mpc are higher than unity, in
  the $L_X$ and $Y_{\rm SZ}$ directions, respectively.
  }
 \label{fig:boost}
\end{figure}

Fig. \ref{fig:boost} shows the resulting boost factors of $Y_{\rm SZ}$
and $L_X$. We use the NICL $Y_{\rm SZ}$ scaling relation.
We adopt the best-fit scaling relations with free slopes.
 The measurement errors of weak-lensing masses, the cluster mass ratios and the probability of the
 projection effect are all taken into account.
Therefore, the errors become larger than those in Figure
\ref{fig:Ysz-M}. Especially, the large error for clusters at
$1.2-2.4h_{70}^{-1}$ is attributed to the error of the probability of
the projection effect. 
The boost factors of all the single clusters and the merging clusters
with peak separations of $1.2-2.4h_{70}^{-1}$ Mpc are consistent
with unity. Hence, once the projection effect is corrected, the
scaling relation holds for the merging clusters when the subcluster is
located outside the main cluster.
On the other hand, the merging clusters with $0.2-0.6h_{70}^{-1}$ Mpc and $0.6-1.2h_{70}^{-1}$ Mpc
show $b_L>1$ and $b_Y>1$, respectively, albeit with a low statistical significance level.
Interestingly, the boost factors in the other scaling relations are $b_Y\sim
1$ at $0.2-0.6h_{70}^{-1}$ Mpc and $b_L\sim 1$ at $0.6-1.2h_{70}^{-1}$
Mpc, respectively. This feature statistically implies that the electron
number density and the electron temperature is enhanced in each
subsample of at $0.2-0.6h_{70}^{-1}$ Mpc and $0.6-1.2h_{70}^{-1}$ Mpc, respectively.
Since the scaling relation we discuss here is based on the
stacked quantities, it does not always indicate that all the clusters in
the subsamples are affected by the merger boost. 
Since each subsample contains clusters both before and after core passages,
we cannot discuss individual boost factors without knowing the fraction of
clusters before and after core passages.
Future systematic X-ray and SZE observations with high angular resolutions
are essential for understanding gas physics involved in cluster mergers.
The follow-up X-ray and SZE observations will enable us to carry out individual cluster analysis and investigate the
merger boost in scaling relations based on individual measurements.

\cite{Krause12} found in cosmological simulations that major mergers
within a Gyr timescale cause an asymmetric scatter in
the $Y_{\rm SZ}$ scaling relation so that the inferred mass of merging
systems is biased low.
They also found that clusters with lower concentration parameters
have a trend to have lower $Y_{\rm SZ}$ at fixed masses. The trend conflicts with our
results (Figs \ref{fig:Chist} and \ref{fig:boost}).
\cite{Shaw08} pointed out, using adiabatic simulations,
a positive correlation between the scatter in concentration and integrated Y
parameter.
\citet{Yu15} reported that the shock propagates out to $\sim r_{200}$
after a few Gyr from the closest encounter and the energy from random
gas motions decays into thermal energy over the next few Gyr. 
Consequently, the thermal $Y_{\rm SZ}$ within $r_{200}$ increase by
$\sim8\%$. Although a fair comparison with their results is difficult,
the trend is likely to be different from our results.
Therefore, future individual cluster analyses for our sample of the
merging clusters are essential to understand understand more precisely
the merger boost and its merger-phase dependence.

\subsection{Non-detection of diffuse radio emissions and collision velocities}

It is well known that major mergers host diffuse radio synchrotron
emission, the so-called radio halos and relics
\citep[e.g.][]{Roettiger99,Govoni04,Venturi07,Venturi08,vanWeeren10,vanWeeren11b,Feretti12}.
The radio halos are diffuse radio emission around cluster centers and
relics are filamentary radio emission at cluster outskirts.
We search diffuse radio emissions in archival NVSS \citep[1.4 GHz;][]{Condon98NVSS}
and TGSS \citep[147.5 MHz;][]{Intema17TGSS} data. 
Following \citet{deGasperin18}, we computed spectral index maps by matching pixel-by-pixel
NVSS and beam-size-matched TGSS maps. Here, we convolve the TGSS images
with a circular beam with the size same as that the NVSS because of the
lower resolution of NVSS. 
The background rms noise of the NVSS is nearly uniform at $0.45\,{\rm
mJy beam}^{-1}$. The rms noise of the TGSS are locally estimated by
$4\sigma$ clippings. The typical rms is $\sim 3\,{\rm mJy beam}^{-1}$.
\citet{deGasperin18} discovered the following feature;
spectral index maps of radio lobes associated with AGNs show steeper
values at its outer edges than at around
its peaks. On the other hand, the spectral index gradient of radio
relics is expected
to be along the minor axis of the filamentary structure.
We cross-match the spectral index map and HSC galaxy maps to identify
radio halos and relics. 
We also compare the NVSS and TGSS radio images with the HSC optical images for an
assessment of whether the radio sources are associated with optical
galaxies or not.
However, we do not find any strong candidates of radio relics and halos
in our sample. Almost all radio sources turn out to be associated with
member galaxies or high redshift sources.

\begin{figure}
\begin{center}
 \includegraphics[width=\hsize]{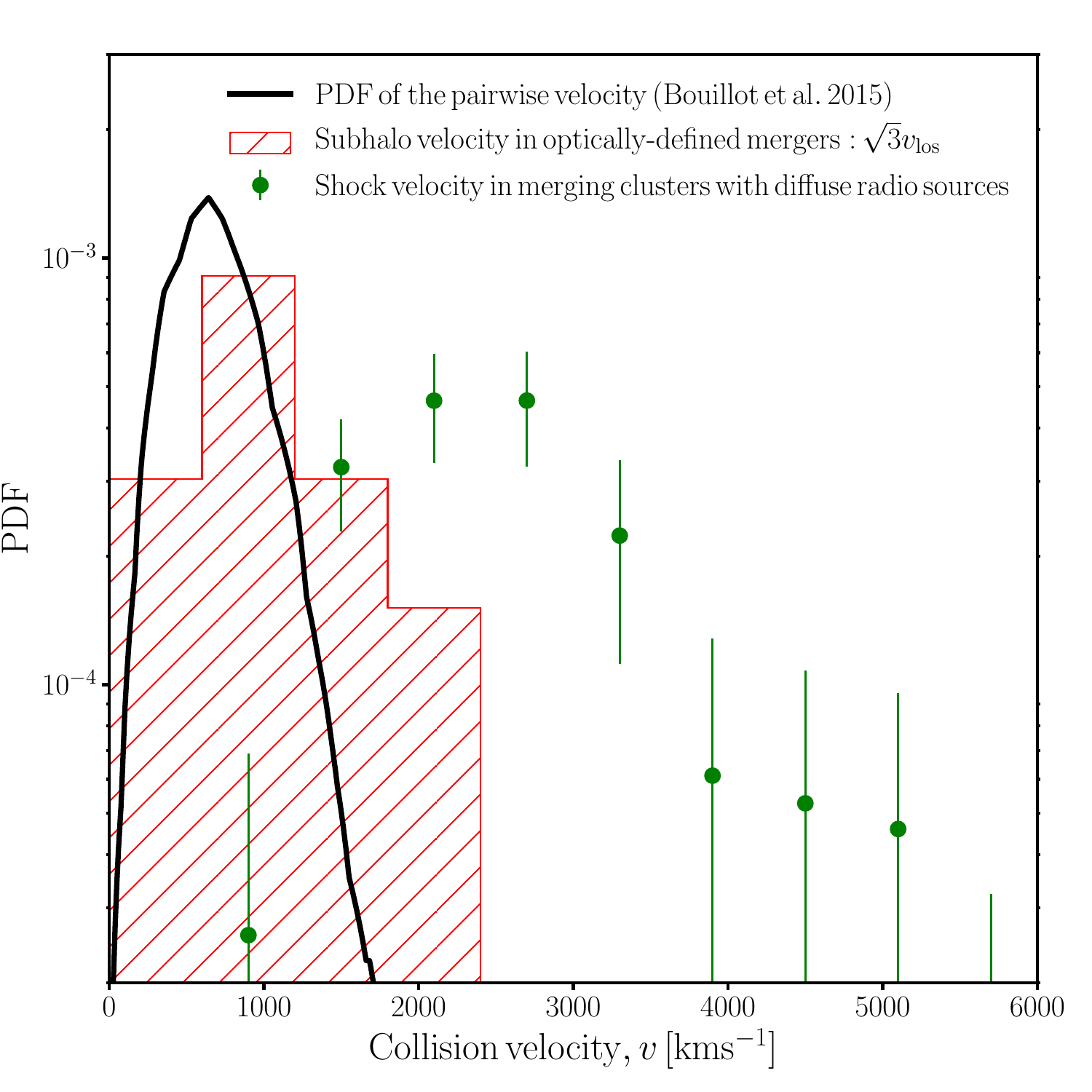}
\end{center}
 \caption{The probability density function (PDF) of collision velocities
 of cluster mergers. The hatched red histogram shows the PDF of
 $\sqrt{3}$-times line-of-sight velocities of the optically-defined
 subhalos. The green circles are the PDF of shock
 velocities of the ICM in merging clusters associated with diffuse radio
 emission. The black
 solid line is a theoretical prediction of the pairwise velocity PDF
 calculated from numerical simulations \citep{Bouillot15}. }
\label{fig:vshock}
\end{figure}

We discuss the non-detection of radio relics and halos from a viewpoint
of the merger collision velocity.
We compare collision velocities of the optical subhalos with shock
velocities in merging clusters hosting diffuse radio emission.
We assume $\sqrt{3}v_{\rm los}$ as the collision velocity of the subhalos,
where $v_{\rm los}$ is the line-of-sight velocity estimated by matching
SDSS spectroscopic redshifts (Sec. \ref{subsec:Sigma_n}).
The shock velocities in massive merging clusters are measured by temperature jumps found in X-ray
observations \citep{Markevitch06,Russell11,Russell12,Akamatsu13,Akamatsu13b,Owers14,Eckert16,Dasadia16,Sarazin16,Emery17}. The shock features are associated with diffuse radio emission.
We assume based on the X-ray morphology that the shock is moving on the
sky plane. The probability density function (PDF) of collision velocities is
shown in Fig. \ref{fig:vshock}. Since the measurement uncertainty of
X-ray shock velocity is relatively large, the mean values and error bars
of shock velocity in each bin are computed from 10000 Monte Carlo redistributions. 
The PDF of collision velocities of the optically-defined mergers are
systematically lower than that of the merging clusters with diffuse radio
emissions. The modes of the collision velocity PDFs differ by a factor of
$2.5$, indicating that the kinetic energy of the optically-defined
mergers is about one order of magnitude lower than that of the merging clusters with
diffuse radio emissions. Assuming a constant conversion from the kinetic
energy of cluster mergers to a synchrotron radio power,
the radio power for optically-defined mergers is expected to be
one order of magnitude lower than the observed values of the radio sources.
It is thus very difficult to detect diffuse radio emission in our
sample in the archival radio data.
The future joint search with the LOFAR Two-metre Sky Survey
\citep[LoTSS][]{Shimwell17,LoTSS18_1} would be powerful to search diffuse radio emissions
 in the merging clusters with the better sensitivity and higher resolution. 
The low-frequency radio telescopes like LOFAR, Murchison Widefield Array
(MWA), the Australian SKA Pathfinder (ASKAP), and Square Kilometre Array
(SKA) will be able to search diffuse radio emission from relativistic
electrons with long lifetimes in merging clusters.

We compare our result with the PDF of the pairwise velocity of cluster mergers in
cosmological simulations \citep{Bouillot15}.
The solid black line in Fig. \ref{fig:vshock} is the PDF at $z=0.5$ retrieved from Fig 12. of
\citet{Bouillot15}. The selection criteria are halo pairs with average
mass $M>10^{14}\hMsol$ and distance separations $<10h^{-1}{\rm Mpc}$ with
$\Lambda$CDM cosmology, which is similar to our sample selection.
The predicted PDF agrees well with our results.
Similar results are also reported by studies of numerical simulations
\citep[e.g.][]{Lee10,Thompson12,Kraljic15}.
\citet{Thompson12} showed that the pair population at a high end of the pairwise velocity distribution
decreases with smaller separations. Specifically, pairs with separations $<2h^{-1}{\rm Mpc}$ have the maximum pairwise velocity on the
order of $\sim 1800\,{\rm kms^{-1}}$ at $z\sim0.3-0.5$. 
However, since the number density of the pairwise velocity at $v\simgt 2000\,{\rm
kms^{-1}}$ is two orders of magnitude lower than that of $v\simlt 1000\,{\rm
kms^{-1}}$, the theoretical prediction in the high probability velocity range
is not significantly changed by a different choice of distance separations.
In contrast to the good agreement between the predicted and observed
subhalo velocities, the shock velocity in the merging clusters with diffuse
radio emissions is systematically higher. One of differences between the
two observing samples is the mass; our sample of clusters has a moderate mass
$M_{200}\sim 3\times10^{14}h_{70}^{-1}\Msol$, whereas the merging
clusters hosting diffuse radio emissions are more massive
\citep[e.g. $\sim 10^{15}h_{70}^{-1}\Msol$;][]{Okabe15b}.
\citet{Kraljic15} showed that the fraction of halo pairs with higher
relative velocities is  larger for massive halos.
A large number of pairs with masses larger than $\sim 10^{15}\hMsol$ and
relative velocities $\simgt 2000\,{\rm kms^{-1}}$ are also found in \citet{Bouillot15}.
Therefore, the discrepancy between the shock and subhalo velocity
distributions can be explained by a selection bias. In other words, the merging
clusters with diffuse radio emission have high collision velocities.

\subsection{Stacked images of low redshift and massive clusters}

\begin{figure*}
\begin{center}
 \includegraphics[width=0.7\hsize]{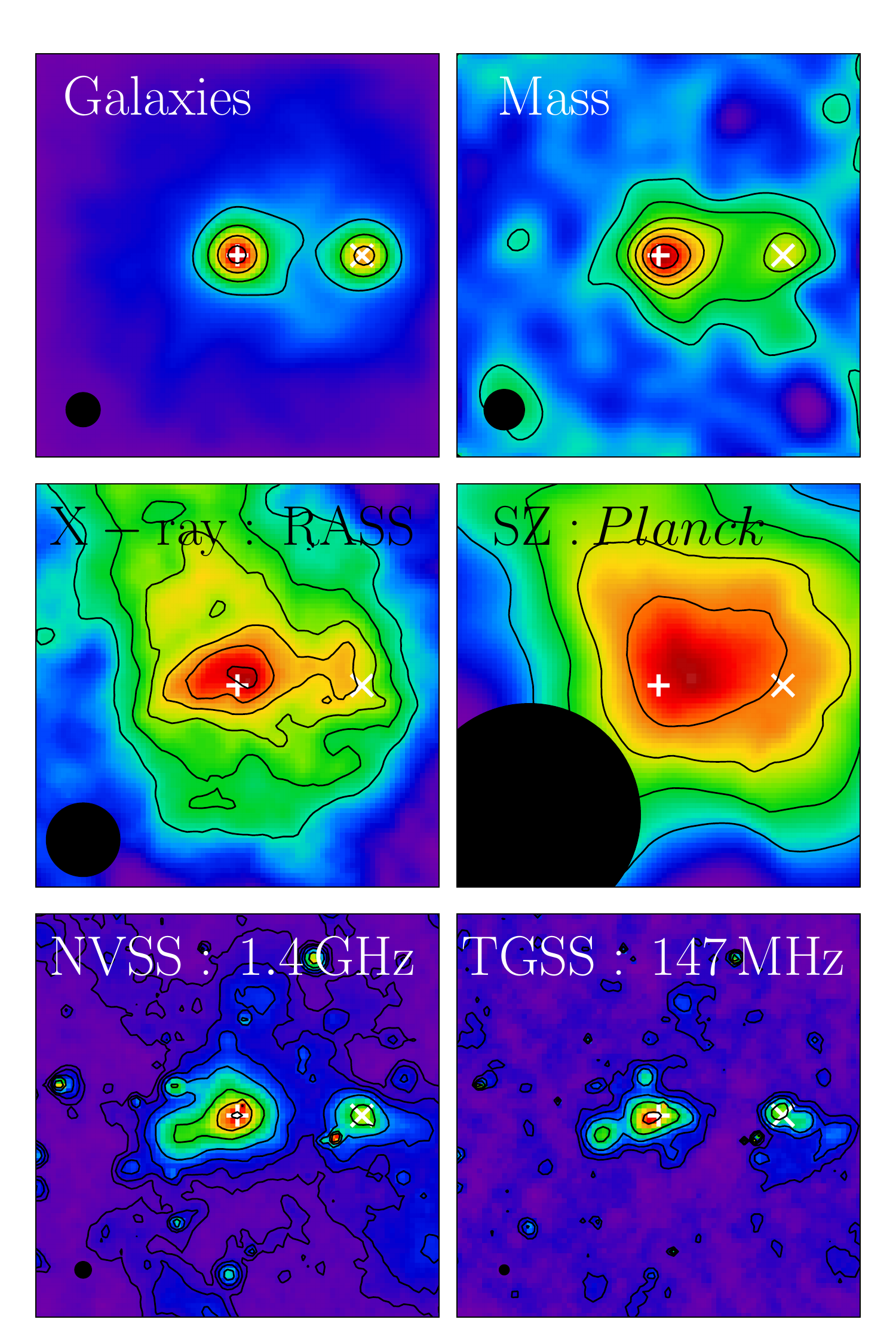}
\end{center}
\caption{Stacked X-ray images (count s$^{-1}$) for 36 merging clusters with
 $z<0.4$ and $N_{\rm cor}>30$. The x-axis is the merger axis.
 The white plus and cross are the
 positions of the highest and second highest peaks in galaxy maps. 
 The black circle (elliptical) at the bottom-left corner of each panel denotes FWHM of each
 stacked image.  {\it Top-left}: The number distribution of luminous red galaxies.
 Overlaid are contours more than 3 galaxies per pixel, stepped by 2 galaxies.
 Two galaxy concentrations are found around the main cluster center and
 the subhalo. {\it Top-right}: The weak lensing mass map.
 Overlaid are contours more than $1\sigma$, stepped by $2\sigma$. Here,
 $\sigma$ is the $1\sigma$ uncertainty of the weak-mass reconstruction.
 {\it Middle-left}: The RASS X-ray map. Overlaid are contours more than
 $2\sigma$, stepped by $1\sigma$, where $\sigma$ is the noise of
 the stacked map. The central gas distribution is
 elongated along the merger axis, whereas the overall gas is oriented
 almost perpendicular to the merger axis. {\it Middle-right}: The {\it Planck} SZ
 map. Overlaid are contours more than $5\sigma$, stepped by
 $2\sigma$. Here, $\sigma$ denotes the noise of stacked $y$ map.
 {\it Bottom-left} The NVSS radio map (1.4 GHz)
 Overlaid are contours are [2,5,10,20,40,60,80]$\times \sigma_{\rm rms}$,
 where $\sigma_{\rm rms}$ is the root mean square (rms) noise in
 the stacked NVSS map.
 {\it Bottom-right} The TGSS radio map (147.5 MHz)
 Overlaid are contours are [2,5,10,20,40]$\times \sigma_{\rm rms}$,
 where $\sigma_{\rm rms}$ is the rms noise in the stacked TGSS map.
We do exclude point sources in X-ray and SZE measurements but do not
 exclude radio point sources to avoid artifact features.}
 \label{fig:aligned_maps}
\end{figure*}

We compute stacked images, by aligning the main-subhalo pairs,
of luminous red galaxies, mass, RASS X-ray, {\it Planck} SZE, NVSS radio
at $1.4$ GHz and TGSS radio at $147$ MHz, in order to understand the interplay
between the baryonic components and dark matter during cluster mergers.
Since the empirical FWHM of the RASS image tracing the gas density is
$\sim3.3$ arcmin, gas structures of low redshift and massive clusters can be
resolved. We therefore select 36 low redshift and massive clusters from the
merging clusters, using the selection criteria of $z<0.4$ and $N_{\rm
cor}>30$. The number of sampling clusters is comparable to the minimum
number of the subsamples in the scaling relation study (Sec. \ref{subsec:Ysz} and table \ref{tab:subsample}).
The maps are rotated such that the cluster and subhalo pair are aligned with
the $x$-axis in the images, with the center of the main peaks, and then rescaled according to their
corresponding angular pair separation.
We denote the $x$-axis as the merger axis.
We excluded known point sources for the X-ray and SZE maps as in the
scaling relation study. As for the NVSS and TGSS images, we do not apply
exclude any point sources or extended radio lobes to avoid
artifact features.
In the X-ray and radio bands, we consider the changes of the
beam (FWHM) sizes of each cluster image, caused by the rotations and
rescalings.
The resulting FWHMs are shown as the black circle at the bottom-left corner
of each panel of Fig. \ref{fig:aligned_maps}.
For a visualization purpose, we do not adopt the weight of lensing
contribution in each map making.

The top-left and top-right panels show the number distribution of luminous red
galaxies and the weak lensing mass map, respectively. The distribution of red galaxies is
concentrated around the main and sub peaks as expected. The
weak-lensing mass map shows two peaks associated with the galaxy peaks
and an elongated distribution bridging between them. The peak-height
ratio between the sub and main halos in the mass map is $\sim0.5$,
slightly lower than that in the galaxy map, $\sim0.8$.

The middle-left and middle-right panels are stacked RASS X-ray and {\it
Planck} SZE maps, respectively.
The X-ray main and second peaks are associated with the main and
subhalos, respectively.
The X-ray flux from the subhalo peak is much smaller than from the main
halo peak, indicating that the gas halo associated with the subhalo is
disrupted by ram-stripping and/or hydrodynamic instabilities.
The peak position of the second X-ray peak is
slightly offset from the subhalo center to the main peak.
Although the trend is not significant, it implies a presence of ram
pressure stripping and that some fraction of the subsample is at the
post merger phase.
The central core of X-ray emission is elongated along the merger axis,
whereas the overall distribution is oriented almost perpendicular to the merger axis.
To quantify the ellipticity, we fit the two dimensional distribution
with an elliptical $\beta$ model convoluted with the PSF size. The
orientation angle of the major axis of the central gas core is $\phi_e=3^{+14}_{-16}$ deg from the
merger axis. On the other hand, the major axis of the overall
distribution is $\phi_e=61_{-4}^{+5}$ deg from the merger axis.
When the image of each cluster is randomly rotated centering the main peak, the
feature disappears. We also compute the stacked images for the
subsamples divided by peak separations.
We found that the elongated gas perpendicular to the merger axis is
significant in the merging clusters with $0.2h_{70}^{-1}\,{\rm Mpc}<d_{\rm sh}<0.6h_{70}^{-1}\,{\rm Mpc}$.
The asymmetrical distribution
suggests that, as the two gas cores are approaching,
the main gas core is stretched along the merger axis, but
the gas halo outside the core is compressed along the merger axis,
and subsequently pushed outwards perpendicular to the axis \citep{Ricker01,ZuHone11,Ha18}.
The expansion is significant after the core passage, which is also
consistent with the offset feature between galaxy and gas due to ram-pressure stripping. 
If there is non-zero spin-parameter, the direction of the expansion is
slightly shifted \citep[e.g.][]{Ricker01}.
Although the elongation feature of main gas cores along the merger axis
and the destruction of gas subhalos were reported in on-going mergers by
X-ray and weak-lensing joint analyses
\citep[e.g.][]{Okabe08,Okabe11,Okabe15b,Medezinski15},
the expanding feature perpendicular to the merger axis is not reported elsewhere.
A caveat is that the optically-defined merging clusters cannot distinguish
between pre- and post- merger phase.
Nevertheless, high X-ray luminosity is found in the
closest-separation merging clusters at $0.2h_{70}^{-1}\,{\rm Mpc}<d_{\rm
sh}<0.6h_{70}^{-1}\,{\rm Mpc}$ (Fig. \ref{fig:Ysz-M}).
The overall distribution of the stacked {\it Planck} SZE map is similar to
the X-ray distribution. Since the {\it Planck} FWHM is much poorer than the
RASS one, we cannot constrain the major axis of the gas halo ellipticity. 
Although the stacked images suggest that the post-merging clusters are
dominant in the low redshift and massive clusters, 
future systematic observations by {\it XMM-Netwon}, {\it eROSITA}, {\it
Chandra}, and ActPol with high angular resolutions will be crucial to
identity cluster dynamical state of each cluster and reveal how gas
features are changed by their merging phase.

The bottom panels show the stacked NVSS and TGSS images.
Very significant radio emissions ($>10\sigma_{\rm rms}$) are found around the main
and sub peaks. The spectral index estimated from the two images is $\sim-1$.
They are radio lobes associated with cluster galaxies.
In the NVSS band, the extended radio distribution at several $\sigma_{\rm
rms}$ is found between the two peaks.
However, since these radio lobes are too bright, less significant radio emissions are
hindered by envelopes of radio lobes, and therefore we 
cannot identify other diffuse emission at outer radii.


\section{Summary} \label{sec:sum}

 We carried out multi-wavelength studies of optically-defined merging
 clusters selected from the Hyper Suprime-Cam Subaru Strategic Program.
 We adopted a peak-finding method of projected distributions of
 luminous red galaxies, and defined  merging clusters with
 multiple-peaks and single clusters with single peak from the HSC CAMIRA cluster catalog.  
Since the number of luminous red galaxies is almost
conserved during cluster mergers, in contrast to a merging cluster
 catalog in X-ray that is significantly affected by the merger boost,
our catalog represents a homogeneous, unbiased sample of cluster mergers.
Our main results are summarized as follows.

\begin{itemize}
\item Our weak lensing analysis has shown that halo concentration for the
      merging clusters is $\sim70\%$ smaller than that of the single
      clusters. The low concentration is in good agreement with predictions
      of numerical simulations \citep[e.g.][]{Neto07,Child18}. The
      concentration parameter decreases as the subhalo distance increases
      within $r_{200}$.

\item We also conducted a two-halo component analysis to interpret the stacked tangential
      shear profile and found that the subhalo mass ratio to the main halo mass is 
      $\simgt 0.1$. Once the subhalo component is modeled separately, the
      concentration of the main halos of the merging clusters is consistent with that of the
      single clusters.
      
 \item The spatial distribution of subhalos is found to be less
       centrally concentrated than the dark matter distribution of
       the main halo, which agrees with theoretical predictions considering
       dynamical friction and tidal stripping of subhalos \citep[e.g.][]{Gao04,Diemand04,Han16}.
      We found that the chance probability of the projection effect of
       surrounding clusters along the line-of-sight is not large at $r\simlt r_{200}$.     

 \item Using the SDSS spectroscopic data, we have shown
       that most of the subhalos are located within the caustics
       boundary in the stacked velocity-space diagram.

\item We estimated red fractions of cluster galaxies based on photometric
      redshifts, and found that the red fractions of the merging
      clusters are comparable to or smaller than those of the single clusters. 

\item We found the tight correlation between the mass and richness
      and no significant deviation of the merging clusters.
      
\item We derived scaling relations of the {\it Planck} SZE ($Y_{\rm SZ}$) and {\it ROSAT} X-ray
      luminosity ($L_X$) with masses and found that their slopes are consistent with and steeper than
      self-similar solutions, respectively.  We estimated the merger boosts on $Y_{\rm
      SZ}$ and $L_X$ taking account of the mass ratio of
      the main and sub clusters and the projection effect.
      The boost factors of $L_X$ and $Y_{\rm SZ}$ are found to be about a factor of
      two for the merging clusters at $0.2-0.6
      h_{70}^{-1}\,{\rm Mpc}$ and $0.6-1.2 h_{70}^{-1}\,{\rm Mpc}$, respectively.

\item  We did not find strong candidates of radio halos and relics in
       our merging cluster sample.

\item  The histogram of collision velocities of our samples is in good
       agreement with cosmological simulations
       \citep[e.g.][]{Bouillot15} and $\sim 2.5$ times lower than
       shock velocities of major mergers with diffuse radio emission. 
      
\item The RASS X-ray stacked image, aligned with the main-subhalo
      pairs, for low redshift and massive clusters, shows that
      the central gas core is elongated along the merger axis and overall gas
      distribution is misaligned by $\sim60$ deg. The {\it Planck} SZE map
      with poor angular resolution shows the similar elongation. 
      The feature indicates that the ICM at large radii are pushed outwards by colliding two clusters.
\end{itemize}

Cross-matching our merging cluster sample 
with the on-going and future surveys ({\it eROSITA}, ACTPol, and LoTSS) and
follow-up X-ray, radio, and spectroscopic observations of merging
clusters at various merging phases
will be crucial to understand the interplay between baryons and dark matter over the
whole process of cluster mergers and constrain the merger boost more precisely. 
Especially, individual cluster analyses will be able to identify merger
phases and their multi-wavelength studies will prove their dynamical
dependence underlying cluster mergers.

Current peak-finding method cannot resolve a transient state of
pericenter of subhalo \citep[e.g.][]{Ludlow12} due to the limitation of
smoothing scale. Further improvement of techniques to find merging
clusters from sparse distributions of member galaxies is vitally important.

\begin{ack}

The Hyper Suprime-Cam (HSC) collaboration includes the astronomical communities of Japan and Taiwan, and Princeton University. The HSC instrumentation and software were developed by the National Astronomical Observatory of Japan (NAOJ), the Kavli Institute for the Physics and Mathematics of the Universe (Kavli IPMU), the University of Tokyo, the High Energy Accelerator Research Organization (KEK), the Academia Sinica Institute for Astronomy and Astrophysics in Taiwan (ASIAA), and Princeton University. Funding was contributed by the FIRST program from Japanese Cabinet Office, the Ministry of Education, Culture, Sports, Science and Technology (MEXT), the Japan Society for the Promotion of Science (JSPS), Japan Science and Technology Agency (JST), the Toray Science Foundation, NAOJ, Kavli IPMU, KEK, ASIAA, and Princeton University. 

This paper makes use of software developed for the Large Synoptic Survey
 Telescope. We thank the LSST Project for making their code available as
 free software at  http://dm.lsst.org.

 The Pan-STARRS1 Surveys (PS1) have been made possible through contributions of the Institute for Astronomy, the University of Hawaii, the Pan-STARRS Project Office, the Max-Planck Society and its participating institutes, the Max Planck Institute for Astronomy, Heidelberg and the Max Planck Institute for Extraterrestrial Physics, Garching, The Johns Hopkins University, Durham University, the University of Edinburgh, Queen’s University Belfast, the Harvard-Smithsonian Center for Astrophysics, the Las Cumbres Observatory Global Telescope Network Incorporated, the National Central University of Taiwan, the Space Telescope Science Institute, the National Aeronautics and Space Administration under Grant No. NNX08AR22G issued through the Planetary Science Division of the NASA Science Mission Directorate, the National Science Foundation under Grant No. AST-1238877, the University of Maryland, and Eotvos Lorand University (ELTE) and the Los Alamos National Laboratory.

Based on data collected at the Subaru Telescope and retrieved from the
 HSC data archive system, which is operated by Subaru Telescope and Astronomy Data Center at National Astronomical Observatory of Japan.

 This work was supported by the Funds for the Development of Human
 Resources in Science and Technology under MEXT, Japan and 
 Core Research for Energetic Universe in Hiroshima University (the MEXT
 program for promoting the enhancement of research universities, Japan).

 M.O. is supported in part by World Premier International Research
 Center Initiative (WPI Initiative), MEXT, Japan, and JSPS KAKENHI Grant
 Number JP15H05892 and JP18K03693.
 H.A acknowledges the support of NWO via a Veni grant. SRON is supported
 financially by NWO, the Netherlands Organization for Scientific Research.

\end{ack}

\bibliographystyle{apj}
\bibliography{my,hsc,hsc_merger,hsc_tech}

\begin{thebibliography}{134}
\expandafter\ifx\csname natexlab\endcsname\relax\def\natexlab#1{#1}\fi

\bibitem[{{Abolfathi} {et~al.}(2018){Abolfathi}, {Aguado}, {Aguilar}, {Allende
  Prieto}, {Almeida}, {Ananna}, {Anders}, {Anderson}, {Andrews}, {Anguiano}, \&
  et~al.}]{Abolfathi18}
{Abolfathi}, B., {Aguado}, D.~S., {Aguilar}, G., {et~al.} 2018, \apjs, 235, 42

\bibitem[{{Aihara} {et~al.}(2018{\natexlab{a}}){Aihara}, {Armstrong},
  {Bickerton}, {Bosch}, {Coupon}, {Furusawa}, {Hayashi}, {Ikeda}, {Kamata},
  {Karoji}, {Kawanomoto}, {Koike}, {Komiyama}, {Lang}, {Lupton}, {Mineo},
  {Miyatake}, {Miyazaki}, {Morokuma}, {Obuchi}, {Oishi}, {Okura}, {Price},
  {Takata}, {Tanaka}, {Tanaka}, {Tanaka}, {Uchida}, {Uraguchi}, {Utsumi},
  {Wang}, {Yamada}, {Yamanoi}, {Yasuda}, {Arimoto}, {Chiba}, {Finet},
  {Fujimori}, {Fujimoto}, {Furusawa}, {Goto}, {Goulding}, {Gunn}, {Harikane},
  {Hattori}, {Hayashi}, {He{\l}miniak}, {Higuchi}, {Hikage}, {Ho}, {Hsieh},
  {Huang}, {Huang}, {Imanishi}, {Iwata}, {Jaelani}, {Jian}, {Kashikawa},
  {Katayama}, {Kojima}, {Konno}, {Koshida}, {Kusakabe}, {Leauthaud}, {Lee},
  {Lin}, {Lin}, {Mandelbaum}, {Matsuoka}, {Medezinski}, {Miyama}, {Momose},
  {More}, {More}, {Mukae}, {Murata}, {Murayama}, {Nagao}, {Nakata}, {Niida},
  {Niikura}, {Nishizawa}, {Oguri}, {Okabe}, {Ono}, {Onodera}, {Onoue}, {Ouchi},
  {Pyo}, {Shibuya}, {Shimasaku}, {Simet}, {Speagle}, {Spergel}, {Strauss},
  {Sugahara}, {Sugiyama}, {Suto}, {Suzuki}, {Tait}, {Takada}, {Terai}, {Toba},
  {Turner}, {Uchiyama}, {Umetsu}, {Urata}, {Usuda}, {Yeh}, \&
  {Yuma}}]{HSC1stDR}
{Aihara}, H., {Armstrong}, R., {Bickerton}, S., {et~al.} 2018{\natexlab{a}},
  \pasj, 70, S8

\bibitem[{{Aihara} {et~al.}(2018{\natexlab{b}}){Aihara}, {Arimoto},
  {Armstrong}, {Arnouts}, {Bahcall}, {Bickerton}, {Bosch}, {Bundy}, {Capak},
  {Chan}, {Chiba}, {Coupon}, {Egami}, {Enoki}, {Finet}, {Fujimori}, {Fujimoto},
  {Furusawa}, {Furusawa}, {Goto}, {Goulding}, {Greco}, {Greene}, {Gunn},
  {Hamana}, {Harikane}, {Hashimoto}, {Hattori}, {Hayashi}, {Hayashi},
  {He{\l}miniak}, {Higuchi}, {Hikage}, {Ho}, {Hsieh}, {Huang}, {Huang},
  {Ikeda}, {Imanishi}, {Inoue}, {Iwasawa}, {Iwata}, {Jaelani}, {Jian},
  {Kamata}, {Karoji}, {Kashikawa}, {Katayama}, {Kawanomoto}, {Kayo}, {Koda},
  {Koike}, {Kojima}, {Komiyama}, {Konno}, {Koshida}, {Koyama}, {Kusakabe},
  {Leauthaud}, {Lee}, {Lin}, {Lin}, {Lupton}, {Mandelbaum}, {Matsuoka},
  {Medezinski}, {Mineo}, {Miyama}, {Miyatake}, {Miyazaki}, {Momose}, {More},
  {More}, {Moritani}, {Moriya}, {Morokuma}, {Mukae}, {Murata}, {Murayama},
  {Nagao}, {Nakata}, {Niida}, {Niikura}, {Nishizawa}, {Obuchi}, {Oguri},
  {Oishi}, {Okabe}, {Okamoto}, {Okura}, {Ono}, {Onodera}, {Onoue}, {Osato},
  {Ouchi}, {Price}, {Pyo}, {Sako}, {Sawicki}, {Shibuya}, {Shimasaku},
  {Shimono}, {Shirasaki}, {Silverman}, {Simet}, {Speagle}, {Spergel},
  {Strauss}, {Sugahara}, {Sugiyama}, {Suto}, {Suyu}, {Suzuki}, {Tait},
  {Takada}, {Takata}, {Tamura}, {Tanaka}, {Tanaka}, {Tanaka}, {Tanaka},
  {Terai}, {Terashima}, {Toba}, {Tominaga}, {Toshikawa}, {Turner}, {Uchida},
  {Uchiyama}, {Umetsu}, {Uraguchi}, {Urata}, {Usuda}, {Utsumi}, {Wang}, {Wang},
  {Wong}, {Yabe}, {Yamada}, {Yamanoi}, {Yasuda}, {Yeh}, {Yonehara}, \&
  {Yuma}}]{HSC1styrOverview}
{Aihara}, H., {Arimoto}, N., {Armstrong}, R., {et~al.} 2018{\natexlab{b}},
  \pasj, 70, S4

\bibitem[{{Akamatsu} {et~al.}(2013){Akamatsu}, {Inoue}, {Sato}, {Matsusita},
  {Ishisaki}, \& {Sarazin}}]{Akamatsu13b}
{Akamatsu}, H., {Inoue}, S., {Sato}, T., {et~al.} 2013, \pasj, 65, 89

\bibitem[{{Akamatsu} \& {Kawahara}(2013)}]{Akamatsu13}
{Akamatsu}, H., \& {Kawahara}, H. 2013, Publications of the Astronomical
  Society of Japan, 65, 16

\bibitem[{{Akamatsu} {et~al.}(2016){Akamatsu}, {Gu}, {Shimwell}, {Mernier},
  {Mao}, {Urdampilleta}, {de Plaa}, {R{\"o}ttgering}, \&
  {Kaastra}}]{Akamatsu16}
{Akamatsu}, H., {Gu}, L., {Shimwell}, T.~W., {et~al.} 2016, \aap, 593, L7

\bibitem[{{Akamatsu} {et~al.}(2017){Akamatsu}, {Mizuno}, {Ota}, {Zhang}, {van
  Weeren}, {Kawahara}, {Fukazawa}, {Kaastra}, {Kawaharada}, {Nakazawa},
  {Ohashi}, {R{\"o}ttgering}, {Takizawa}, {Vink}, \& {Zandanel}}]{Akamatsu17}
{Akamatsu}, H., {Mizuno}, M., {Ota}, N., {et~al.} 2017, \aap, 600, A100

\bibitem[{{Anderson} {et~al.}(2015){Anderson}, {Gaspari}, {White}, {Wang}, \&
  {Dai}}]{Anderson15}
{Anderson}, M.~E., {Gaspari}, M., {White}, S.~D.~M., {Wang}, W., \& {Dai}, X.
  2015, \mnras, 449, 3806

\bibitem[{{Arnaud} {et~al.}(2010){Arnaud}, {Pratt}, {Piffaretti},
  {B{\"o}hringer}, {Croston}, \& {Pointecouteau}}]{Arnaud10}
{Arnaud}, M., {Pratt}, G.~W., {Piffaretti}, R., {et~al.} 2010, \aap, 517, A92

\bibitem[{{Bekki}(1999)}]{Bekki99}
{Bekki}, K. 1999, \apjl, 510, L15

\bibitem[{{Belsole} {et~al.}(2004){Belsole}, {Pratt}, {Sauvageot}, \&
  {Bourdin}}]{Belsole04}
{Belsole}, E., {Pratt}, G.~W., {Sauvageot}, J.-L., \& {Bourdin}, H. 2004, \aap,
  415, 821

\bibitem[{{Bhattacharya} {et~al.}(2013){Bhattacharya}, {Habib}, {Heitmann}, \&
  {Vikhlinin}}]{Bhattacharya13}
{Bhattacharya}, S., {Habib}, S., {Heitmann}, K., \& {Vikhlinin}, A. 2013, \apj,
  766, 32

\bibitem[{{B{\"o}hringer} {et~al.}(2007){B{\"o}hringer}, {Schuecker}, {Pratt},
  {Arnaud}, {Ponman}, {Croston}, {Borgani}, {Bower}, {Briel}, {Collins},
  {Donahue}, {Forman}, {Finoguenov}, {Geller}, {Guzzo}, {Henry}, {Kneissl},
  {Mohr}, {Matsushita}, {Mullis}, {Ohashi}, {Pedersen}, {Pierini}, {Quintana},
  {Raychaudhury}, {Reiprich}, {Romer}, {Rosati}, {Sabirli}, {Temple}, {Viana},
  {Vikhlinin}, {Voit}, \& {Zhang}}]{Bohringer07}
{B{\"o}hringer}, H., {Schuecker}, P., {Pratt}, G.~W., {et~al.} 2007, \aap, 469,
  363

\bibitem[{{Boller} {et~al.}(2016){Boller}, {Freyberg}, {Tr{\"u}mper}, {Haberl},
  {Voges}, \& {Nandra}}]{Bolle16}
{Boller}, T., {Freyberg}, M.~J., {Tr{\"u}mper}, J., {et~al.} 2016, \aap, 588,
  A103

\bibitem[{{Bosch} {et~al.}(2018){Bosch}, {Armstrong}, {Bickerton}, {Furusawa},
  {Ikeda}, {Koike}, {Lupton}, {Mineo}, {Price}, {Takata}, {Tanaka}, {Yasuda},
  {AlSayyad}, {Becker}, {Coulton}, {Coupon}, {Garmilla}, {Huang}, {Krughoff},
  {Lang}, {Leauthaud}, {Lim}, {Lust}, {MacArthur}, {Mandelbaum}, {Miyatake},
  {Miyazaki}, {Murata}, {More}, {Okura}, {Owen}, {Swinbank}, {Strauss},
  {Yamada}, \& {Yamanoi}}]{Bosch18HSC}
{Bosch}, J., {Armstrong}, R., {Bickerton}, S., {et~al.} 2018, \pasj, 70, S5

\bibitem[{{Bouillot} {et~al.}(2015){Bouillot}, {Alimi}, {Corasaniti}, \&
  {Rasera}}]{Bouillot15}
{Bouillot}, V.~R., {Alimi}, J.-M., {Corasaniti}, P.-S., \& {Rasera}, Y. 2015,
  \mnras, 450, 145

\bibitem[{{Brunetti} {et~al.}(2008){Brunetti}, {Giacintucci}, {Cassano},
  {Lane}, {Dallacasa}, {Venturi}, {Kassim}, {Setti}, {Cotton}, \&
  {Markevitch}}]{Brunetti08}
{Brunetti}, G., {Giacintucci}, S., {Cassano}, R., {et~al.} 2008, \nat, 455, 944

\bibitem[{{Bullock} {et~al.}(2001){Bullock}, {Kolatt}, {Sigad}, {Somerville},
  {Kravtsov}, {Klypin}, {Primack}, \& {Dekel}}]{Bullock01}
{Bullock}, J.~S., {Kolatt}, T.~S., {Sigad}, Y., {et~al.} 2001, \mnras, 321, 559

\bibitem[{{Butcher} \& {Oemler}(1984)}]{Butcher84}
{Butcher}, H., \& {Oemler}, Jr., A. 1984, \apj, 285, 426

\bibitem[{{Carrasco Kind} \& {Brunner}(2014)}]{MLZ14}
{Carrasco Kind}, M., \& {Brunner}, R.~J. 2014, \mnras, 438, 3409

\bibitem[{{Child} {et~al.}(2018){Child}, {Habib}, {Heitmann}, {Frontiere},
  {Finkel}, {Pope}, \& {Morozov}}]{Child18}
{Child}, H.~L., {Habib}, S., {Heitmann}, K., {et~al.} 2018, \apj, 859, 55

\bibitem[{{Cibirka} {et~al.}(2017){Cibirka}, {Cypriano}, {Brimioulle}, {Gruen},
  {Erben}, {van Waerbeke}, {Miller}, {Finoguenov}, {Kirkpatrick}, {Henry},
  {Rykoff}, {Rozo}, {Dupke}, {Kneib}, {Shan}, \& {Spinelli}}]{Cibirka17}
{Cibirka}, N., {Cypriano}, E.~S., {Brimioulle}, F., {et~al.} 2017, \mnras, 468,
  1092

\bibitem[{{Condon} {et~al.}(1998){Condon}, {Cotton}, {Greisen}, {Yin},
  {Perley}, {Taylor}, \& {Broderick}}]{Condon98NVSS}
{Condon}, J.~J., {Cotton}, W.~D., {Greisen}, E.~W., {et~al.} 1998, \aj, 115,
  1693

\bibitem[{{Coupon} {et~al.}(2018){Coupon}, {Czakon}, {Bosch}, {Komiyama},
  {Medezinski}, {Miyazaki}, \& {Oguri}}]{Coupon18HSC}
{Coupon}, J., {Czakon}, N., {Bosch}, J., {et~al.} 2018, \pasj, 70, S7

\bibitem[{{Dasadia} {et~al.}(2016){Dasadia}, {Sun}, {Sarazin}, {Morandi},
  {Markevitch}, {Wik}, {Feretti}, {Giovannini}, {Govoni}, \&
  {Vacca}}]{Dasadia16}
{Dasadia}, S., {Sun}, M., {Sarazin}, C., {et~al.} 2016, \apjl, 820, L20

\bibitem[{{de Gasperin} {et~al.}(2018){de Gasperin}, {Intema}, \&
  {Frail}}]{deGasperin18}
{de Gasperin}, F., {Intema}, H.~T., \& {Frail}, D.~A. 2018, \mnras, 474, 5008

\bibitem[{{Deshev} {et~al.}(2017){Deshev}, {Finoguenov}, {Verdugo}, {Ziegler},
  {Park}, {Hwang}, {Haines}, {Kamphuis}, {Tamm}, {Einasto}, {Hwang}, \&
  {Park}}]{Deshev17}
{Deshev}, B., {Finoguenov}, A., {Verdugo}, M., {et~al.} 2017, \aap, 607, A131

\bibitem[{{Diaferio} \& {Geller}(1997)}]{Diaferio97}
{Diaferio}, A., \& {Geller}, M.~J. 1997, \apj, 481, 633

\bibitem[{{Diemand} {et~al.}(2004){Diemand}, {Moore}, \& {Stadel}}]{Diemand04}
{Diemand}, J., {Moore}, B., \& {Stadel}, J. 2004, \mnras, 352, 535

\bibitem[{{Diemer} \& {Kravtsov}(2015)}]{Diemer15}
{Diemer}, B., \& {Kravtsov}, A.~V. 2015, \apj, 799, 108

\bibitem[{{Dressler} \& {Gunn}(1983)}]{Dressler83}
{Dressler}, A., \& {Gunn}, J.~E. 1983, \apj, 270, 7

\bibitem[{{Duffy} {et~al.}(2008){Duffy}, {Schaye}, {Kay}, \& {Dalla
  Vecchia}}]{Duffy08}
{Duffy}, A.~R., {Schaye}, J., {Kay}, S.~T., \& {Dalla Vecchia}, C. 2008,
  \mnras, 390, L64

\bibitem[{{Eckert} {et~al.}(2016){Eckert}, {Jauzac}, {Vazza}, {Owers}, {Kneib},
  {Tchernin}, {Intema}, \& {Knowles}}]{Eckert16}
{Eckert}, D., {Jauzac}, M., {Vazza}, F., {et~al.} 2016, \mnras, 461, 1302

\bibitem[{{Emery} {et~al.}(2017){Emery}, {Bogd{\'a}n}, {Kraft},
  {Andrade-Santos}, {Forman}, {Hardcastle}, \& {Jones}}]{Emery17}
{Emery}, D.~L., {Bogd{\'a}n}, {\'A}., {Kraft}, R.~P., {et~al.} 2017, \apj, 834,
  159

\bibitem[{{Evrard}(1991)}]{Evrard91}
{Evrard}, A.~E. 1991, \mnras, 248, 8P

\bibitem[{{Feretti} {et~al.}(2012){Feretti}, {Giovannini}, {Govoni}, \&
  {Murgia}}]{Feretti12}
{Feretti}, L., {Giovannini}, G., {Govoni}, F., \& {Murgia}, M. 2012, \aapr, 20,
  54

\bibitem[{{Ferrari} {et~al.}(2005){Ferrari}, {Benoist}, {Maurogordato},
  {Cappi}, \& {Slezak}}]{Ferrari05}
{Ferrari}, C., {Benoist}, C., {Maurogordato}, S., {Cappi}, A., \& {Slezak}, E.
  2005, \aap, 430, 19

\bibitem[{{Fujita} {et~al.}(1999){Fujita}, {Takizawa}, {Nagashima}, \&
  {Enoki}}]{Fujita99b}
{Fujita}, Y., {Takizawa}, M., {Nagashima}, M., \& {Enoki}, M. 1999, \pasj, 51,
  L1

\bibitem[{{Fujita} {et~al.}(2003){Fujita}, {Takizawa}, \& {Sarazin}}]{Fujita03}
{Fujita}, Y., {Takizawa}, M., \& {Sarazin}, C.~L. 2003, \apj, 584, 190

\bibitem[{{Fujita} {et~al.}(2015){Fujita}, {Takizawa}, {Yamazaki}, {Akamatsu},
  \& {Ohno}}]{Fujita15}
{Fujita}, Y., {Takizawa}, M., {Yamazaki}, R., {Akamatsu}, H., \& {Ohno}, H.
  2015, \apj, 815, 116

\bibitem[{{Furusawa} {et~al.}(2018){Furusawa}, {Koike}, {Takata}, {Okura},
  {Miyatake}, {Lupton}, {Bickerton}, {Price}, {Bosch}, {Yasuda}, {Mineo},
  {Yamada}, {Miyazaki}, {Nakata}, {Koshida}, {Komiyama}, {Utsumi},
  {Kawanomoto}, {Jeschke}, {Noumaru}, {Schubert}, {Iwata}, {Finet},
  {Fujiyoshi}, {Tajitsu}, {Terai}, \& {Lee}}]{Furusawa18HSC}
{Furusawa}, H., {Koike}, M., {Takata}, T., {et~al.} 2018, \pasj, 70, S3

\bibitem[{{Gao} {et~al.}(2004){Gao}, {White}, {Jenkins}, {Stoehr}, \&
  {Springel}}]{Gao04}
{Gao}, L., {White}, S.~D.~M., {Jenkins}, A., {Stoehr}, F., \& {Springel}, V.
  2004, \mnras, 355, 819

\bibitem[{{G{\'o}rski} {et~al.}(2005){G{\'o}rski}, {Hivon}, {Banday},
  {Wandelt}, {Hansen}, {Reinecke}, \& {Bartelmann}}]{HEALPix05}
{G{\'o}rski}, K.~M., {Hivon}, E., {Banday}, A.~J., {et~al.} 2005, \apj, 622,
  759

\bibitem[{{Govoni} {et~al.}(2004){Govoni}, {Markevitch}, {Vikhlinin}, {van
  Speybroeck}, {Feretti}, \& {Giovannini}}]{Govoni04}
{Govoni}, F., {Markevitch}, M., {Vikhlinin}, A., {et~al.} 2004, \apj, 605, 695

\bibitem[{{Ha} {et~al.}(2018){Ha}, {Ryu}, \& {Kang}}]{Ha18}
{Ha}, J.-H., {Ryu}, D., \& {Kang}, H. 2018, \apj, 857, 26

\bibitem[{{Han} {et~al.}(2016){Han}, {Cole}, {Frenk}, \& {Jing}}]{Han16}
{Han}, J., {Cole}, S., {Frenk}, C.~S., \& {Jing}, Y. 2016, \mnras, 457, 1208

\bibitem[{{Hirata} \& {Seljak}(2003)}]{Hirata03}
{Hirata}, C., \& {Seljak}, U. 2003, \mnras, 343, 459

\bibitem[{{Huang} {et~al.}(2018){Huang}, {Leauthaud}, {Murata}, {Bosch},
  {Price}, {Lupton}, {Mandelbaum}, {Lackner}, {Bickerton}, {Miyazaki},
  {Coupon}, \& {Tanaka}}]{Haung18HSC}
{Huang}, S., {Leauthaud}, A., {Murata}, R., {et~al.} 2018, \pasj, 70, S6

\bibitem[{{Hurier} {et~al.}(2013){Hurier}, {Mac{\'{\i}}as-P{\'e}rez}, \&
  {Hildebrandt}}]{Hurier13}
{Hurier}, G., {Mac{\'{\i}}as-P{\'e}rez}, J.~F., \& {Hildebrandt}, S. 2013,
  \aap, 558, A118

\bibitem[{{Intema} {et~al.}(2017){Intema}, {Jagannathan}, {Mooley}, \&
  {Frail}}]{Intema17TGSS}
{Intema}, H.~T., {Jagannathan}, P., {Mooley}, K.~P., \& {Frail}, D.~A. 2017,
  \aap, 598, A78

\bibitem[{{Johnston} {et~al.}(2007){Johnston}, {Sheldon}, {Wechsler}, {Rozo},
  {Koester}, {Frieman}, {McKay}, {Evrard}, {Becker}, \& {Annis}}]{Johnston07}
{Johnston}, D.~E., {Sheldon}, E.~S., {Wechsler}, R.~H., {et~al.} 2007, ArXiv
  e-prints,0709.1159

\bibitem[{{Kaiser}(1987)}]{Kaiser87}
{Kaiser}, N. 1987, \mnras, 227, 1

\bibitem[{{Kapferer} {et~al.}(2006){Kapferer}, {Ferrari}, {Domainko}, {Mair},
  {Kronberger}, {Schindler}, {Kimeswenger}, {van Kampen}, {Breitschwerdt}, \&
  {Ruffert}}]{Kapferer06}
{Kapferer}, W., {Ferrari}, C., {Domainko}, W., {et~al.} 2006, \aap, 447, 827

\bibitem[{{Kawanomoto} {et~al.}(2018){Kawanomoto}, {Uraguchi}, {Komiyama},
  {Miyazaki}, {Furusawa}, {Finet}, {Hattori}, {Wang}, {Yasuda}, \&
  {Suzuki}}]{Kawanomoto18HSC}
{Kawanomoto}, S., {Uraguchi}, F., {Komiyama}, Y., {et~al.} 2018, \pasj, 70, 66

\bibitem[{{Kay} {et~al.}(2012){Kay}, {Peel}, {Short}, {Thomas}, {Young},
  {Battye}, {Liddle}, \& {Pearce}}]{Kay12}
{Kay}, S.~T., {Peel}, M.~W., {Short}, C.~J., {et~al.} 2012, \mnras, 422, 1999

\bibitem[{{Koester} {et~al.}(2007){Koester}, {McKay}, {Annis}, {Wechsler},
  {Evrard}, {Bleem}, {Becker}, {Johnston}, {Sheldon}, {Nichol}, {Miller},
  {Scranton}, {Bahcall}, {Barentine}, {Brewington}, {Brinkmann}, {Harvanek},
  {Kleinman}, {Krzesinski}, {Long}, {Nitta}, {Schneider}, {Sneddin}, {Voges},
  \& {York}}]{Koester07}
{Koester}, B.~P., {McKay}, T.~A., {Annis}, J., {et~al.} 2007, \apj, 660, 239

\bibitem[{{Komiyama} {et~al.}(2018){Komiyama}, {Chiba}, {Tanaka}, {Tanaka},
  {Kirihara}, {Miki}, {Mori}, {Lupton}, {Guhathakurta}, {Kalirai}, {Gilbert},
  {Kirby}, {Lee}, {Jang}, {Sharma}, \& {Hayashi}}]{Komiyama18HSC}
{Komiyama}, Y., {Chiba}, M., {Tanaka}, M., {et~al.} 2018, \apj, 853, 29

\bibitem[{{Koyama} {et~al.}(2010){Koyama}, {Kodama}, {Shimasaku}, {Hayashi},
  {Okamura}, {Tanaka}, \& {Tokoku}}]{Koyama10}
{Koyama}, Y., {Kodama}, T., {Shimasaku}, K., {et~al.} 2010, \mnras, 403, 1611

\bibitem[{{Kraljic} \& {Sarkar}(2015)}]{Kraljic15}
{Kraljic}, D., \& {Sarkar}, S. 2015, \jcap, 4, 050

\bibitem[{{Krause} {et~al.}(2012){Krause}, {Pierpaoli}, {Dolag}, \&
  {Borgani}}]{Krause12}
{Krause}, E., {Pierpaoli}, E., {Dolag}, K., \& {Borgani}, S. 2012, \mnras, 419,
  1766

\bibitem[{{Kronberger} {et~al.}(2008){Kronberger}, {Kapferer}, {Ferrari},
  {Unterguggenberger}, \& {Schindler}}]{Kronberger08}
{Kronberger}, T., {Kapferer}, W., {Ferrari}, C., {Unterguggenberger}, S., \&
  {Schindler}, S. 2008, \aap, 481, 337

\bibitem[{{Lavery} \& {Henry}(1988)}]{Lavery88}
{Lavery}, R.~J., \& {Henry}, J.~P. 1988, \apj, 330, 596

\bibitem[{{Lee} \& {Komatsu}(2010)}]{Lee10}
{Lee}, J., \& {Komatsu}, E. 2010, \apj, 718, 60

\bibitem[{{Ludlow} {et~al.}(2014){Ludlow}, {Navarro}, {Angulo},
  {Boylan-Kolchin}, {Springel}, {Frenk}, \& {White}}]{Ludlow14}
{Ludlow}, A.~D., {Navarro}, J.~F., {Angulo}, R.~E., {et~al.} 2014, \mnras, 441,
  378

\bibitem[{{Ludlow} {et~al.}(2012){Ludlow}, {Navarro}, {Li}, {Angulo},
  {Boylan-Kolchin}, \& {Bett}}]{Ludlow12}
{Ludlow}, A.~D., {Navarro}, J.~F., {Li}, M., {et~al.} 2012, \mnras, 427, 1322

\bibitem[{{Ludlow} {et~al.}(2009){Ludlow}, {Navarro}, {Springel}, {Jenkins},
  {Frenk}, \& {Helmi}}]{Ludlow09}
{Ludlow}, A.~D., {Navarro}, J.~F., {Springel}, V., {et~al.} 2009, \apj, 692,
  931

\bibitem[{{Mamon} {et~al.}(2013){Mamon}, {Biviano}, \& {Bou{\'e}}}]{Mamon13}
{Mamon}, G.~A., {Biviano}, A., \& {Bou{\'e}}, G. 2013, \mnras, 429, 3079

\bibitem[{{Mandelbaum} {et~al.}(2006){Mandelbaum}, {Seljak}, {Kauffmann},
  {Hirata}, \& {Brinkmann}}]{Mandelbaum06}
{Mandelbaum}, R., {Seljak}, U., {Kauffmann}, G., {Hirata}, C.~M., \&
  {Brinkmann}, J. 2006, \mnras, 368, 715

\bibitem[{{Mandelbaum} {et~al.}(2017){Mandelbaum}, {Lanusse}, {Leauthaud},
  {Armstrong}, {Simet}, {Miyatake}, {Meyers}, {Bosch}, {Miyazaki}, \&
  {Tanaka}}]{Mandelbaum17}
{Mandelbaum}, R., {Lanusse}, F., {Leauthaud}, A., {et~al.} 2017,
  arXiv:171000885

\bibitem[{{Mandelbaum} {et~al.}(2018){Mandelbaum}, {Miyatake}, {Hamana},
  {Oguri}, {Simet}, {Armstrong}, {Bosch}, {Murata}, {Lanusse}, {Leauthaud},
  {Coupon}, {More}, {Takada}, {Miyazaki}, {Speagle}, {Shirasaki}, {Sif{\'o}n},
  {Huang}, {Nishizawa}, {Medezinski}, {Okura}, {Okabe}, {Czakon}, {Takahashi},
  {Coulton}, {Hikage}, {Komiyama}, {Lupton}, {Strauss}, {Tanaka}, \&
  {Utsumi}}]{HSCWL1styr}
{Mandelbaum}, R., {Miyatake}, H., {Hamana}, T., {et~al.} 2018, \pasj, 70, S25

\bibitem[{{Markevitch}(2006)}]{Markevitch06}
{Markevitch}, M. 2006, in ESA Special Publication, Vol. 604, The X-ray Universe
  2005, ed. A.~{Wilson}, 723

\bibitem[{{Medezinski} {et~al.}(2015){Medezinski}, {Umetsu}, {Okabe}, {Nonino},
  {Molnar}, {Massey}, {Dupke}, \& {Merten}}]{Medezinski15}
{Medezinski}, E., {Umetsu}, K., {Okabe}, N., {et~al.} 2015, ArXiv e-prints

\bibitem[{{Medezinski} {et~al.}(2018){Medezinski}, {Oguri}, {Nishizawa},
  {Speagle}, {Miyatake}, {Umetsu}, {Leauthaud}, {Murata}, {Mandelbaum},
  {Sif{\'o}n}, {Strauss}, {Huang}, {Simet}, {Okabe}, {Tanaka}, \&
  {Komiyama}}]{Medezinski18}
{Medezinski}, E., {Oguri}, M., {Nishizawa}, A.~J., {et~al.} 2018, \pasj, 70, 30

\bibitem[{{Meneghetti} {et~al.}(2014){Meneghetti}, {Rasia}, {Vega}, {Merten},
  {Postman}, {Yepes}, {Sembolini}, {Donahue}, {Ettori}, {Umetsu}, {Balestra},
  {Bartelmann}, {Benitez}, {Biviano}, {Bouwens}, {Bradley}, {Broadhurst},
  {Coe}, {Czakon}, {De Petris}, {Ford}, {Giocoli}, {Gottloeber}, {Grillo},
  {Infante}, {Jouvel}, {Kelson}, {Koekemoer}, {Lahav}, {Lemze}, {Medezinski},
  {Melchior}, {Mercurio}, {Molino}, {Moscardini}, {Monna}, {Moustakas},
  {Moustakas}, {Nonino}, {Rhodes}, {Rosati}, {Sayers}, {Seitz}, {Zheng}, \&
  {Zitrin}}]{Meneghetti14}
{Meneghetti}, M., {Rasia}, E., {Vega}, J., {et~al.} 2014, ArXiv e-prints

\bibitem[{{Merten} {et~al.}(2015){Merten}, {Meneghetti}, {Postman}, {Umetsu},
  {Zitrin}, {Medezinski}, {Nonino}, {Koekemoer}, {Melchior}, {Gruen},
  {Moustakas}, {Bartelmann}, {Host}, {Donahue}, {Coe}, {Molino}, {Jouvel},
  {Monna}, {Seitz}, {Czakon}, {Lemze}, {Sayers}, {Balestra}, {Rosati},
  {Ben{\'{\i}}tez}, {Biviano}, {Bouwens}, {Bradley}, {Broadhurst}, {Carrasco},
  {Ford}, {Grillo}, {Infante}, {Kelson}, {Lahav}, {Massey}, {Moustakas},
  {Rasia}, {Rhodes}, {Vega}, \& {Zheng}}]{Merten15}
{Merten}, J., {Meneghetti}, M., {Postman}, M., {et~al.} 2015, \apj, 806, 4

\bibitem[{{Miyaoka} {et~al.}(2018){Miyaoka}, {Okabe}, {Kitaguchi}, {Oguri},
  {Fukazawa}, {Mandelbaum}, {Medezinski}, {Babazaki}, {Nishizawa}, {Hamana},
  {Lin}, {Akamatsu}, {Chiu}, {Fujita}, {Ichinohe}, {Komiyama}, {Sasaki},
  {Takizawa}, {Ueda}, {Umetsu}, {Coupon}, {Hikage}, {Hoshino}, {Leauthaud},
  {Matsushita}, {Mitsuishi}, {Miyatake}, {Miyazaki}, {More}, {Nakazawa}, {Ota},
  {Sato}, {Spergel}, {Tamura}, {Tanaka}, {Tanaka}, \& {Utsumi}}]{Miyaoka18}
{Miyaoka}, K., {Okabe}, N., {Kitaguchi}, T., {et~al.} 2018, \pasj, 70, S22

\bibitem[{{Miyatake} {et~al.}(2018){Miyatake}, {Battaglia}, {Hilton},
  {Medezinski}, {Nishizawa}, {More}, {Aiola}, {Bahcall}, {Bond}, {Calabrese},
  {Choi}, {Devlin}, {Dunkley}, {Dunner}, {Fuzia}, {Gallardo}, {Gralla},
  {Hasselfield}, {Halpern}, {Hikage}, {Hill}, {Hincks}, {Hlo{\v z}ek},
  {Huffenberger}, {Hughes}, {Koopman}, {Kosowsky}, {Louis}, {Madhavacheril},
  {McMahon}, {Mandelbaum}, {Marriage}, {Maurin}, {Miyazaki}, {Moodley},
  {Murata}, {Naess}, {Newburgh}, {Niemack}, {Nishimichi}, {Okabe}, {Oguri},
  {Osato}, {Page}, {Partridge}, {Robertson}, {Sehgal}, {Shirasaki}, {Sievers},
  {Sif{\'o}n}, {Simon}, {Sherwin}, {Spergel}, {Staggs}, {Stein}, {Takada},
  {Trac}, {Umetsu}, {van Engelen}, \& {Wollack}}]{Miyatake18}
{Miyatake}, H., {Battaglia}, N., {Hilton}, M., {et~al.} 2018, arXiv:1804.05873

\bibitem[{{Miyazaki} {et~al.}(2015){Miyazaki}, {Oguri}, {Hamana}, {Tanaka},
  {Miller}, {Utsumi}, {Komiyama}, {Furusawa}, {Sakurai}, {Kawanomoto},
  {Nakata}, {Uraguchi}, {Koike}, {Tomono}, {Lupton}, {Gunn}, {Karoji},
  {Aihara}, {Murayama}, \& {Takada}}]{Miyazaki15}
{Miyazaki}, S., {Oguri}, M., {Hamana}, T., {et~al.} 2015, \apj, 807, 22

\bibitem[{{Miyazaki} {et~al.}(2018{\natexlab{a}}){Miyazaki}, {Oguri}, {Hamana},
  {Shirasaki}, {Koike}, {Komiyama}, {Umetsu}, {Utsumi}, {Okabe}, {More},
  {Medezinski}, {Lin}, {Miyatake}, {Murayama}, {Ota}, \&
  {Mitsuishi}}]{Miyazaki18}
---. 2018{\natexlab{a}}, \pasj, 70, S27

\bibitem[{{Miyazaki} {et~al.}(2018{\natexlab{b}}){Miyazaki}, {Komiyama},
  {Kawanomoto}, {Doi}, {Furusawa}, {Hamana}, {Hayashi}, {Ikeda}, {Kamata},
  {Karoji}, {Koike}, {Kurakami}, {Miyama}, {Morokuma}, {Nakata}, {Namikawa},
  {Nakaya}, {Nariai}, {Obuchi}, {Oishi}, {Okada}, {Okura}, {Tait}, {Takata},
  {Tanaka}, {Tanaka}, {Terai}, {Tomono}, {Uraguchi}, {Usuda}, {Utsumi},
  {Yamada}, {Yamanoi}, {Aihara}, {Fujimori}, {Mineo}, {Miyatake}, {Oguri},
  {Uchida}, {Tanaka}, {Yasuda}, {Takada}, {Murayama}, {Nishizawa}, {Sugiyama},
  {Chiba}, {Futamase}, {Wang}, {Chen}, {Ho}, {Liaw}, {Chiu}, {Ho}, {Lai},
  {Lee}, {Jeng}, {Iwamura}, {Armstrong}, {Bickerton}, {Bosch}, {Gunn},
  {Lupton}, {Loomis}, {Price}, {Smith}, {Strauss}, {Turner}, {Suzuki},
  {Miyazaki}, {Muramatsu}, {Yamamoto}, {Endo}, {Ezaki}, {Ito}, {Kawaguchi},
  {Sofuku}, {Taniike}, {Akutsu}, {Dojo}, {Kasumi}, {Matsuda}, {Imoto}, {Miwa},
  {Suzuki}, {Takeshi}, \& {Yokota}}]{Miyazaki18HSC}
{Miyazaki}, S., {Komiyama}, Y., {Kawanomoto}, S., {et~al.} 2018{\natexlab{b}},
  \pasj, 70, S1

\bibitem[{{Miyazaki} {et~al.}(2018{\natexlab{c}}){Miyazaki}, {Komiyama},
  {Kawanomoto}, {Doi}, {Furusawa}, {Hamana}, {Hayashi}, {Ikeda}, {Kamata},
  {Karoji}, {Koike}, {Kurakami}, {Miyama}, {Morokuma}, {Nakata}, {Namikawa},
  {Nakaya}, {Nariai}, {Obuchi}, {Oishi}, {Okada}, {Okura}, {Tait}, {Takata},
  {Tanaka}, {Tanaka}, {Terai}, {Tomono}, {Uraguchi}, {Usuda}, {Utsumi},
  {Yamada}, {Yamanoi}, {Aihara}, {Fujimori}, {Mineo}, {Miyatake}, {Oguri},
  {Uchida}, {Tanaka}, {Yasuda}, {Takada}, {Murayama}, {Nishizawa}, {Sugiyama},
  {Chiba}, {Futamase}, {Wang}, {Chen}, {Ho}, {Liaw}, {Chiu}, {Ho}, {Lai},
  {Lee}, {Jeng}, {Iwamura}, {Armstrong}, {Bickerton}, {Bosch}, {Gunn},
  {Lupton}, {Loomis}, {Price}, {Smith}, {Strauss}, {Turner}, {Suzuki},
  {Miyazaki}, {Muramatsu}, {Yamamoto}, {Endo}, {Ezaki}, {Ito}, {Kawaguchi},
  {Sofuku}, {Taniike}, {Akutsu}, {Dojo}, {Kasumi}, {Matsuda}, {Imoto}, {Miwa},
  {Suzuki}, {Takeshi}, \& {Yokota}}]{HSCcam}
---. 2018{\natexlab{c}}, \pasj, 70, S1

\bibitem[{{Nagai} \& {Kravtsov}(2005)}]{Nagai05}
{Nagai}, D., \& {Kravtsov}, A.~V. 2005, \apj, 618, 557

\bibitem[{{Navarro} {et~al.}(1996){Navarro}, {Frenk}, \& {White}}]{NFW96}
{Navarro}, J.~F., {Frenk}, C.~S., \& {White}, S.~D.~M. 1996, \apj, 462, 563

\bibitem[{{Neto} {et~al.}(2007){Neto}, {Gao}, {Bett}, {Cole}, {Navarro},
  {Frenk}, {White}, {Springel}, \& {Jenkins}}]{Neto07}
{Neto}, A.~F., {Gao}, L., {Bett}, P., {et~al.} 2007, \mnras, 381, 1450

\bibitem[{{Nishizawa} {et~al.}(2018){Nishizawa}, {Oguri}, {Oogi}, {More},
  {Nishimichi}, {Nagashima}, {Lin}, {Mandelbaum}, {Takada}, {Bahcall},
  {Coupon}, {Huang}, {Jian}, {Komiyama}, {Leauthaud}, {Lin}, {Miyatake},
  {Miyazaki}, \& {Tanaka}}]{Nishizawa18}
{Nishizawa}, A.~J., {Oguri}, M., {Oogi}, T., {et~al.} 2018, \pasj, 70, S24

\bibitem[{{Oguri}(2014)}]{Oguri14b}
{Oguri}, M. 2014, \mnras, 444, 147

\bibitem[{{Oguri} {et~al.}(2012){Oguri}, {Bayliss}, {Dahle}, {Sharon},
  {Gladders}, {Natarajan}, {Hennawi}, \& {Koester}}]{Oguri12}
{Oguri}, M., {Bayliss}, M.~B., {Dahle}, H., {et~al.} 2012, \mnras, 420, 3213

\bibitem[{{Oguri} {et~al.}(2018){Oguri}, {Lin}, {Lin}, {Nishizawa}, {More},
  {More}, {Hsieh}, {Medezinski}, {Miyatake}, {Jian}, {Lin}, {Takada}, {Okabe},
  {Speagle}, {Coupon}, {Leauthaud}, {Lupton}, {Miyazaki}, {Price}, {Tanaka},
  {Chiu}, {Komiyama}, {Okura}, {Tanaka}, \& {Usuda}}]{Oguri18}
{Oguri}, M., {Lin}, Y.-T., {Lin}, S.-C., {et~al.} 2018, \pasj, 70, S20

\bibitem[{{Okabe} {et~al.}(2015){Okabe}, {Akamatsu}, {Kakuwa}, {Fujita},
  {Zhang}, {Tanaka}, \& {Umetsu}}]{Okabe15b}
{Okabe}, N., {Akamatsu}, H., {Kakuwa}, J., {et~al.} 2015, \pasj

\bibitem[{{Okabe} {et~al.}(2011){Okabe}, {Bourdin}, {Mazzotta}, \&
  {Maurogordato}}]{Okabe11}
{Okabe}, N., {Bourdin}, H., {Mazzotta}, P., \& {Maurogordato}, S. 2011, \apj,
  741, 116

\bibitem[{{Okabe} {et~al.}(2014){Okabe}, {Futamase}, {Kajisawa}, \&
  {Kuroshima}}]{Okabe14a}
{Okabe}, N., {Futamase}, T., {Kajisawa}, M., \& {Kuroshima}, R. 2014, \apj,
  784, 90

\bibitem[{{Okabe} \& {Smith}(2016)}]{Okabe16b}
{Okabe}, N., \& {Smith}, G.~P. 2016, \mnras, 461, 3794

\bibitem[{{Okabe} {et~al.}(2013){Okabe}, {Smith}, {Umetsu}, {Takada}, \&
  {Futamase}}]{Okabe13}
{Okabe}, N., {Smith}, G.~P., {Umetsu}, K., {Takada}, M., \& {Futamase}, T.
  2013, \apjl, 769, L35

\bibitem[{{Okabe} \& {Umetsu}(2008)}]{Okabe08}
{Okabe}, N., \& {Umetsu}, K. 2008, \pasj, 60, 345

\bibitem[{{Okabe} {et~al.}(2010){Okabe}, {Zhang}, {Finoguenov}, {Takada},
  {Smith}, {Umetsu}, \& {Futamase}}]{Okabe10c}
{Okabe}, N., {Zhang}, Y.-Y., {Finoguenov}, A., {et~al.} 2010, \apj, 721, 875

\bibitem[{{Owers} {et~al.}(2013){Owers}, {Baldry}, {Bauer}, {Bland-Hawthorn},
  {Brown}, {Cluver}, {Colless}, {Driver}, {Edge}, {Hopkins}, {van Kampen},
  {Lara-Lopez}, {Liske}, {Loveday}, {Pimbblet}, {Ponman}, \&
  {Robotham}}]{Owers13}
{Owers}, M.~S., {Baldry}, I.~K., {Bauer}, A.~E., {et~al.} 2013, \apj, 772, 104

\bibitem[{{Owers} {et~al.}(2014){Owers}, {Nulsen}, {Couch}, {Ma}, {David},
  {Forman}, {Hopkins}, {Jones}, \& {van Weeren}}]{Owers14}
{Owers}, M.~S., {Nulsen}, P.~E.~J., {Couch}, W.~J., {et~al.} 2014, \apj, 780,
  163

\bibitem[{{Piffaretti} {et~al.}(2011){Piffaretti}, {Arnaud}, {Pratt},
  {Pointecouteau}, \& {Melin}}]{Piffaretti11}
{Piffaretti}, R., {Arnaud}, M., {Pratt}, G.~W., {Pointecouteau}, E., \&
  {Melin}, J.-B. 2011, \aap, 534, A109

\bibitem[{{Planck Collaboration} {et~al.}(2011){Planck Collaboration},
  {Aghanim}, {Arnaud}, {Ashdown}, {Aumont}, {Baccigalupi}, {Balbi}, {Banday},
  {Barreiro}, {Bartelmann}, \& et~al.}]{Planck11SZE}
{Planck Collaboration}, {Aghanim}, N., {Arnaud}, M., {et~al.} 2011, \aap, 536,
  A12

\bibitem[{{Planck Collaboration} {et~al.}(2016){Planck Collaboration},
  {Aghanim}, {Arnaud}, {Ashdown}, {Aumont}, {Baccigalupi}, {Banday},
  {Barreiro}, {Bartlett}, {Bartolo}, \& et~al.}]{Planck16ymap}
---. 2016, \aap, 594, A22

\bibitem[{{Ragagnin} {et~al.}(2018){Ragagnin}, {Dolag}, {Moscardini},
  {Biviano}, \& {D'Onofrio}}]{Ragagnin18}
{Ragagnin}, A., {Dolag}, K., {Moscardini}, L., {Biviano}, A., \& {D'Onofrio},
  M. 2018, arXiv:1810.08212

\bibitem[{{Remazeilles} {et~al.}(2013){Remazeilles}, {Aghanim}, \&
  {Douspis}}]{Remazeilles13}
{Remazeilles}, M., {Aghanim}, N., \& {Douspis}, M. 2013, \mnras, 430, 370

\bibitem[{{Ricker} \& {Sarazin}(2001)}]{Ricker01}
{Ricker}, P.~M., \& {Sarazin}, C.~L. 2001, \apj, 561, 621

\bibitem[{{Roettiger} {et~al.}(1999){Roettiger}, {Burns}, \&
  {Stone}}]{Roettiger99}
{Roettiger}, K., {Burns}, J.~O., \& {Stone}, J.~M. 1999, \apj, 518, 603

\bibitem[{{Ruggiero} \& {Lima Neto}(2017)}]{Ruggiero17}
{Ruggiero}, R., \& {Lima Neto}, G.~B. 2017, \mnras, 468, 4107

\bibitem[{{Russell} {et~al.}(2011){Russell}, {van Weeren}, {Edge}, {McNamara},
  {Sanders}, {Fabian}, {Baum}, {Canning}, {Donahue}, \& {O'Dea}}]{Russell11}
{Russell}, H.~R., {van Weeren}, R.~J., {Edge}, A.~C., {et~al.} 2011, \mnras,
  417, L1

\bibitem[{{Russell} {et~al.}(2012){Russell}, {McNamara}, {Sanders}, {Fabian},
  {Nulsen}, {Canning}, {Baum}, {Donahue}, {Edge}, {King}, \&
  {O'Dea}}]{Russell12}
{Russell}, H.~R., {McNamara}, B.~R., {Sanders}, J.~S., {et~al.} 2012, \mnras,
  423, 236

\bibitem[{{Rykoff} {et~al.}(2008){Rykoff}, {Evrard}, {McKay}, {Becker},
  {Johnston}, {Koester}, {Nord}, {Rozo}, {Sheldon}, {Stanek}, \&
  {Wechsler}}]{Rykoff08}
{Rykoff}, E.~S., {Evrard}, A.~E., {McKay}, T.~A., {et~al.} 2008, \mnras, 387,
  L28

\bibitem[{{Sarazin}(2002)}]{Sarazin02}
{Sarazin}, C.~L. 2002, in Astrophysics and Space Science Library, Vol. 272,
  Merging Processes in Galaxy Clusters, ed. L.~{Feretti}, I.~M. {Gioia}, \&
  G.~{Giovannini}, 1--38

\bibitem[{{Sarazin} {et~al.}(2016){Sarazin}, {Finoguenov}, {Wik}, \&
  {Clarke}}]{Sarazin16}
{Sarazin}, C.~L., {Finoguenov}, A., {Wik}, D.~R., \& {Clarke}, T.~E. 2016,
  ArXiv:160607433

\bibitem[{{Schneider} {et~al.}(1998){Schneider}, {van Waerbeke}, {Jain}, \&
  {Kruse}}]{Schneider98}
{Schneider}, P., {van Waerbeke}, L., {Jain}, B., \& {Kruse}, G. 1998, \mnras,
  296, 873

\bibitem[{{Shaw} {et~al.}(2008){Shaw}, {Holder}, \& {Bode}}]{Shaw08}
{Shaw}, L.~D., {Holder}, G.~P., \& {Bode}, P. 2008, \apj, 686, 206

\bibitem[{{Shimwell} {et~al.}(2017){Shimwell}, {R{\"o}ttgering}, {Best},
  {Williams}, {Dijkema}, {de Gasperin}, {Hardcastle}, {Heald}, {Hoang},
  {Horneffer}, {Intema}, {Mahony}, {Mandal}, {Mechev}, {Morabito}, {Oonk},
  {Rafferty}, {Retana-Montenegro}, {Sabater}, {Tasse}, {van Weeren},
  {Br{\"u}ggen}, {Brunetti}, {Chy{\.z}y}, {Conway}, {Haverkorn}, {Jackson},
  {Jarvis}, {McKean}, {Miley}, {Morganti}, {White}, {Wise}, {van Bemmel},
  {Beck}, {Brienza}, {Bonafede}, {Calistro Rivera}, {Cassano}, {Clarke},
  {Cseh}, {Deller}, {Drabent}, {van Driel}, {Engels}, {Falcke}, {Ferrari},
  {Fr{\"o}hlich}, {Garrett}, {Harwood}, {Heesen}, {Hoeft}, {Horellou},
  {Israel}, {Kapi{\'n}ska}, {Kunert-Bajraszewska}, {McKay}, {Mohan},
  {Orr{\'u}}, {Pizzo}, {Prandoni}, {Schwarz}, {Shulevski}, {Sipior}, {Smith},
  {Sridhar}, {Steinmetz}, {Stroe}, {Varenius}, {van der Werf}, {Zensus}, \&
  {Zwart}}]{Shimwell17}
{Shimwell}, T.~W., {R{\"o}ttgering}, H.~J.~A., {Best}, P.~N., {et~al.} 2017,
  \aap, 598, A104

\bibitem[{{Shimwell} {et~al.}(2018){Shimwell}, {Tasse}, {Hardcastle}, {Mechev},
  {Williams}, {Best}, {R{\"o}ttgering}, {Callingham}, {Dijkema}, {de Gasperin},
  {Hoang}, {Hugo}, {Mirmont}, {Oonk}, {Prandoni}, {Rafferty}, {Sabater},
  {Smirnov}, {van Weeren}, {White}, {Atemkeng}, {Bester}, {Bonnassieux},
  {Br{\"u}ggen}, {Brunetti}, {Chy{\.z}y}, {Cochrane}, {Conway}, {Croston},
  {Danezi}, {Duncan}, {Haverkorn}, {Heald}, {Iacobelli}, {Intema}, {Jackson},
  {Jamrozy}, {Jarvis}, {Lakhoo}, {Mevius}, {Miley}, {Morabito}, {Morganti},
  {Nisbet}, {Orr{\'u}}, {Perkins}, {Pizzo}, {Schrijvers}, {Smith}, {Vermeulen},
  {Wise}, {Alegre}, {Bacon}, {van Bemmel}, {Beswick}, {Bonafede}, {Botteon},
  {Bourke}, {Brienza}, {Calistro Rivera}, {Cassano}, {Clarke}, {Conselice},
  {Dettmar}, {Drabent}, {Dumba}, {Emig}, {En{\ss}lin}, {Ferrari}, {Garrett},
  {G{\'e}nova-Santos}, {Goyal}, {G{\"u}rkan}, {Hale}, {Harwood}, {Heesen},
  {Hoeft}, {Horellou}, {Jackson}, {Kokotanekov}, {Kondapally},
  {Kunert-Bajraszewska}, {Mahatma}, {Mahony}, {Mandal}, {McKean}, {Merloni},
  {Mingo}, {Miskolczi}, {Mooney}, {Nikiel-Wroczy{\'n}ski}, {O'Sullivan},
  {Quinn}, {Reich}, {Roskowi{\'n}ski}, {Rowlinson}, {Savini}, {Saxena},
  {Schwarz}, {Shulevski}, {Sridhar}, {Stacey}, {Urquhart}, {van der Wiel},
  {Varenius}, {Webster}, \& {Wilber}}]{LoTSS18_1}
{Shimwell}, T.~W., {Tasse}, C., {Hardcastle}, M.~J., {et~al.} 2018, ArXiv
  e-prints

\bibitem[{{Stroe} {et~al.}(2017){Stroe}, {Sobral}, {Paulino-Afonso}, {Alegre},
  {Calhau}, {Santos}, \& {van Weeren}}]{Stroe17}
{Stroe}, A., {Sobral}, D., {Paulino-Afonso}, A., {et~al.} 2017, \mnras, 465,
  2916

\bibitem[{{Stroe} {et~al.}(2015){Stroe}, {Sobral}, {Dawson}, {Jee}, {Hoekstra},
  {Wittman}, {van Weeren}, {Br{\"u}ggen}, \& {R{\"o}ttgering}}]{Stroe15b}
{Stroe}, A., {Sobral}, D., {Dawson}, W., {et~al.} 2015, \mnras, 450, 646

\bibitem[{{Takizawa} \& {Naito}(2000)}]{Takizawa00}
{Takizawa}, M., \& {Naito}, T. 2000, \apj, 535, 586

\bibitem[{{Tanaka} {et~al.}(2018){Tanaka}, {Coupon}, {Hsieh}, {Mineo},
  {Nishizawa}, {Speagle}, {Furusawa}, {Miyazaki}, \& {Murayama}}]{HSCPhotoz17}
{Tanaka}, M., {Coupon}, J., {Hsieh}, B.-C., {et~al.} 2018, \pasj, 70, S9

\bibitem[{{Taylor} \& {Babul}(2005)}]{Taylor05b}
{Taylor}, J.~E., \& {Babul}, A. 2005, \mnras, 364, 515

\bibitem[{{Thompson} \& {Nagamine}(2012)}]{Thompson12}
{Thompson}, R., \& {Nagamine}, K. 2012, \mnras, 419, 3560

\bibitem[{{Tinker} {et~al.}(2008){Tinker}, {Kravtsov}, {Klypin}, {Abazajian},
  {Warren}, {Yepes}, {Gottl{\"o}ber}, \& {Holz}}]{Tinker08}
{Tinker}, J., {Kravtsov}, A.~V., {Klypin}, A., {et~al.} 2008, \apj, 688, 709

\bibitem[{{Tormen} {et~al.}(2004){Tormen}, {Moscardini}, \&
  {Yoshida}}]{Tormen04}
{Tormen}, G., {Moscardini}, L., \& {Yoshida}, N. 2004, \mnras, 350, 1397

\bibitem[{{Truemper}(1982)}]{Truemper82}
{Truemper}, J. 1982, Advances in Space Research, 2, 241

\bibitem[{{Umetsu} {et~al.}(2014){Umetsu}, {Medezinski}, {Nonino}, {Merten},
  {Postman}, {Meneghetti}, {Donahue}, {Czakon}, {Molino}, {Seitz}, {Gruen},
  {Lemze}, {Balestra}, {Ben{\'{\i}}tez}, {Biviano}, {Broadhurst}, {Ford},
  {Grillo}, {Koekemoer}, {Melchior}, {Mercurio}, {Moustakas}, {Rosati}, \&
  {Zitrin}}]{Umetsu14}
{Umetsu}, K., {Medezinski}, E., {Nonino}, M., {et~al.} 2014, \apj, 795, 163

\bibitem[{{van Weeren} {et~al.}(2011){van Weeren}, {Br{\"u}ggen},
  {R{\"o}ttgering}, {Hoeft}, {Nuza}, \& {Intema}}]{vanWeeren11b}
{van Weeren}, R.~J., {Br{\"u}ggen}, M., {R{\"o}ttgering}, H.~J.~A., {et~al.}
  2011, \aap, 533, A35

\bibitem[{{van Weeren} {et~al.}(2010){van Weeren}, {R{\"o}ttgering},
  {Br{\"u}ggen}, \& {Hoeft}}]{vanWeeren10}
{van Weeren}, R.~J., {R{\"o}ttgering}, H.~J.~A., {Br{\"u}ggen}, M., \& {Hoeft},
  M. 2010, Science, 330, 347

\bibitem[{{van Weeren} {et~al.}(2017){van Weeren}, {Andrade-Santos}, {Dawson},
  {Golovich}, {Lal}, {Kang}, {Ryu}, {Br{\`i}ggen}, {Ogrean}, {Forman}, {Jones},
  {Placco}, {Santucci}, {Wittman}, {Jee}, {Kraft}, {Sobral}, {Stroe}, \&
  {Fogarty}}]{vanWeeren17}
{van Weeren}, R.~J., {Andrade-Santos}, F., {Dawson}, W.~A., {et~al.} 2017,
  Nature Astronomy, 1, 0005

\bibitem[{{Venturi} {et~al.}(2007){Venturi}, {Giacintucci}, {Brunetti},
  {Cassano}, {Bardelli}, {Dallacasa}, \& {Setti}}]{Venturi07}
{Venturi}, T., {Giacintucci}, S., {Brunetti}, G., {et~al.} 2007, \aap, 463, 937

\bibitem[{{Venturi} {et~al.}(2008){Venturi}, {Giacintucci}, {Dallacasa},
  {Cassano}, {Brunetti}, {Bardelli}, \& {Setti}}]{Venturi08}
{Venturi}, T., {Giacintucci}, S., {Dallacasa}, D., {et~al.} 2008, \aap, 484,
  327

\bibitem[{{Vikhlinin} {et~al.}(2009){Vikhlinin}, {Burenin}, {Ebeling},
  {Forman}, {Hornstrup}, {Jones}, {Kravtsov}, {Murray}, {Nagai}, {Quintana}, \&
  {Voevodkin}}]{Vikhlinin09a}
{Vikhlinin}, A., {Burenin}, R.~A., {Ebeling}, H., {et~al.} 2009, \apj, 692,
  1033

\bibitem[{{Wen} {et~al.}(2012){Wen}, {Han}, \& {Liu}}]{Wen12}
{Wen}, Z.~L., {Han}, J.~L., \& {Liu}, F.~S. 2012, \apjs, 199, 34

\bibitem[{{Yang} {et~al.}(2006){Yang}, {Mo}, {van den Bosch}, {Jing},
  {Weinmann}, \& {Meneghetti}}]{Yang06}
{Yang}, X., {Mo}, H.~J., {van den Bosch}, F.~C., {et~al.} 2006, \mnras, 373,
  1159

\bibitem[{{Yu} {et~al.}(2015){Yu}, {Nelson}, \& {Nagai}}]{Yu15}
{Yu}, L., {Nelson}, K., \& {Nagai}, D. 2015, \apj, 807, 12

\bibitem[{{ZuHone}(2011)}]{ZuHone11}
{ZuHone}, J.~A. 2011, \apj, 728, 54

\end{thebibliography}

\end{document}